\documentclass[12pt,preprint]{aastex}

\begin{document}
\shorttitle{Membership of the Orion Nebula population from COUP}
\shortauthors{Getman et al.} \slugcomment{Accepted for publication
in the Astrophysical Journal Supplement, the COUP Special Issue}

\title{Membership of the Orion Nebula population from the Chandra
Orion Ultradeep Project}

\author{Konstantin V.\ Getman\altaffilmark{1}, Eric D.\ Feigelson\altaffilmark{1},
Nicolas Grosso\altaffilmark{2}, Mark J.\
McCaughrean\altaffilmark{3,4}, Giusi Micela\altaffilmark{5},
Patrick Broos\altaffilmark{1}, Gordon Garmire\altaffilmark{1},
Leisa Townsley\altaffilmark{1}}

\altaffiltext{1}{Department of Astronomy \& Astrophysics, 525 Davey
Laboratory, Pennsylvania State University, University Park PA 16802}

\altaffiltext{2}{Laboratoire d'Astrophysique de Grenoble,
Universit\'e Joseph Fourier, BP 53, 38041 Grenoble Cedex 9,
France}

\altaffiltext{3}{University of Exeter, School of Physics, Stocker
Road, Exeter EX4 4QL, Devon, UK}

\altaffiltext{4}{Astrophysikalisches Institut Potsdam, An der
Sternwarte 16, D-14482 Potsdam, Germany}

\altaffiltext{5}{INAF, Osservatorio Astronomico di Palermo G. S.
Vaiana, Piazza del Parlamento 1, I-90134 Palermo, Italy}

\begin{abstract}

The $Chandra$ Orion Ultradeep project (COUP) observation described
in a companion paper by Getman et al. provides an exceptionally
deep X-ray survey of the Orion Nebula Cluster and associated
embedded young stellar objects. Membership of the region is
important for studies of the stellar IMF, cluster dynamics, and
star formation. The COUP study detected 1616 X-ray sources.

In this study we confirm cloud membership for 1315 stars, identify
16 probable foreground field stars having optical counterparts
with discrepant proper motions, and classify the remaining 285
X-ray sources, of which 51 are lightly and 234 heavily obscured.

The 51 lightly obscured sources without known counterparts fall
into three groups. (i) Sixteen are likely new members of the Orion
Nebula Cluster. (ii) Two with unusually soft and non-flaring X-ray
emission appear to be associated with nebular shocks, and may be
new examples of X-rays produced at the bow shocks of Herbig-Haro
outflows. (iii) The remaining thirty three are very weak uncertain
sources, possibly spurious.

Out of 234 heavily absorbed sources without optical or
near-infrared counterparts 75 COUP sources are likely new embedded
cloud members (with membership for 42 confirmed by powerful X-ray
flares), and the remaining 159 are likely extragalactic AGN seen
through the molecular cloud, as argued by a careful simulation of
the extragalactic background population.

Finally, a few new binary companions to Orion stars may have been
found, but most cases of proximate COUP sources can be attributed
to chance superpositions in this crowded field.

\end{abstract}


\section{Introduction}

The Orion Nebula Cluster (ONC), along with the Pleiades and
Hyades, has served as the fundamental calibrator and prototype for
young stellar clusters \citep{ODell01}.  They are critical for
understanding early stellar evolution, the stellar Initial Mass
Function, the dynamical history of clusters and multiple star
systems, and the origins of planetary systems.  The ONC has the
best-characterized stellar population with age around 1 Myr
\citep{ODell01}, and great efforts have been made to determine the
cluster membership from its most massive member, the $\simeq 45$
M$_\odot$ O7 star $\theta^1$ Ori C, to substellar objects down to
several Jupiter masses. Behind the ONC on the same line of sight
lies the stellar population of the first Orion Molecular Core,
OMC-1, which hosts the Orion hot molecular core and the nearest
site of massive star formation \citep{Kurtz00}. The stellar
population of the ONC has been carefully identified by studies of
proper motion \citep{Jones88}, optical spectra
\citep{Hillenbrand97}, optical imaging \citep[e.g.]{Prosser94,
Bally00}, infrared photometry \citep[e.g.][McCaughrean et al.\, in
preparation]{Hillenbrand00, Lada04}, infrared spectroscopy
\citep{Slesnick04}, and radio continuum emission \citep{Zapata04}.
Over 2000 ONC members, and a score of embedded OMC-1 sources, are
catalogued.

High-sensitivity and high-resolution X-ray surveys obtained with
the $Chandra$ X-ray Observatory are also effective in detecting a
large fraction of the Orion population, both in the lightly
absorbed ONC and the heavily absorbed OMC-1
populations\footnote{Other X-ray surveys are mentioned in the
introduction to the first paper \citep{Getman05}.}. X-ray surveys
are complementary to optical and infrared surveys because they
trace magnetic activity (mainly plasma heated in violent magnetic
reconnection flares) rather than photospheric or circumstellar
disk blackbody emission. The survey methods thus have different
sensitivities; for example, $Chandra$ COUP sources can penetrate
up to $A_V \simeq 500$ mag into the cloud, considerably deeper
than existing near-infrared surveys in the $JHKL$ bands. $Chandra$
is also often effective in resolving multiple systems on arcsecond
scales, as the mirror/detector system provide excellent dynamic
range. Surveys in different bands are also subject to different
types of contamination. Foreground and background Galactic stars
confuse optical and near-infrared (ONIR) studies, but have much
less impact on X-ray studies, as magnetic activity in pre-main
sequence (PMS) stars is elevated $10^{1}-10^{4}$ above main
sequence levels \citep{Preibisch05a}. X-ray studies are thus
particularly effective in uncovering heavily obscured low-mass
cloud populations and in discriminating cloud pre-main sequence
populations from unrelated older stars. However, it is more
difficult to remove extragalactic active galactic nuclei (AGN)
from high-sensitivity X-ray images - their X-ray properties can be
similar to those of absorbed PMS stars, and their number rapidly
grows with increasing sensitivity of X-ray images. We carefully
treat this problem here.

Two half-day exposures on the Orion Nebula obtained in 1999-2000
revealed 1075 X-ray sources; 974 were associated with known Orion
stars while 101 were unidentified X-ray sources, most of which
were suspected new Orion members \citep{Feigelson02}. A $\simeq
10$ day $Chandra$ exposure of the Orion Nebula was obtained in
January 2003, providing the basis for the $Chandra$ Orion
Ultradeep Project (\dataset[ADS/Sa.CXO#obs/COUP]{COUP}). The
observations, data processing, source detection, photon
extraction, and basic source properties are described by
\citet{Getman05}. We discuss here the COUP findings with respect
to the Orion stellar population.

One principal result is immediately evident: with $\simeq 10$
times the sensitivity of the earlier exposure, only 50\% more
sources are seen (1616 $vs.$ 1075).  We find few new lightly
obscured members of the ONC, confirming the near-completeness of
earlier membership catalogs.  COUP contributions to Orion
population studies mainly involve heavily absorbed stars which
must be carefully discriminated from background AGN based on their
variability and spatial distributions (\S
\ref{unidentified_section}). A few sources may be non-stellar
X-ray emission regions arising from shocks in Herbig-Haro outflows
(\S \ref{HH_srcs.sec}). Double sources, a few of which may be
physically related multiple star systems, are presented in \S
\ref{x_ray binaries_section}.  Stellar contaminants unrelated to
the Orion cloud are treated in \S \ref{field_stars_section}. A
distance of 450 pc to the Orion Nebula is assumed throughout
\citep{ODell01}.

\section{Discriminating new cloud members from extragalactic
contaminants \label{unidentified_section}}

With the extremely deep net exposure of 838 ks, the COUP field
undoubtedly suffers significant contamination by AGN and other
extragalactic sources.  From megasecond exposures of high-Galactic
latitude region \citep[e.g.][]{Brandt01}, we can estimate that
$\simeq 500-600$ extragalactic sources would be detectable in the
COUP field in the hard $2-8$ keV band which penetrates most cloud
column densities. Thus, ideally without taking into consideration
the molecular cloud distribution and even more importantly
ignoring the effects of the complex elevated COUP instrumental
background, about 30-40\% of the 1616 COUP sources are expected to
be extragalactic contaminants. This is a much greater fraction
than the $\leq2$\% contamination estimated for the shorter
1999-2000 $Chandra$ exposures \citep{Feigelson02, Flaccomio03}
because the extragalactic source density climbs as $\log N \propto
\log S^{-\alpha}$ where $1.0 < \alpha < 1.5$ while the ONC X-ray
luminosity function is steeply declining below $\log L_t \sim 29$
erg s$^{-1}$ \citep{Feigelson02}.

Here we apply a more sophisticated analysis of the extragalactic
contamination than these rough estimates. First, the ONIR
counterparts of Orion stars are nearly always brighter than the
counterparts to extragalactic sources.  Second, prior studies of
the variability characteristics of AGN and Orion population stars
show that only the latter exhibit high-amplitude fast-rise flares.
This criterion is more effective in the COUP study than in other
$Chandra$ observations of young stellar clusters because the long
exposure improves the chances of capturing stellar flares. Third,
we construct realistic Monte Carlo simulations of the
extragalactic source surface density and flux distribution as a
function of location in the COUP field, taking into account
spectral properties of the AGN, spatial variations in molecular
cloud absorption and the variations in detector background
peculiar to the COUP field. The resulting predicted extragalactic
populations are compared to COUP sources without ONIR
counterparts.

\subsection{Candidate new members \label{candidates_section}}

Our consideration of extragalactic contaminants and new cloud
members is restricted to the 285 COUP sources without known ONIR
counterparts \citep[see Tables 9,10][]{Getman05}\footnote{In
\citet{Getman05} the number of sources without known ONIR
counterparts is actually 273, not 285. Five of the discrepant
sources are explained here in \S \ref{additional
binaries_section}. For the remaining seven COUP sources
(\object[COUP 0356]{356}, \object[COUP 0577]{577}, \object[COUP
0598]{598}, \object[COUP 0635]{635}, \object[COUP 0703]{703},
\object[COUP 0704]{704}, and \object[COUP 0748]{748}), their 2MASS
registrations within the VLT FOV reported in \citet{Getman05} were
incorrect. As confirmed by VLT data, either 2MASS-Chandra
separations are greater than 0.8\arcsec and thus not considered to
be true coincidences, or the 2MASS detection algorithm peaked on
bright knots in the nebular emission rather than true stars.}.
These 285 sources are listed in
Tables~\ref{flaring_table}-\ref{non_flaring_table} and their
locations are plotted in Figure \ref{spat_distrib_fig}. The first
nine columns of the tables give information extracted from the
COUP source tables of \citet{Getman05}; see table notes for
details. If the source lies in the Orion Nebula, a flux of $F_h =
1 \times 10^{-15}$ erg s$^{-1}$ cm$^{-2}$ corresponds to a
hard-band luminosity of $L_h = 2.3 \times 10^{28}$ erg s$^{-1}$.
Column 10 gives the column density through the molecular cloud
derived from the velocity-integrated intensity of $^{13}$CO from
the single-dish map obtained by \citet{Bally87}. The $^{13}$CO
species is optically thin in most directions of the OMC
\citep{Bally87}, and therefore its brightness is a probe of the
column density of molecular gas through the entire cloud. This map
is shown in the background of Figure \ref{spat_distrib_fig}. The
hydrogen column is estimated by the approximate correspondence of
$A_{V} \sim 5$ mag to 5 K km s$^{-1}$ and the conversion $\log N_H
= 21.2 + \log A_V$ cm$^{-2}$ \citep{Vuong03}\footnote{Vuong's
relation, the latest published conversion between $\log N_H$ and
$A_V$, is based primarily on data from $\rho$ Oph cloud with only
six data points from Orion, of which five have only moderate
extinctions of $\log N_H < 21.8$ cm$^{-2}$. This relation has been
extrapolated here to higher column densities. Examination of $\log
N_H - A_V$ relation up to $A_V \sim 200$ mag, gas-to-dust ratio,
and depletion effects is a future planned effort of the COUP
project.}.

Column 11 gives our proposed membership classification of these
sources without ONIR counterparts as follows:
\begin{description}

\item [OMC = Orion Molecular Core]  These are 42 COUP sources
without ONIR counterparts which exhibit one or more high-amplitude
X-ray flares, a property characteristic of Orion cloud members but
not extragalactic AGNs . These sources also exhibit two other
properties not directly selected: they are all heavily absorbed
($\log N_H \geq 22.0$ cm$^{-2}$, and almost all are spatially
associated with the dense OMC-1 cores, the dense molecular
filament which extends northwards from OMC-1 to OMC-2/3. We
therefore classify these stars as members of the OMC, and list
them separately in Table \ref{flaring_table}.

\item [EG = Extragalactic] There are 192 heavily absorbed sources
without ONIR counterparts which do not exhibit flares (Table
\ref{non_flaring_table}). We establish in \S
\ref{EG_simul_section} that most, but not all, of these are
extragalactic sources. By comparing the observed and predicted
local surface densities of these sources, in light of the
requirement that extragalactic sources will be isotropically
distributed, we provide in column 12 an estimated probability of
membership.   We label 33 of these sources `OMC or EG?'.  Most of
these are probably new obscured Orion cloud members leaving 159
probably extragalactic (`EG') sources.

\item [ONC = Orion Nebula Cluster]  There are 18 sources
without ONIR counterparts which exhibit lower X-ray absorptions
than expected if they lie in or beyond the molecular cloud. None
of these exhibit X-ray flares. The nature of these sources is
uncertain and is discussed in \S \ref{non_flaring
unabsorbed_section}. Two of them may be associated with
Herbig-Haro outflows from Orion protostars, while the other 16 may
be new very low mass members of the ONC.

\item [HH = Herbig-Haro objects]  COUP sees possible two
cases where the X-ray emission is unusually soft, constant and
faint, and is associated with a nebular shock rather than a star.
These are cases where the bow shocks of high velocity Herbig-Haro
ejecta are sufficiently hot to produce soft X-rays.

\item [Unc = Uncertain]  There are 33 cases of lightly absorbed
sources with $\la 20$ net counts. While the situation for these
sources individually is uncertain, we suspect that, in most cases,
the X-ray source detection is spurious and no source is actually
present.

\end{description}

We now develop this classification in more detail.

\subsection{$K_s$ band magnitude distributions}

The star-galaxy discrimination benefits greatly by their disjoint
distributions of ONIR magnitudes.  This is best shown in the $K_s$
band where the VLT has covered the inner $7.4\arcmin\ \times
7.4\arcmin$ of the Orion region to $K_s \simeq 20$ (McCaughrean et
al., in preparation) and the 2MASS survey covers the outer region
to $K_s \simeq 15$. In the ONC, the substellar limit occurs around
$K_s \simeq 15$ \citep{Muench02} and only 14 COUP sources have
counterparts with $16 < K_s < 20$ \citep{Getman05}.  In contrast,
for extragalactic X-ray populations only $\sim$1\% have
counterparts with $K_s < 18$ \citep{Barger02} and most of these
are resolved galaxies.

We are therefore confident that very few ($<$10 and possibly zero)
of the COUP sources with ONIR counterparts listed by
\citet{Getman05} are extragalactic. Nearly every COUP source with
an optical or NIR counterpart will be a true member of the ONC, an
obscured member of the background molecular cloud, or possibly a
foreground or background interloper from the Galactic field star
population.

\subsection{Flaring members}\label{flaring_members_section}

Perhaps the most definitive criterion for distinguishing young
Orion stars from background objects is the presence of strong
X-ray flares.  These typically have rise times $\leq 1$ hour and
decay times of $10-100$ hours, though a wide diversity of flare
morphologies is seen \citep{Wolk05}.  This flare emission is
attributed to plasma trapped in closed magnetic structures and
violently heated by magnetic reconnection events \citep{Favata05}.
Such X-ray flares are not expected to be seen either from older
background main sequence stars that are $10^{1}-10^{4}$ times less
powerful than Orion stars or from extragalactic AGN which
typically exhibit slow and mild variability over the $\simeq 2$
week COUP exposure time \citep{Bauer03, Paolillo04}.

\begin{figure}
\centering
\includegraphics[angle=0.,width=0.9\textwidth]{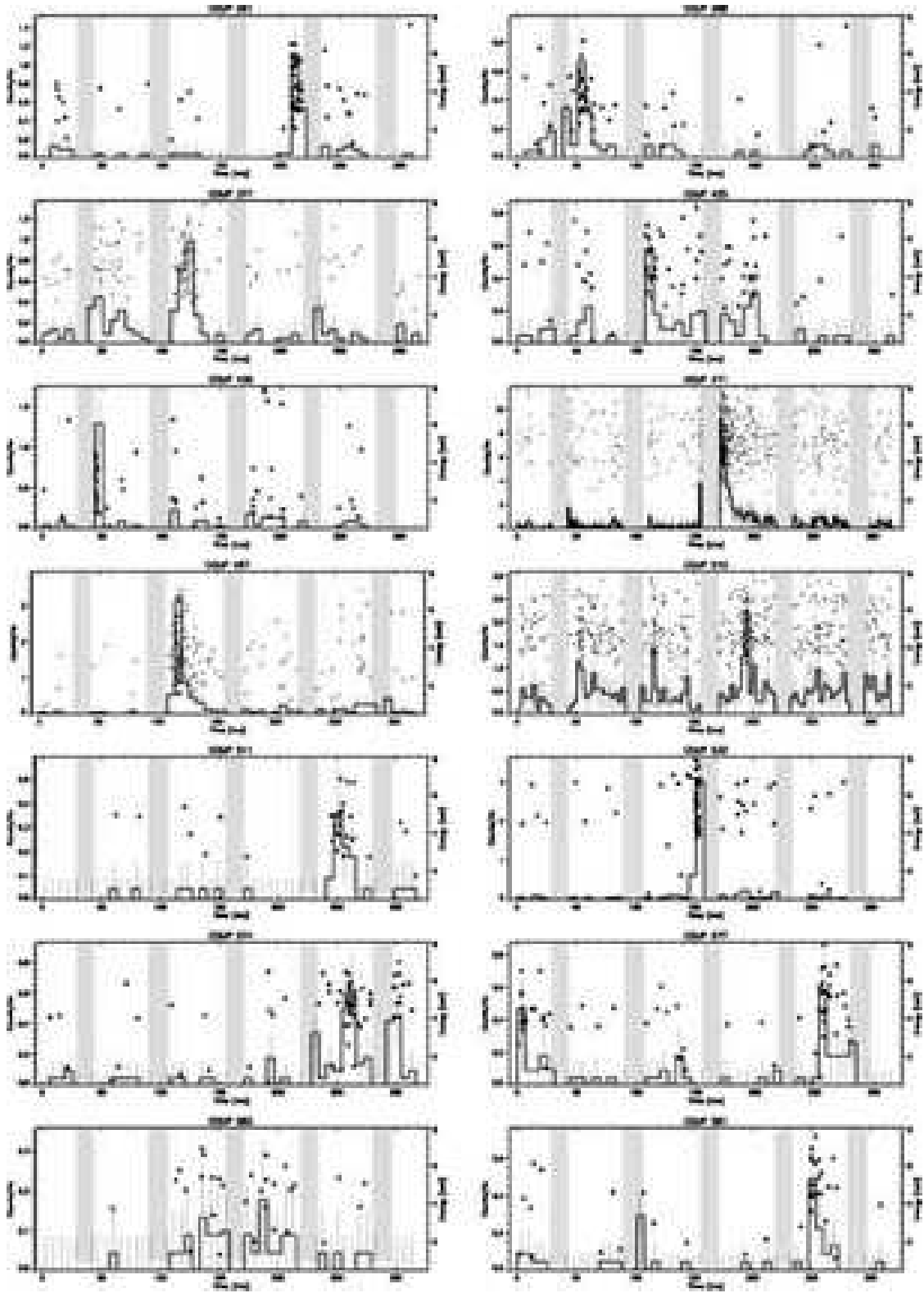}
\hspace{0.00in} \caption{Lightcurves (line) and photon arrival
diagrams (dots) for 42 X-ray flaring newly discovered members of
the Orion Nebula region. The histograms show binned count rates
(left ordinate) in the total $(0.5-8.0)$ keV energy band.  Dots
show individual photon arrival times with vertical position
indicating their photon energies (right ordinate).  Abscissae show
the time after the beginning of the COUP observation in hours.
Grey bars indicate the time gaps in the COUP exposure due to
perigees of the $Chandra$ satellite orbit.\label{40_flares_fig}}
\end{figure}

Based on results of the Kolmogorov-Smirnov and Bayesian Blocks
analysis described by \citet{Getman05}, and visual inspection of
source lightcurves and photon arrival diagrams, we find that 42 of
285 unidentified sources have strong flares characteristic of
Orion stars.  To quantify their variability, all 42 have $\log
P_{{\rm KS}} <2$ or ${\rm BBNum \geq 2}$ and all except three
(COUP \object[COUP 0582]{582}, \object[COUP 0656]{656},
\object[COUP 0678]{678}) have ${\rm BBmax/BBmin}
> 4$ \citep[Section 8]{Getman05}. These 42 sources are classified
as `OMC' (i.e. Orion Molecular Cloud members) in Table
\ref{non_flaring_table} and their lightcurves are shown in Figure
\ref{40_flares_fig}\footnote{In addition OMC membership for COUP
678 and 681 is confirmed by \citet{Grosso05} who report that they
are associated with anonymous NICMOS sources in the image of
\citet{Stolovy98}.}.

Although these sources were selected only on the basis of flaring
lightcurves, they exhibit two additional indicators of cloud
membership.  First, their spatial distribution is strongly
clustered around the dense OMC-1 cores (Figure
\ref{spat_distrib_fig}), inconsistent with the expected
distributions of unrelated stellar or extragalactic populations.
Second, the absorbing column densities derived from {\it XSPEC}
spectral fits for the 42 flaring sources (median $\log N_H \sim
23.2$ cm$^{-2}$) are very high, excluding the possibility that
they lie in front of the cloud cores.  Indeed, their X-ray
absorptions appear systematically higher than the average depth in
OMC cores (median $\log N_H(^{13}CO) \sim 22.8$ cm$^{-2}$). This
may be explained by poor resolution of $^{13}$CO maps or
inaccurate conversion between $^{13}$CO and total hydrogen columns
but, if real, may indicate that these sources are very young with
the local absorption in an envelope or disk.

\subsection{Extragalactic population simulations \label{EG_simul_section}}

\begin{figure}
\centering
\includegraphics[angle=0.,width=6.5in]{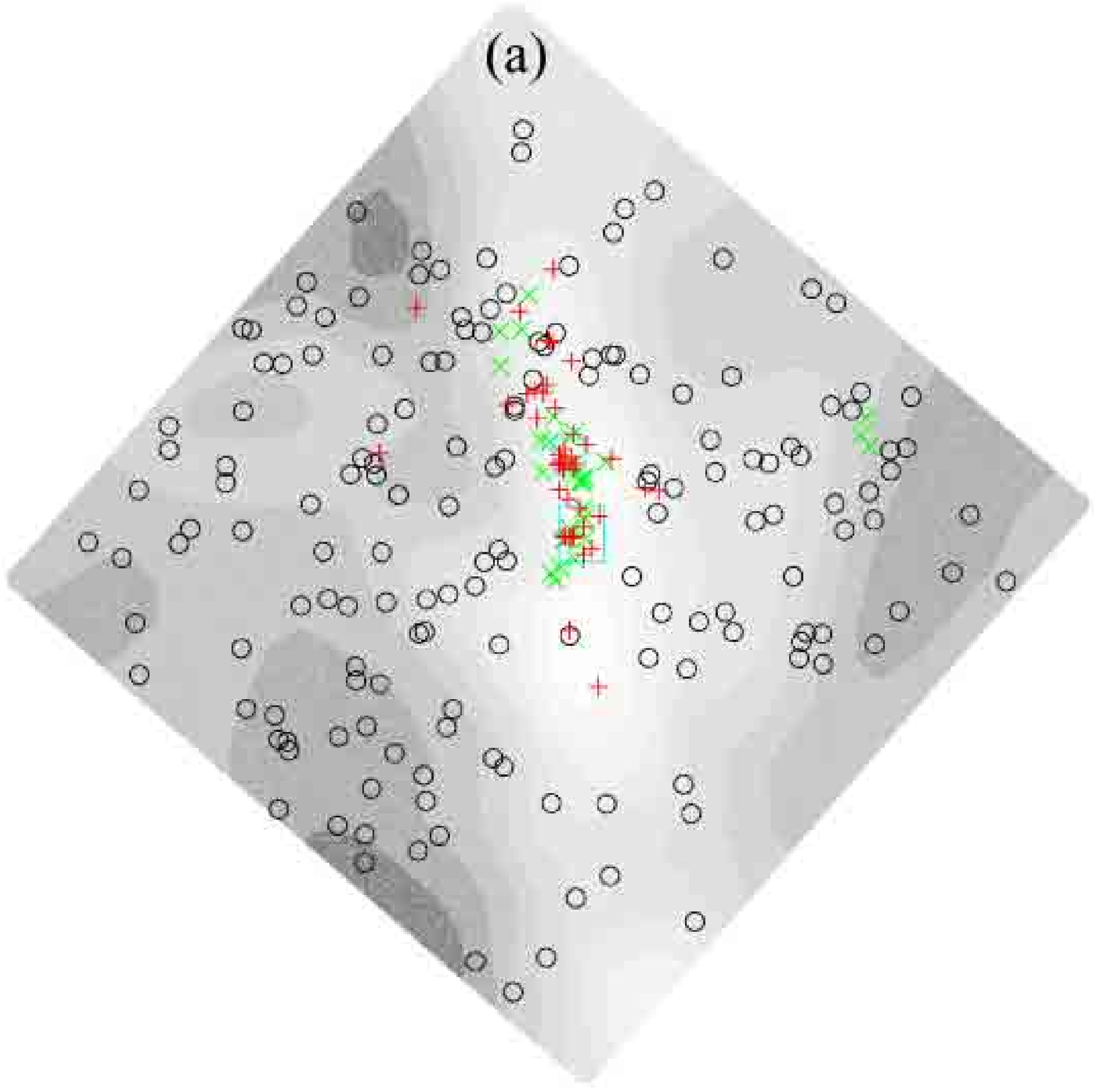}
\caption{The locations of hard (=absorbed) COUP sources without
optical or near-infrared counterparts are plotted on a map of
molecular line emission in the Orion Nebula region: (a) the full
$17\arcmin \times 17\arcmin$ COUP field centered at $(\alpha,
\delta) = (83.82101, -5.39440)$; and (b) a close-up of the
$5\arcmin \times 10\arcmin$ OMC-1 cores region. Two small cyan
boxes indicate BN/KL and OMC1-S regions discussed in detail by
\citep{Grosso05}. Forty-two flaring unidentified X-ray sources are
marked by red $+$. From 192 hard unidentified non-flaring COUP
sources, 159 classified as extragalactic are marked by black
circles, and 33 likely additional cloud members are marked by
green $\times$. Both the 42 flaring and 33 non-flaring sources
cluster around the OMC cores. \label{spat_distrib_fig}}
\end{figure}

To evaluate the expected contamination of extragalactic X-ray
sources in the COUP ACIS-I field, we constructed Monte-Carlo
simulations of the extragalactic population by placing artificial
sources randomly across the field.  The entire analysis of this
subsection is confined to the COUP hard band ($2-8$ keV). The
incident fluxes of individual simulated sources were drawn from
the observed X-ray background $\log N - \log S$ distribution
described by \citet{Moretti03}, assuming a power law photon index
distribution consistent with flux-dependencies described in Table
3 of \citet{Brandt01}. The spectrum of each individual simulated
source was convolved with absorption obtained from the $^{13}$CO
column map shown in Figure \ref{spat_distrib_fig} at the source
position. To account for the observed COUP background for each
simulated source, which is nonuniform due to the readout trails
and highly populated wings of bright ONC X-ray sources, we applied
the background noise from the location of the nearest COUP source
(reported in Table 4 of \citet{Getman05}) and removed very weak
extragalactic sources when they would not have been detected above
the local background level.  The resulting detection threshold
ranged from 3.3 to 5.0 times the root-mean-square local noise
level, representing the signal-to-noise range of the faintest COUP
sources.

\begin{figure}
\centering
\includegraphics[angle=0.,width=6.5in]{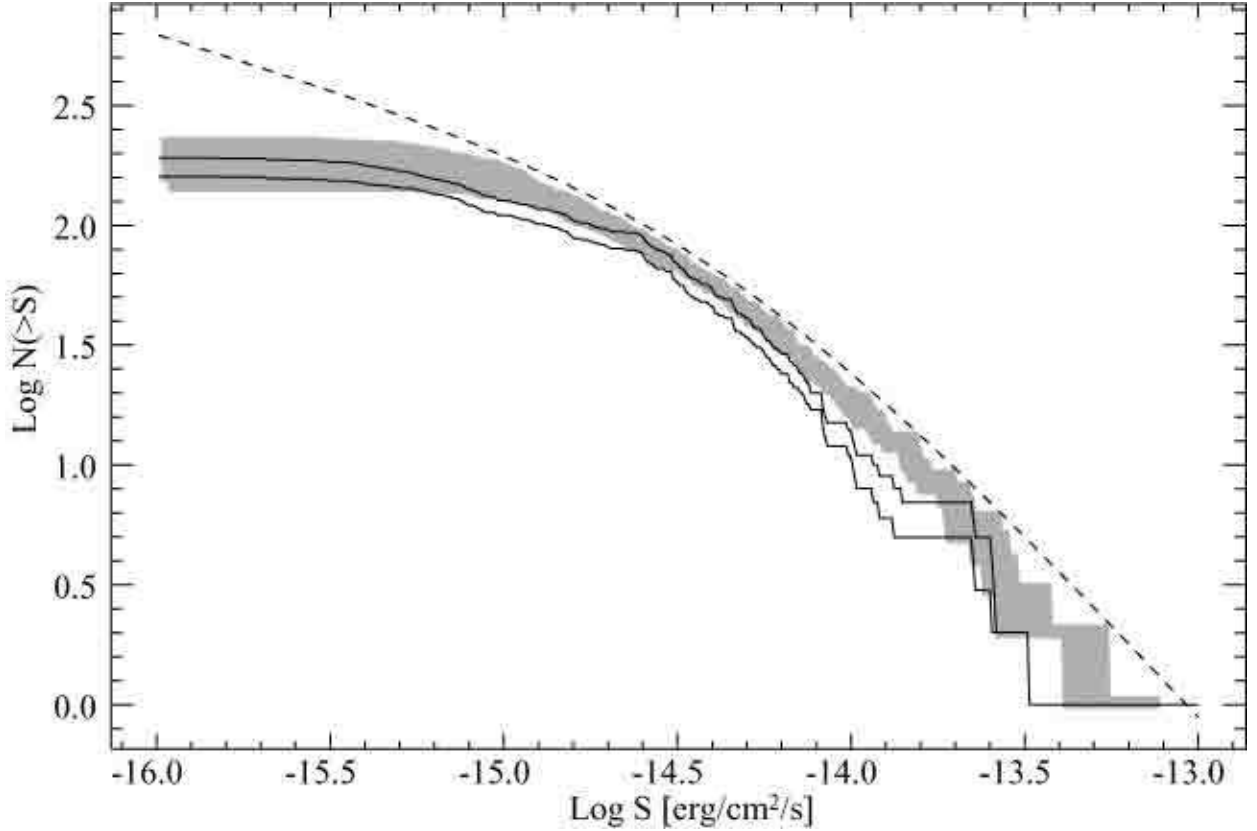}
\caption{Evaluation of extragalactic source contamination using
the X-ray source $\log N - \log S$ diagram. $N$ gives the number
of sources seen or calculated in the COUP field of view as a
function of source flux in the hard $2-8$ keV band.  Upper solid
line: Observed flux distributions of the 192 heavily absorbed
non-flaring COUP sources without ONIR counterparts (`EG' and `OMC
or EG?' sources in Table \ref{non_flaring_table}). Lower solid
line: Same for the 159 sources with low probability of cloud
membership (`EG' sources only). Dashed line: Expected unobscured
extragalactic $\log N - \log S$ distribution \citep{Moretti03}.
Grey band: Range of simulated distributions of this extragalactic
population for the COUP field corrected for source spectral
variations, spatially variable absorption by the Orion Molecular
Cloud, and variations in the COUP background.
\label{logn_logs_fig}}
\end{figure}

The results of a number of such simulations in the form of a $\log
N - \log S$ X-ray source surface density plot are shown as a grey
band in Figure \ref{logn_logs_fig}. A range of $140-240$
extragalactic sources are predicted to have been detected by the
COUP processing, which agrees well with the observed sample of 192
hard-band COUP sources without ONIR counterparts (in Table
\ref{non_flaring_table}, upper solid line in Figure
\ref{logn_logs_fig}).  Considering only the total population,
these results are consistent with no new Orion cloud members among
the 192 sources.

However, we also know the spatial distribution of both the
observed unidentified COUP sources and the predicted extragalactic
population in each simulation.  About $15$\% of the 192 sources
are clustered in the immediate vicinity of the dense OMC-1 cores
where most of the new 42 X-ray flaring members are found (Table
\ref{flaring_table}). This region has a $\simeq 4-$fold higher
concentration of unidentified sources than the rest of the COUP
field ($\sim 2-3$ vs. $0.5-0.8$ sources arcmin$^{-2}$).  This is
precisely the region where the simulated extragalactic populations
have a deficit of sources due to heavy absorption by the molecular
cloud cores.

To identify these clustered sources in an objective fashion, we
applied an adaptive kernel smoothing algorithm to both the
simulated extragalactic population and the 192 unidentified hard
COUP sources.  The kernel width was chosen to accumulate at least
10 COUP sources around each source, resulting in kernel areas
$\sim 2$ arcmin$^{2}$ in the center of the field and larger
towards the edges. We define an approximate membership probability
for each observed source $P = (N_{COUP} - N_{AGN})/N_{COUP}$,
where $N_{COUP}$ is the number of hard unidentified COUP sources
around the current COUP source within the kernel area and
$N_{AGN}$ is the number of expected simulated AGNs within the same
kernel area averaged over the simulations and assuming a
4.4$\sigma$ detection threshold.   These membership probabilities
are listed in Column 12 in Table \ref{non_flaring_table}. The
concentration around the OMC-1 cores corresponds roughly to $P
\geq 0.60$, and we use this criterion to define the `OMC or EG?'
classification in the table and Figure \ref{spat_distrib_fig}.

The $P = 0.60$ threshold is clearly {\it ad hoc} and is unlikely
to provide a clean discrimination between cloud members and
extragalactic AGN.  Nonetheless, we suspect that most of the 33
`OMC or EG?' sources are in fact new obscured Orion cloud members
similar to the `OMC' flaring sources in Table \ref{flaring_table}.
The `OMC or EG?' sources (green $\times$ symbols in Figure
\ref{spat_distrib_fig}) are also spatially clustered around the
OMC-1 cores similar to the 42 new flaring sources (red {\bf +}
symbols).  The median column density of the 33 sources is $\log
N_H \sim 23.3$ cm$^{-2}$, nearly identical to that of flaring
sources, but with fewer counts (median of 43 counts $vs.$ 114 for
the flaring sources).  The `OMC or EG?' sources are either too
weak to show flares, or happened not to exhibit a powerful flare
during the COUP observation.

In this case, the contaminating extragalactic AGN population is
restricted to be no more than 159 sources (192 minus 33 probably
OMC members) in Table \ref{non_flaring_table}.  These sources are
distributed across the field in a roughly random fashion except
for the expected avoidance of the absorbing cloud cores (Fig
\ref{spat_distrib_fig}). Their $\log N - \log S$ distribution is
reasonably consistent with the extragalactic population
simulations (Fig \ref{logn_logs_fig}).

\subsection{Lightly absorbed COUP sources without ONIR counterparts
\label{non_flaring unabsorbed_section}}

About one-fifth of the sources in Table \ref{non_flaring_table}
have X-ray absorptions that appear to be significantly below
values expected if they lie deeply embedded in or behind the Orion
molecular cloud.  A rough criterion for this would be $\log N_H <
\log N_H(^{13}CO) - 0.6$ cm$^{-2}$ where 0.6 is the typical
uncertainty on logarithm $N_H$ seen in Table 6 of
\citet{Getman05}.  We consider the 18 of these sources with $\log
N_H < 22.0$ cm$^{-2}$ to be 'lightly absorbed'; 8 of these have
$\log N_H \leq 20.0$ cm$^{-2}$ which is the lower limit of
absorption detectable with the $Chandra$ instrument.  We thus
consider these 18 sources to lie in front of (or only slightly
embedded in) the molecular cloud.  Since most of the ONC similarly
lies in front of the cloud with only light absorption
\citep{ODell01}, we tentatively classify 16 of these sources as
`ONC' members  and 2 as possible `HH' objects (see \S
\ref{HH_srcs.sec}) in Table \ref{non_flaring_table}.

However, the `ONC' class is probably not homogeneous and we
consider several separate subclasses.
\begin{enumerate}

\item Five of the sources (COUP \object[COUP 0053]{53}, \object[COUP 0135]{135}, \object[COUP 0156]{156}, \object[COUP 1033]{1033} and \object[COUP 1092]{1092}) are
weak X-ray sources with $<20$ net counts located out of the VLT
field of view. From the $L_t-L_{bol}$ relationship
\citep{Preibisch05b}, their photospheric brightness can easily be
below the $K_s \simeq 15$ 2MASS completeness limit. These are
likely new low mass members on the outskirts of the ONC.

\item Two cases may represent associations between the X-ray
position and known ONIR cluster members which were missed by the
automated source identification procedures of \citet{Getman05}.
COUP \object[COUP 0185]{185} and \object[COUP 0257]{257} are
located at edges of the COUP field so the centroid of these point
spread functions can not be accurately found. They are probably
associated with V403 Ori (13\arcsec\/ to the NW) and V387 Ori
(14\arcsec\/ to the SW), respectively.

\item Five cases (COUP \object[COUP 0084]{84}, \object[COUP 0400]{400}, \object[COUP 0908]{908}, \object[COUP 1237]{1237} and \object[COUP 1238]{1238}) lie very
close to known ONC members and may be new binary companions. See
\S \ref{x_ray binaries_section} for details.

\item Four cases (COUP \object[COUP 0495]{495}, \object[COUP 0911]{911}, \object[COUP 1042]{1042}, \object[COUP 1192]{1192}) are located
within the VLT FOV. COUP \object[COUP 0911]{911} is a very faint
X-ray source lying 23\arcsec\/ ENE of $\theta^1$ Ori C; it may be
a new very-low-mass ONC member. COUP \object[COUP 1042]{1042} is a
very faint X-ray source lying 10\arcsec\/ W of the $V \simeq 12$
classical T Tauri G star TU Ori which is already known to have a
flare star companion \citep{Hambarian88}. The nebular emission is
bright in these regions, perhaps inhibiting VLT detection of an
X-ray emitting low-mass ONC member.

\end{enumerate}

Finally, we find that the remaining 33 of 51 unidentified
unabsorbed sources are very weak ($\la 20$ net counts in all cases
and $< 10$ net counts in most cases), have low source
significance, are undetected with the PWDetect source detection
algorithm and (for 27 of the 33) flagged as `uncertain' source
existence in Table 2 of \citet{Getman05}. We indicate their class
as `Unc' here in Table \ref{non_flaring_table}.  We suspect that
their identification as X-ray sources by the procedures described
in Getman et al.\ was mistaken and that most of these are not real
X-ray sources.

\section{X-ray sources associated with nebular shocks
\label{HH_srcs.sec}}

Two sources, COUP \object[COUP 0703]{703} and \object[COUP
0704]{704}, are very faint, show constant X-ray lightcurves
without flares, have unusually soft ACIS spectra, and have no ONIR
stellar counterpart. They coincide with HH\,210, which is one of
the fingers of the molecular flows emerging from the massive
star(s) in the Orion Hot Molecular Core. These are likely new
discoveries of X-ray emission from Herbig-Haro outflow bow shocks,
similar to those seen in the HH 1/2 and HH 80/81 outflows
\citep{Pravdo01, Pravdo04}. We designate these sources as `HH' in
Table \ref{non_flaring_table}. The difference between $\log N_H
\sim 21.9 \pm 0.2$ cm$^{-2}$ of COUP \object[COUP 0703]{703} and
$\log N_H \sim 20.0 \pm 2.5$ cm$^{-2}$ of COUP \object[COUP
0704]{704} can be explained by the high uncertainty of the column
density for COUP \object[COUP 0704]{704} \citep[see Table 6
of][]{Getman05}. Accurate astrometry study and details regarding
X-ray emission from HH\,210 will be presented in Grosso et al. (in
preparation).

\section{Double sources}\label{x_ray binaries_section}

\subsection{Double X-ray sources with separations $<3$\arcsec}

The ONC is an important laboratory for investigating the origin
and evolution of binary star systems.  For binaries with
separations $\simeq 25-500$ AU ($\simeq 0.06\arcsec-1.1$\arcsec),
the ONC has been shown to be strongly deficient in binaries
compared to low-density star formation regions such as the
Taurus-Auriga clouds with a binarity fraction similar to that seen
in main sequence field stars \citep[][and the review by
McCaughrean 2001]{Prosser94, Padgett97, Petr98}.

Searches for wide binaries are dominated by chance superpositions
of stars in the crowded cluster core.  No wide binaries with
$1000-5000$ AU ($2.3\arcsec-11$\arcsec) separations are found
among $M \geq 0.2$ M$_\odot$ cluster members when random
coincidences and proper motions are taken into account
\citep{Scally99}.  In contrast, 15\% of field stars are binaries
in this range \citep{Duquennoy91}. Wide binaries are dynamically
fragile and may be disrupted by gravitational encounters with
nearby stars. $N$-body simulations of the ONC indicate that most
wide binaries will be quickly ($t < 0.4$ Myr) disrupted if the
cluster is in virial equilibrium, but many should have survived if
the cluster is expanding due to ejection of natal molecular gas
\citep{Kroupa99}.  The census of wide binaries in the ONC thus
gives valuable information regarding the dynamical history of the
cluster.

The COUP sample of X-ray emitting stars in the ONC provides a new
capability for measuring ONC wide binaries for three reasons.
First, the sample has different selection effects than traditional
optical samples:  COUP measures magnetic activity rather than
bolometric luminosity diminished by obscuration.  This results in
several dozen new pre-main sequence members, though most of these
are heavily obscured in the cloud behind the cluster core (\S
\ref{flaring_members_section}).  Second, it is sometimes easier to
detect a lower-mass companion to a higher-mass primary in X-rays
because the contrast in X-ray luminosities is smaller than in
bolometric luminosities. This is particularly true for
intermediate-mass ($1.5 < M < 10$ M$_\odot$) primaries.  Finally,
a statistical link between binarity and X-ray emission has been
reported in main sequence field stars \citep{Makarov02} which
would assist in the detection of new companions, though no such
effect has been found in the Taurus-Auriga pre-main sequence
population \citep{Konig01}.

Table \ref{binaries_table} lists the 61 pairs of COUP sources
lying within 3\arcsec\/ of each other. The closest pair has
separation 0.7\arcsec, but the minimum detectable separation is a
function of off-axis location in the $Chandra$ field.  The first
six columns of the table give X-ray information from
\citet{Getman05}, the next four columns give $JHK_s$ and
optical-band counterparts from Getman et al., and the last two
columns give membership classification based on results obtained
in this study.  The lower COUP sequence number always refers to
the western (W) component of the double.

Small `postage-stamp' images of these 61 doubles are shown in
Figure \ref{close_binaries_fig}. For each pair we give the
original COUP data (top panel), a maximum-likelihood
reconstruction of the region (middle panel), and a reproduction of
the best available $K_s$-band image (bottom panel). The
reconstruction was performed using the Lucy-Richardson algorithm
\citep{Lucy74} with $100-200$ iterations based on the local
$Chandra$ point spread function generated by the CIAO program {\it
mkpsf}.  The procedure is run within the {\it ACIS Extract} data
reduction environment\footnote{Descriptions and codes for {\rm{\it
acis\_extract}} can be found at
\url{http://www.astro.psu.edu/xray/docs/TARA/ae\_users\_guide.html}.}
\citep{Getman05} which uses the {\it max\_likelihood.pro}
procedure provided by the ASTRO library of the IDL software
package.

As demonstrated by \citet{Scally99}, it is not easy to
differentiate true physical binaries from unrelated cluster
members in chance proximity.  For 48 of the 61 COUP doubles, both
components appear in the VLT $JHK_s$ catalog (McCaughrean et al.\,
in preparation).  The physical reality of these doubles is best
discussed in a future study of the full VLT sample. In 19 of these
cases, both components have different bright optical counterparts
in the \citet{Jones88} or \citet{Prosser94} catalogs (optical star
designations 0-9999) and thus were included in earlier ONC
binarity studies \citep{McCaughrean01}.

To evaluate the importance of chance proximity, we performed Monte
Carlo simulations of COUP source positions by randomly locating
point sources within circular annuli of 0.5\arcmin\/ or
0.25\arcmin\/ width with the number of simulated sources equal to
the COUP source counts in each annulus. Simulations show that,
within the central 8\arcmin area of the COUP field, the expected
number of random pairs with separations $< 3\arcsec$ ranges from
70 to 90.  The observed and simulated off-axis angle and source
separation distribution functions are also similar.  Thus, chance
proximity between physically unrelated COUP sources can explain
the presence of all 61 COUP wide-double sources listed in Table 3.
The difference between the 61 observed and 70$-$90 predicted
chance pairs can be attributed to incompleteness in the COUP
detection procedure in close proximity to bright X-ray sources.

We note however that 11 of the COUP doubles are new in the sense
that one or both components have no counterpart in any optical or
infrared survey.  These are classified as `NM' or `NM?' in Table
\ref{binaries_table}.  The doubles COUP \object[COUP
0589]{589}-\object[COUP 0590]{590}, \object[COUP
0621]{621}-\object[COUP 0628]{628}, and \object[COUP
0678]{678}-\object[COUP 0681]{681} are potentially physical
binaries because of similarities in the X-ray absorption in the
spectra of the two components.

We conclude that, when X-ray/X-ray source doubles are considered,
the COUP survey has uncovered no more than three, but quite
possibly zero, new physical binary systems with separations
between 0.7\arcsec\/ and 3.0\arcsec\/ ($250-1000$ AU projected
separation). Forty-eight other doubles are independently found in
COUP and in the new VLT merged near-infrared catalog of
McCaughrean et al., but these are consistent with chance proximity
of unrelated Orion stars.

\subsection{Additional binary findings}
\label{additional binaries_section}

When single COUP sources are examined for proximity to previously
known stars, several additional noteworthy cases emerge. In at
least 8 cases, COUP has detected new sources proximate to known
Orion cloud members, and some of these may be physical binaries.
Five examples (COUP \object[COUP 0599]{599}, \object[COUP
0678]{678}, \object[COUP 0683]{683}, \object[COUP 0940]{940},
\object[COUP 1237]{1237}) occur in the central part of the field
where VLT NIR counterparts reported in \citet{Getman05} have
unusually high positional offsets of $>0.6\arcsec$.  The
complicated case of the massive, heavily obscured
Becklin-Neugebauer (BN) Object is discussed by \citep{Grosso05}.
Here, an X-ray bright new star (\object[COUP 0599]{599} in the
COUP catalog of Getman et al. but now designated COUP 599a) lies
0.9\arcsec\/ from the BN Object, which appears to be coincident
with a fainter X-ray source designated COUP 599b which was not
found in the standard processing of Getman et al. Similarly, we
report here that COUP \object[COUP 0683]{683} is probably
associated with a new companion to VLT 484 with $0.9\arcsec$
positional offset. COUP source \object[COUP 0678]{678} is probably
a new deeply embedded star with $\log N_H = 22.9$ cm$^{-2}$ and
not associated with VLT 476 with offset $0.9\arcsec$.  COUP
\object[COUP 0940]{940} was listed in \citet{Getman05} as having
the $L-$band counterpart FLWO 890 \citep{Muench02}, but this match
is unlikely with a positional offset of $1.3\arcsec$ (see footnote
to Table \ref{binaries_table}).

Four additional likely new companions to known Orion members lie
outside the VLT field of view. COUP \object[COUP 0084]{84} is a
faint, lightly absorbed source lying 6.5\arcmin\/ off-axis. Its
point spread function overlaps the $V\simeq 15$ G star JW 141, but
with the X-ray centroid lying 5\arcsec\/ E of the star, it is
probably a new low-mass companion or an unrelated star. Three
bolometrically faint X-ray sources (COUP \object[COUP 0400]{400},
\object[COUP 0908]{908}, and \object[COUP 1238]{1238}) are
detected in the vicinity of three bright off-axis ONC members.
$J-$band 2MASS images provide us with a hint for the presence of
their very weak IR counterparts in the vicinity of bright
primaries.

New low-mass companions around intermediate- and high-mass members
of the ONC are found in the COUP dataset.  COUP source
\object[COUP 1237]{1237} lies in the point spread function wings
of the bright source COUP \object[COUP 1232]{1232} associated with
the O9-B2 star $\theta^2$ Ori A. This and similar sources, and the
implication for X-ray emission from hot young stars, are discussed
in a separate COUP study by \citet{Stelzer05}.

\section{Field stars \label{field_stars_section} }

X-ray surveys are effective in discriminating old field stars from
the cluster population due to the high X-ray emissivity of young
stars \citep[e.g.][]{Feigelson04}. The ratio $L_x/L_{bol}$ for the
Orion population is typically a factor of $10^2-10^3$ larger
compared to that of $0.5-2$ M$_\odot$ old disk stars, though the
difference is smaller for younger ZAMS stars and for lower mass
stars \citep{Preibisch05a, Preibisch05b}. In the COUP field, the
situation is confused by the wide range of obscuration: both Orion
stars in the nebula region and foreground stars suffer little
absorption, while both embedded Orion stars and background disk
stars are heavily absorbed.  Many stars also do not have
sufficient optical photometry and spectroscopy to directly obtain
$A_V$ and $L_{bol}$.

We proceed to examine COUP stars for possible field star
candidates using three sources of information: the proper motion
survey of \citet{Jones88}; the $JHK_s$ NIR colors from the VLT
survey in the inner $7.4\arcmin \times 7.4\arcmin$ region
\citep{McCaughrean04} and the 2MASS survey in the outer region;
and the Besan\c{c}on stellar population synthesis model of the
Galactic disk population \citep{Robin03}.

First, we create a subsample of 33 COUP stars with proper motions
inconsistent with cluster membership shown in Table
\ref{33_possible_field_table}.  From the COUP optical sample of
977 stars, 35 COUP stars have membership probabilities $P < 90$\%
based on the proper motion study of \citet{Jones88}.  Two of them
are omitted here: COUP \object[COUP 0279]{279}, a $P=22$\% star
with an imaged disk (proplyd) that is clearly a cluster member
\citep{Kastner05}; and COUP \object[COUP 1232]{1232} with $P =
80$\% which is $\theta^2$ Ori A, a O 9.5 star that is the
second-most massive member of the Trapezium cluster
\citep{Stelzer05}.

\begin{figure}
\centering
\includegraphics[angle=0.,width=0.9\textwidth]{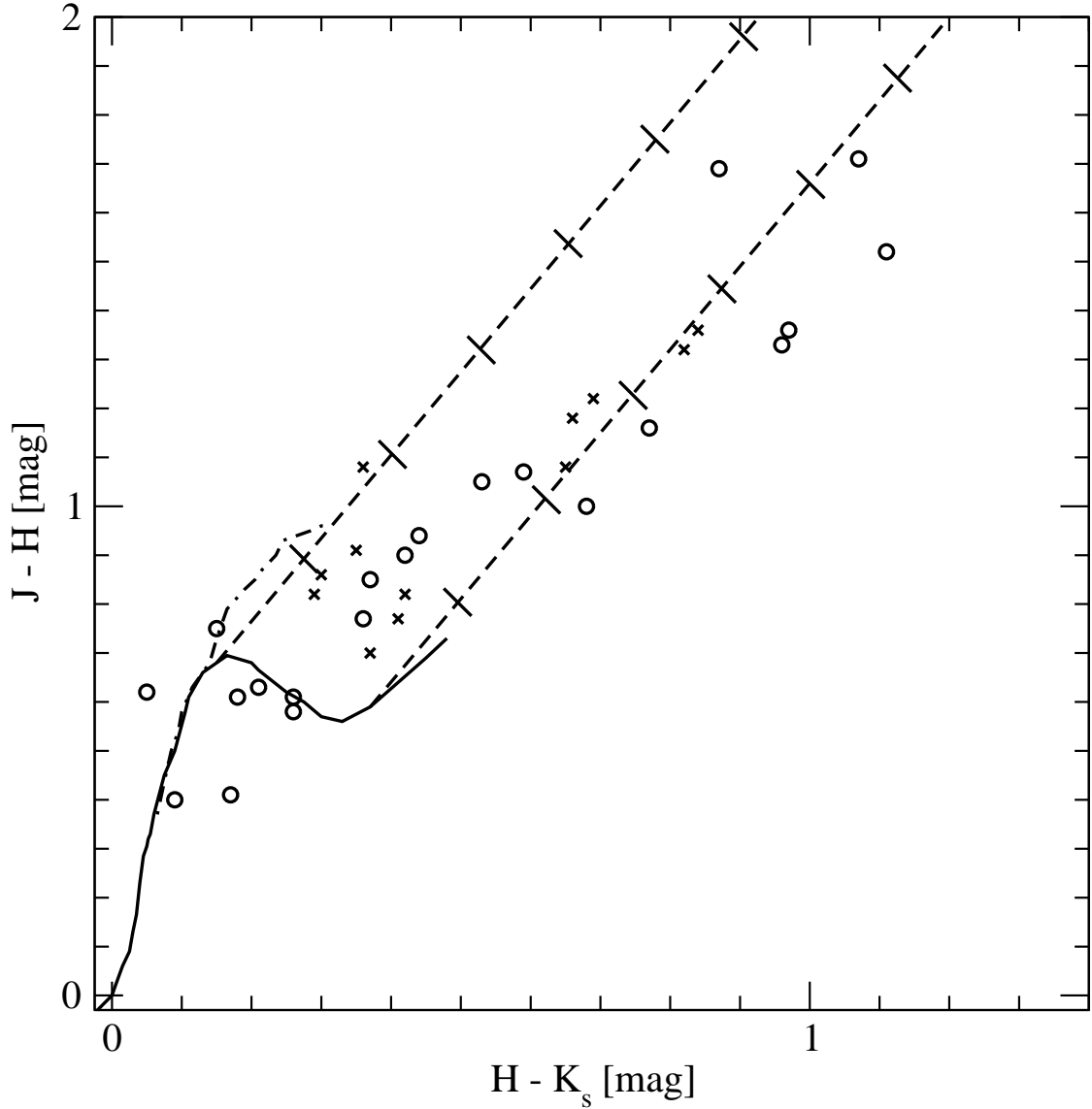}

\caption{Near-infrared color-color diagram of 33 COUP sources with
cluster membership probabilities $P< 90\%$ based on proper
motions. Sixteen lie close to the locus of unreddened main
sequence stars ($J-H < 0.9$). Circles indicate sources with $P <
50\%$. The solid and dot-dashed curves show sites of intrinsic
$JHK_s$ colors of main sequence and giant stars, respectively. The
dashed lines are reddening vectors originating at M0 V (left line)
and M6.5 V (right line), and marked at intervals of $A_V = 2$ mag.
\label{33_field_candidates_cc}}
\end{figure}

Discrepant \citet{Jones88} proper motion measurements alone may
not be a definitive indicator of non-membership, as they are based
on decades of photographic images which may be affected by
binarity, disks, or dynamical events such as stellar ejection
after close 3-body encounters \citep{Tan04}.  We therefore combine
NIR color information, shown in Figure
\ref{33_field_candidates_cc}, with proper motion measurements to
select field star candidates from this subsample.  We adopt the
criterion that field stars should show little reddening ($J-H <
0.9$ roughly equivalent to $A_V < 2-3$) and should not exhibit
$K_s-$band excesses attributable to circumstellar dust.  Sixteen
sources satisfying these criteria, most of which have very low
proper motion membership probabilities ($P<50$\%) are marked by
`F' in column 16 of Table \ref{33_possible_field_table}. All
except one have $K_s < 13$ mag.

We supplement these 16 candidate X-ray-detected field stars with
similar stars which are undetected in the COUP survey\footnote{We
also checked the $JHK_s$ color-color diagram for 356 COUP sources
with IR but without optical counterparts and found that none of
them lies  at the locus of unreddened main sequence stars.  We
consider them as likely cluster members.}. Adopting a limit $K_s <
13$, 159 bright stars (79 in the inner VLT survey region and 80 in
the outer 2MASS region) are not detected in X-ray. The majority of
those ($\sim 70$\%) are subject to factors that hinder X-ray
detection: low local ACIS exposure time, high local background due
to readout streaks from bright ONC sources, and contamination from
the point spread function wings of very nearby bright X-ray
sources. In other cases, 2MASS sources have contamination flags
and may be bright knots in the nebular emission rather than true
stars. Requiring that the sources have $JHK_s$ colors consistent
with unreddened main sequence stars \citep{Bessell88}, we find 6
X-ray-undetected $K_s<13$ stars in the COUP field of view
presented in Table \ref{bright_K_undetected_tab}.  Thus our sample
of possible field stars with the limiting $K_s$ of 13 mag consists
of 22 stars, 16 with COUP detections marked `F' in Table
\ref{33_possible_field_table} and 6 without COUP detection listed
in Table \ref{bright_K_undetected_tab}.

\begin{figure}
\centering
\includegraphics[angle=0.,width=0.9\textwidth]{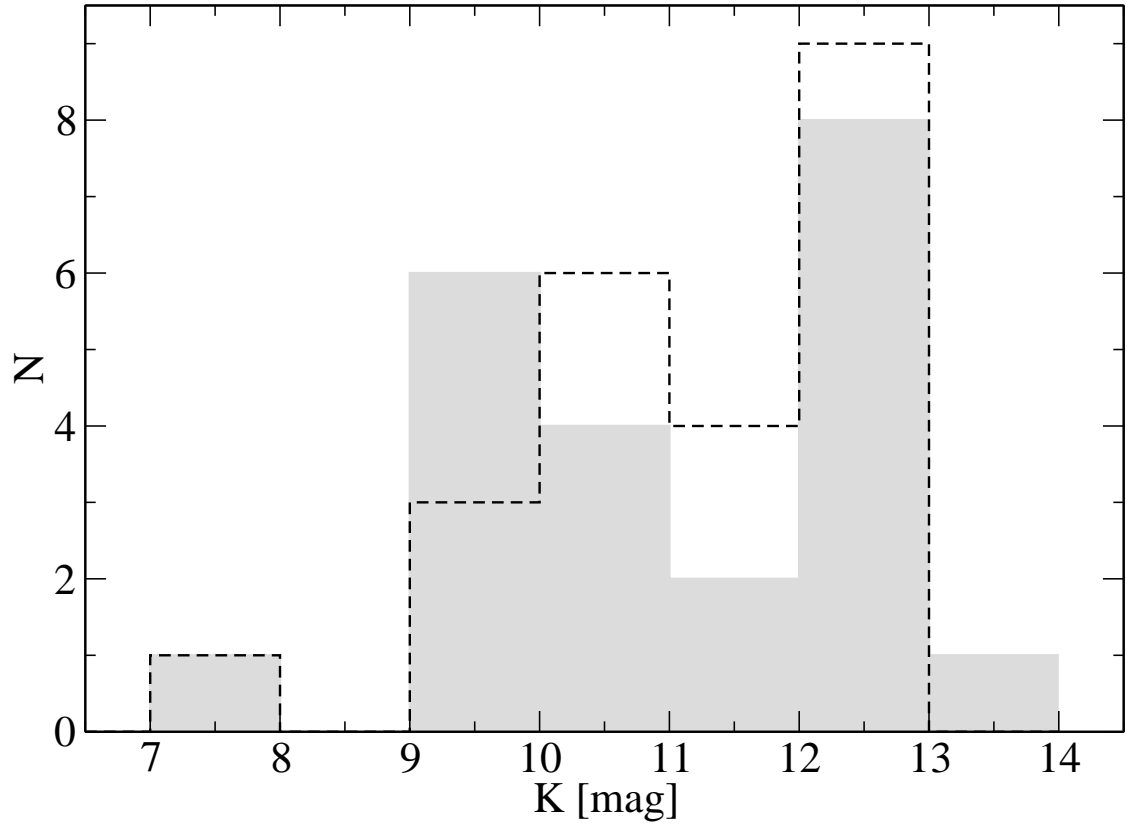}
\caption{$K_s-$band distribution functions for 22 suggested field
star candidates (grey bars), including both detected and
undetected stars in the COUP survey, and 23 field stars predicted
from the Besan\c{c}on stellar population synthesis model (dashed
histogram). \label{KLF_comparison_fig}}
\end{figure}

This sample of 22 field star candidates is compatible with the
predictions of the stellar population synthesis model of the
Galaxy for the sky position of the Orion region out to a distance
of 450 pc \citep{Robin03}\footnote{The Monte Carlo simulation of
the stellar population in the cone subtended by the COUP field was
computed with the Web service provided by the Besan\c{c}on group
at \url{http://bison.obs-besancon.fr/modele/}}. The model predicts
about 23 foreground stars with $K_s < 13$ mag in the COUP field of
view. All except $\simeq 2$ are expected to be main sequence
dwarfs: about 10\% A$-$F stars, 20\% G stars, 50\% K stars, and
20\% M stars.  Two-thirds of the simulated field stars have ages
greater than 2 Gyr with the remaining ages ranging from 0.15 to 2
Gyr. About 60\% of simulated field stars occupy the volume between
300 pc and 450 pc.  The $K_s-$band distribution of the predicted
23 field stars are compared to the distribution for our 22
candidate field stars in Figure \ref{KLF_comparison_fig}.

The number, brightness distribution and other properties of the
predicted and candidate field star population are consistent with
each other.  However, in light of the large number of stars in the
field excluded from consideration here (due to proximity of nearby
X-ray-bright stars), the stars listed in Tables
\ref{33_possible_field_table}-\ref{bright_K_undetected_tab} can
only be considered suggestions rather than clear identifications.

\section{Summary}\label{summary_section}

The COUP observation provides an exceptionally deep X-ray survey
of the nearest rich young (age $\sim$1 Myr) stellar cluster, the
Orion Nebula Cluster and associated embedded young stellar
objects. Membership of the region is important for studies of the
stellar IMF, cluster dynamics, and star formation.  ONC membership
has been intensively investigated at ONIR wavelengths.  COUP
provides the best opportunity to examine the benefits and
limitations of X-ray selection to such membership studies.  In
addition to straightforward associations of X-ray sources with
ONIR stars, we have evaluated the levels of contamination by
stellar and extragalactic sources unrelated to the Orion star
forming cloud using detailed simulations of X-ray emitting
populations.  Our main findings are:

1.  One thousand three hundred thirty one COUP sources have ONIR
counterparts.  Except for about 16 suggested field stars unrelated
to the Orion cloud (Table 4), all of these are true members of the
ONC or are obscured members of the background molecular cloud.
Only 16 of those sources (Table 4, `F') with discrepant proper
motions are suggested field stars based on a stellar population
synthesis model of the Galactic disk.

2.  The long exposure of COUP optimizes the chance of capturing
powerful X-ray flares and thereby distinguishing absorbed young
stellar objects from the background Galactic or extragalactic
populations. We found that 42 heavily absorbed sources without
ONIR identifications exhibit one or more high-amplitude X-ray
flares, which convincingly demonstrates that they are newly
discovered Orion cloud members (Table 1).  They are all heavily
absorbed and almost all are spatially associated with the two
well-known OMC-1 cores and the dense molecular filament which
extends northwards from OMC-1 to OMC-2/3.  Thirty-three additional
obscured COUP sources that do not exhibit X-ray flares lie in this
region and are also likely to be OMC members (Table 2, `OMC or
EG?').

3.  Based on detailed simulations of the extragalactic background
population, including instrumental background and obscuration
effects by the molecular cloud, we find that about 159 COUP
sources are probably extragalactic AGN unrelated to the Orion star
forming region (Table 2, `EG').

4.  We find 16 lightly obscured sources without ONIR counterparts,
likely new members of the Orion Nebular Cluster (`ONC' in Table
2). Some are close neighbors of bright Orion stars.

5.  Among unidentified sources we see two where the X-ray emission
is unusually soft, constant and faint, and is associated with a
nebular shock rather than a star (Table 2, `HH'). These and
possible other similar sources will be studied in detail by Grosso
et al.\ (in preparation).

6.  One hundred twenty-two COUP sources lie within 3\arcsec\/ of
another COUP source, forming 61 apparent double sources.
Simulations indicate that most of these are random pairings rather
than physical wide binaries.  Eleven of these doubles are new with
one or both components not previously found in ONIR studies.

We thank the anonymous referee for his time and many useful
comments that improved this work. COUP is supported by {\it
Chandra} guest observer grant SAO GO3-4009A (E.\ D.\ Feigelson,
PI).  This work was also supported by the ACIS Team contract
NAS8-38252. This publication makes use of data products from the
Two Micron All Sky Survey, which is a joint project of the
University of Massachusetts and the Infrared Processing and
Analysis Center/California Institute of Technology, funded by the
National Aeronautics and Space Administration and the National
Science Foundation.

Facility: \facility{CXO(ASIS)}

\newpage

\clearpage
\newpage

\begin{figure}
\centering
\includegraphics[angle=0.,width=4in]{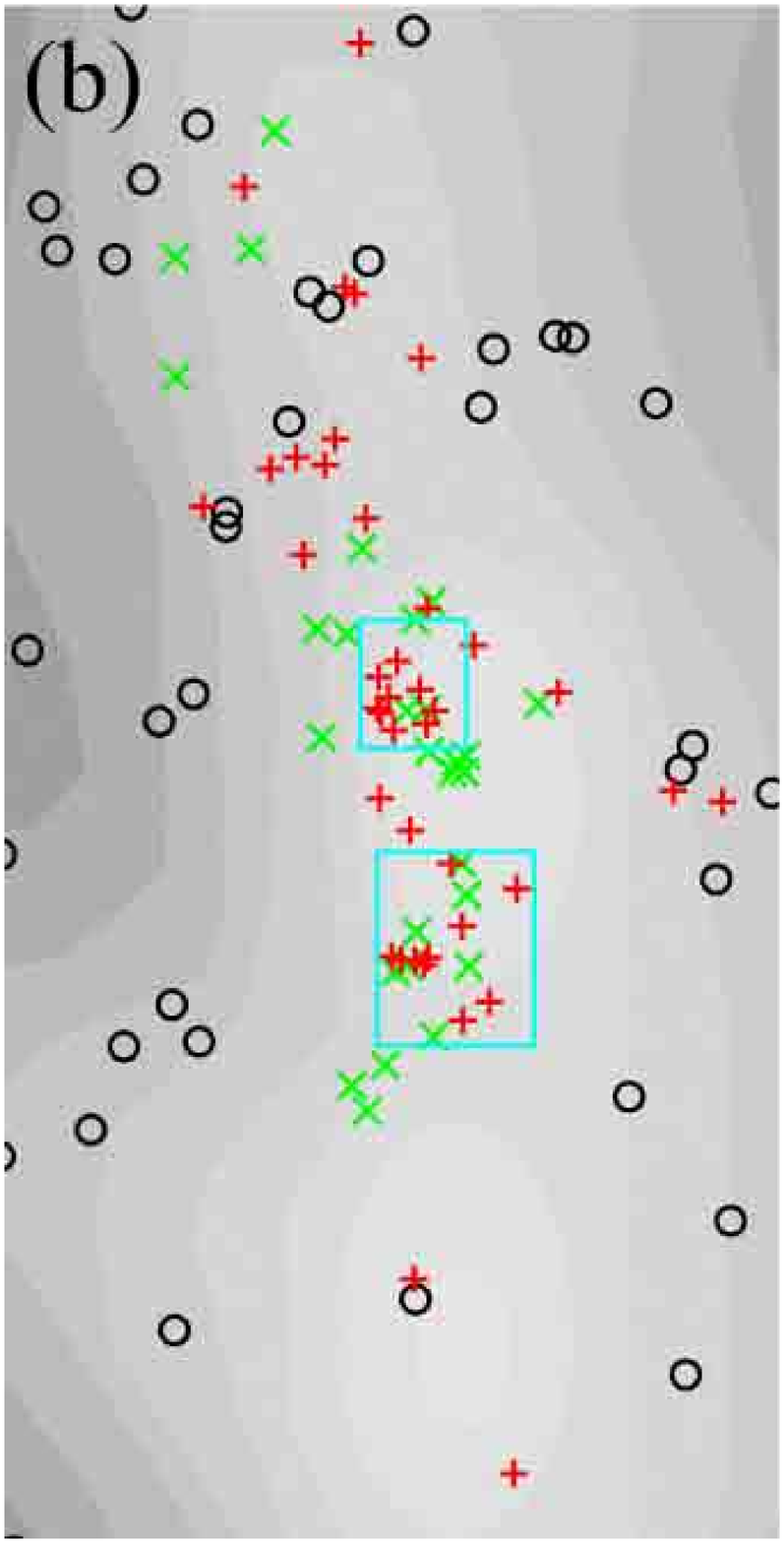}
\end{figure}

\clearpage
\newpage

\clearpage
\newpage

\begin{figure}
\centering
\includegraphics[angle=0.,width=0.9\textwidth]{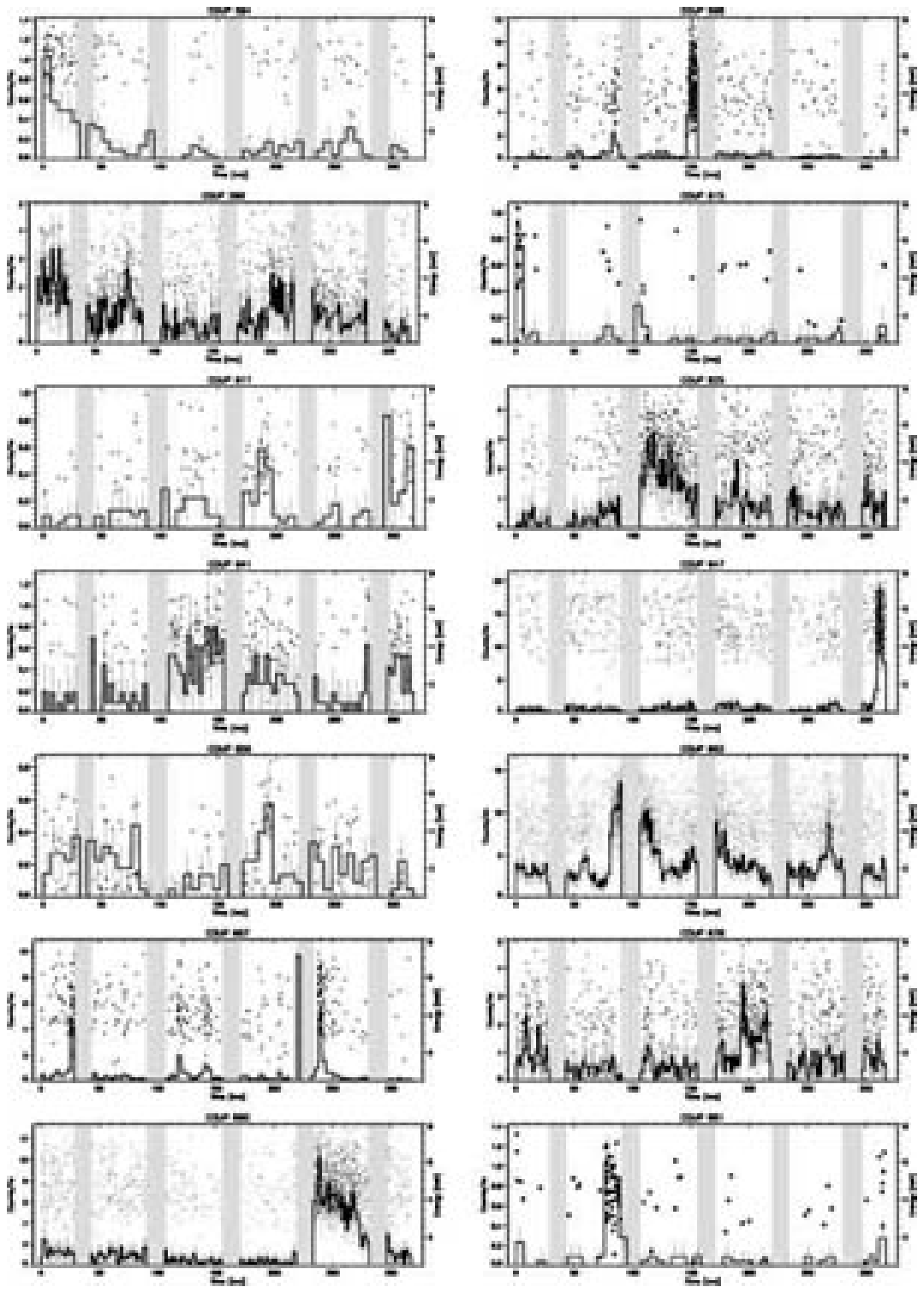}
\hspace{0.00in}
\end{figure}

\clearpage
\newpage

\begin{figure}
\centering
\includegraphics[angle=0.,width=0.9\textwidth]{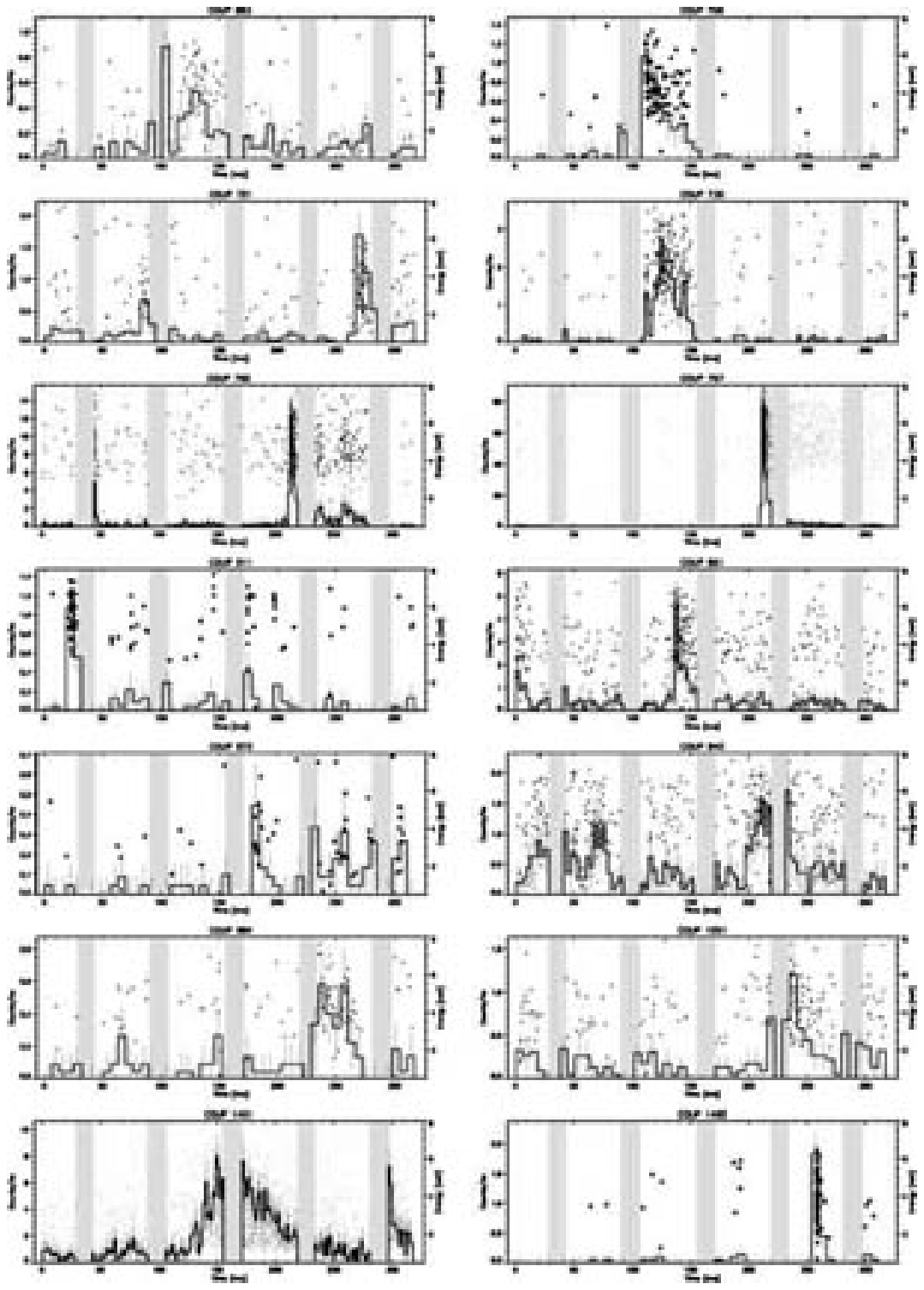}
\hspace{0.00in}
\end{figure}

\clearpage
\newpage

\clearpage
\newpage

\begin{figure}
\begin{minipage}[t]{1.0\textwidth}
  \centering
  \includegraphics[width=0.95\textwidth]{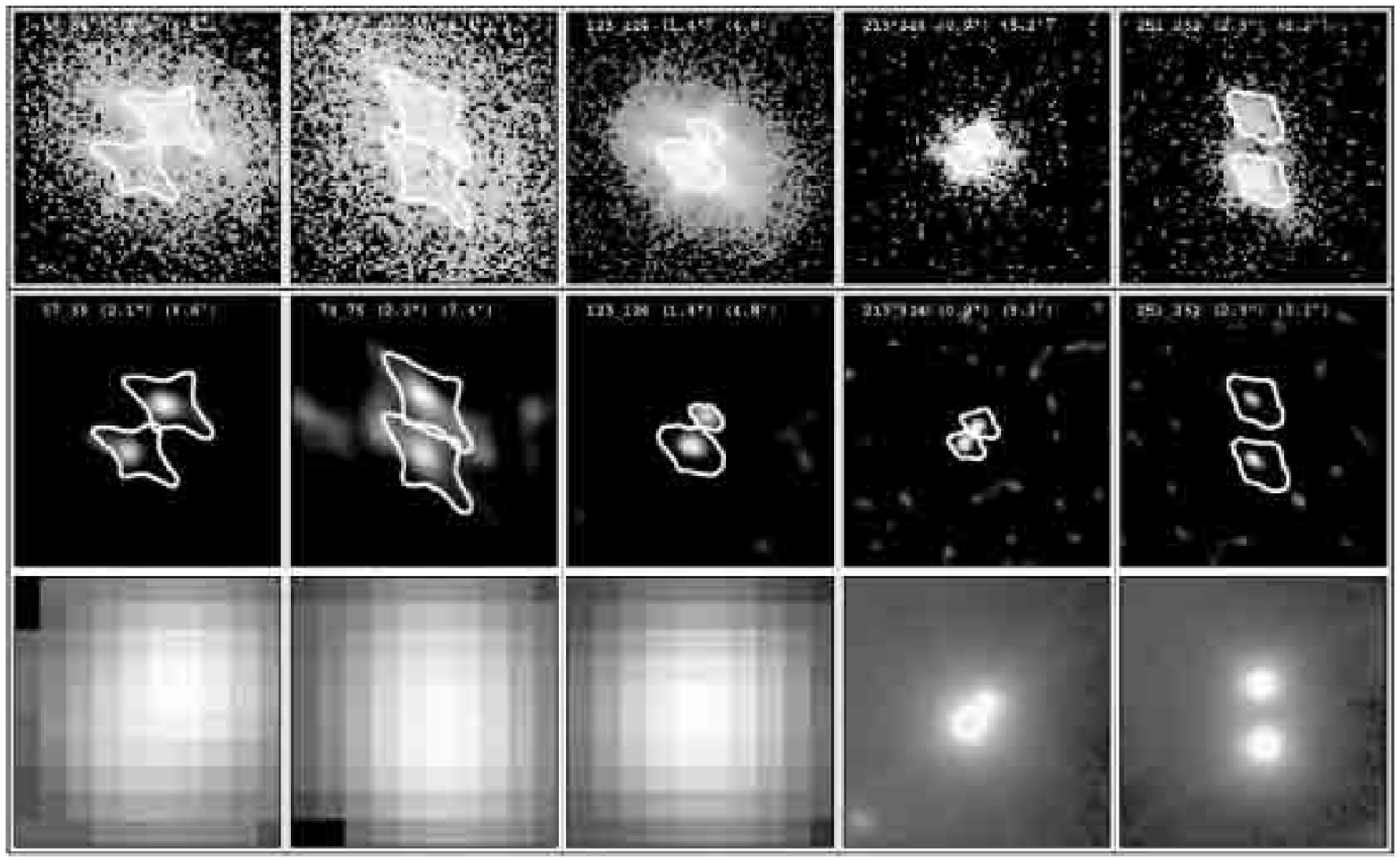} \hspace{0.00in}
\end{minipage}
\begin{minipage}[t]{1.0\textwidth}
  \centering
  \includegraphics[angle=0.,width=0.95\textwidth]{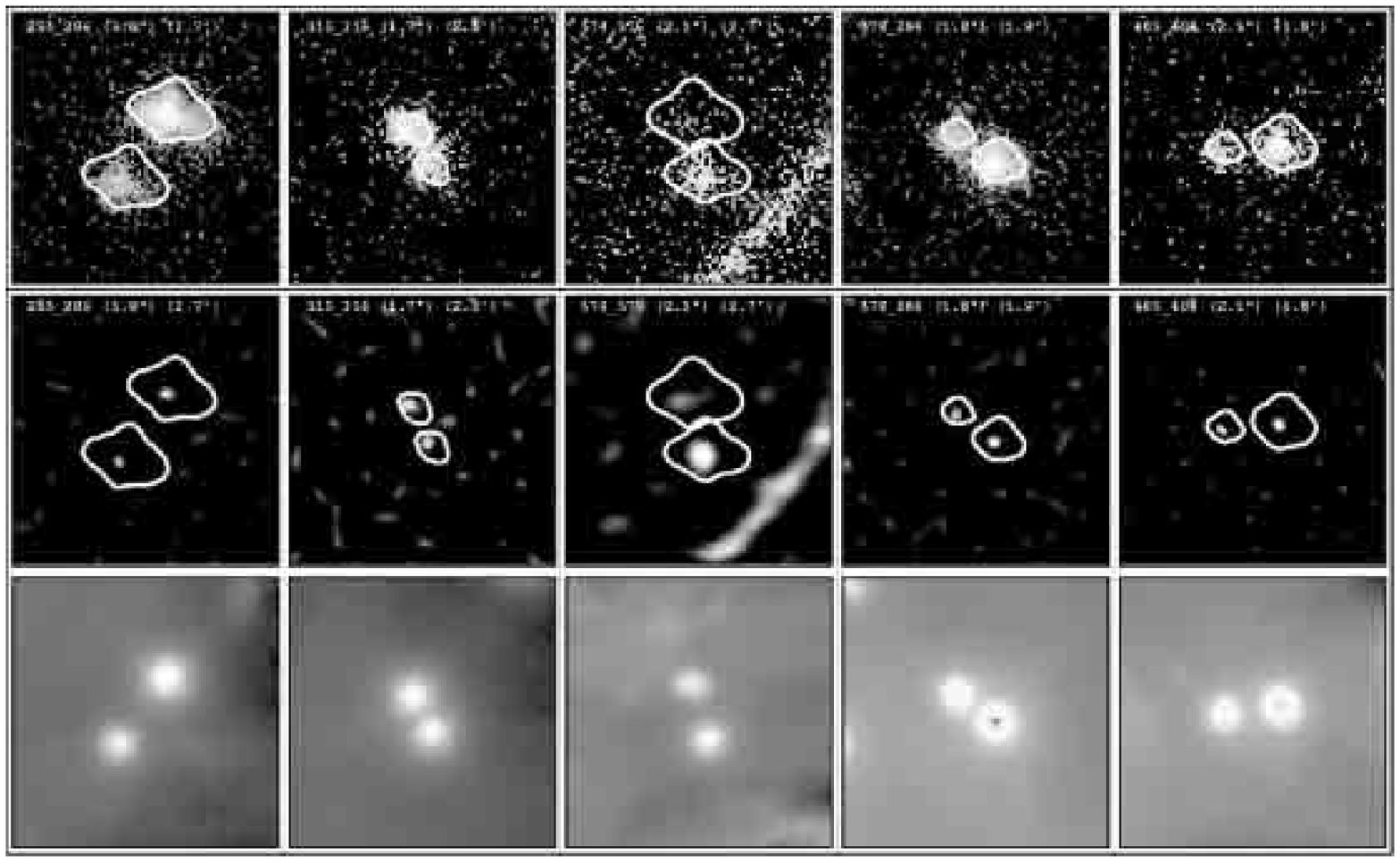} \hspace{0.00in}
\caption{Three postage-stamp images for each of 61 COUP double
sources with separations $<3$\arcsec:  (Top) $10\arcsec \times
10\arcsec$ raw COUP image binned at high resolution with
$0.125\arcsec$ pixel$^{-1}$; (Middle) maximum likelihood
reconstruction of the COUP data; and (Bottom) $K_s$-band image
from the Very Large Telescope or (for far-off axis sources) 2MASS
surveys. Dark centers in some $JHK_s$ stars are detector
saturation effects. The labels give the COUP numbers of binary
components, the component separation, and the off-axis angle.  The
closed curves show the extraction regions used in COUP analysis
which are matched to the local $Chandra$ point spread function and
the proximity of the companion. \label{close_binaries_fig}}
\end{minipage}
\end{figure}

\clearpage
\newpage

\begin{figure}
\centering
\begin{minipage}[t]{1.0\textwidth}
  \centering
  \includegraphics[width=0.95\textwidth]{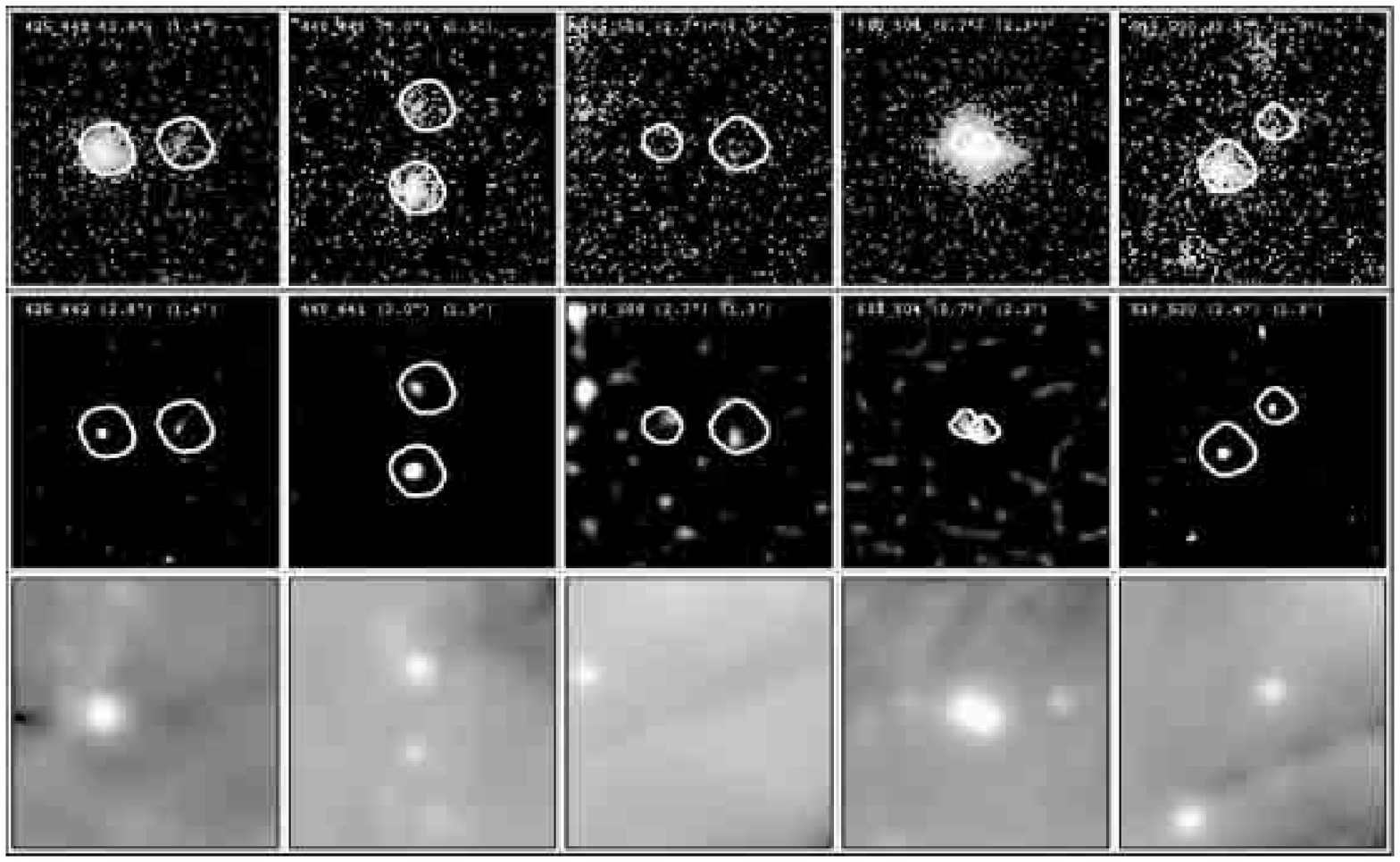} \hspace{0.00in}
\end{minipage}
\begin{minipage}[t]{1.0\textwidth}
  \centering
  \includegraphics[angle=0.,width=0.95\textwidth]{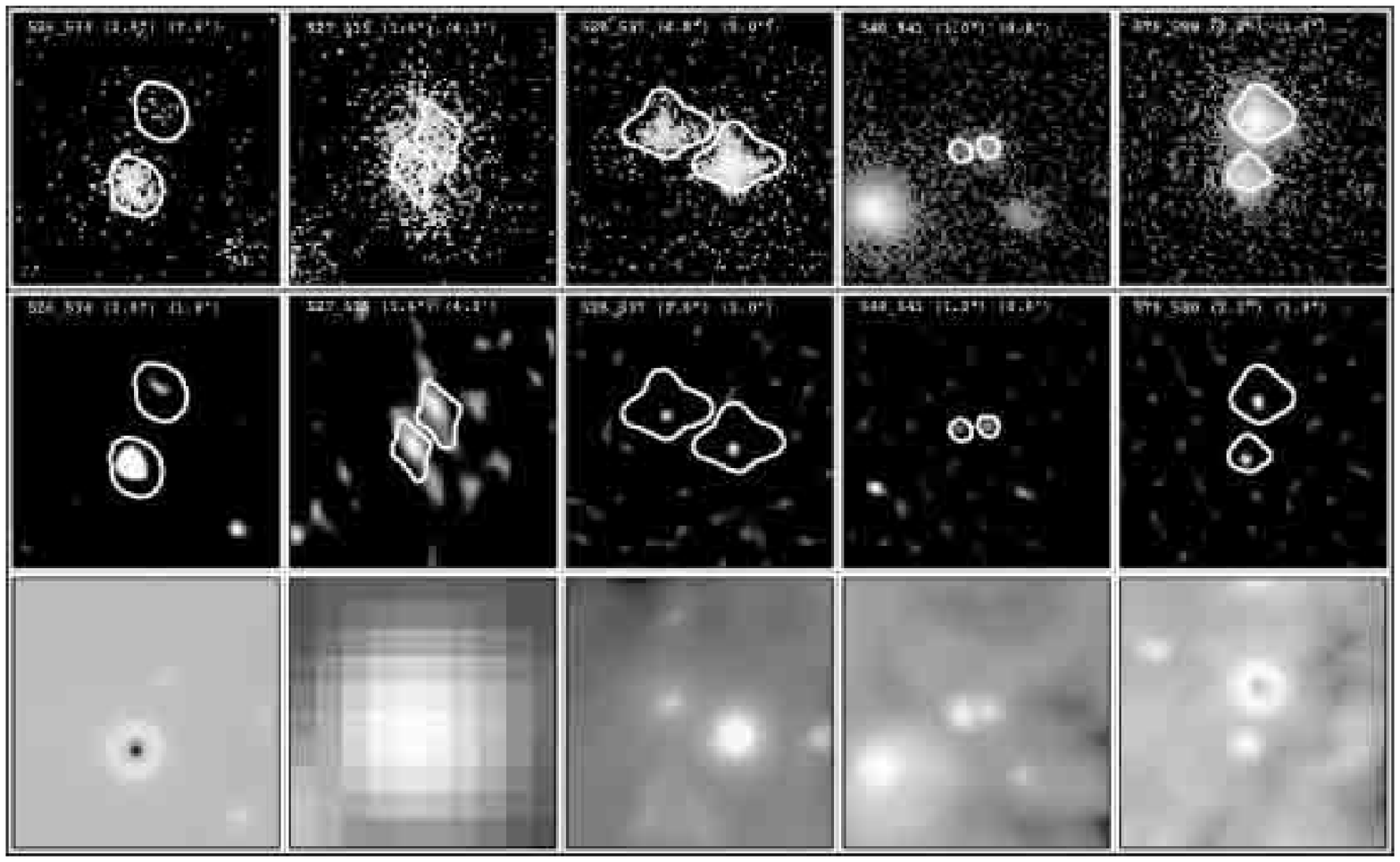} \hspace{0.00in}
\end{minipage}
\end{figure}

\clearpage
\newpage

\begin{figure}
\centering
\begin{minipage}[t]{1.0\textwidth}
  \centering
  \includegraphics[width=0.95\textwidth]{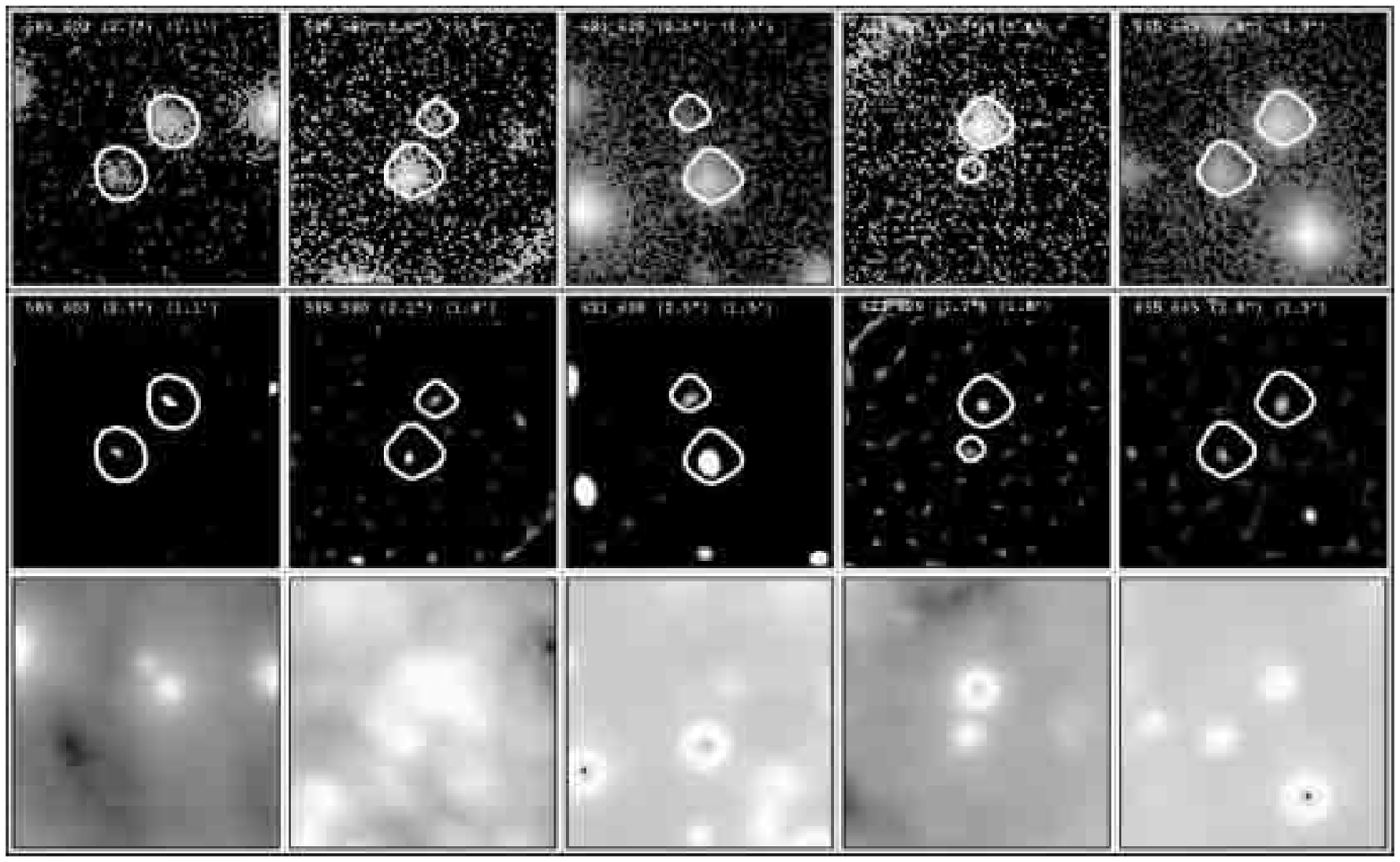} \hspace{0.00in}
\end{minipage}
\begin{minipage}[t]{1.0\textwidth}
  \centering
  \includegraphics[angle=0.,width=0.95\textwidth]{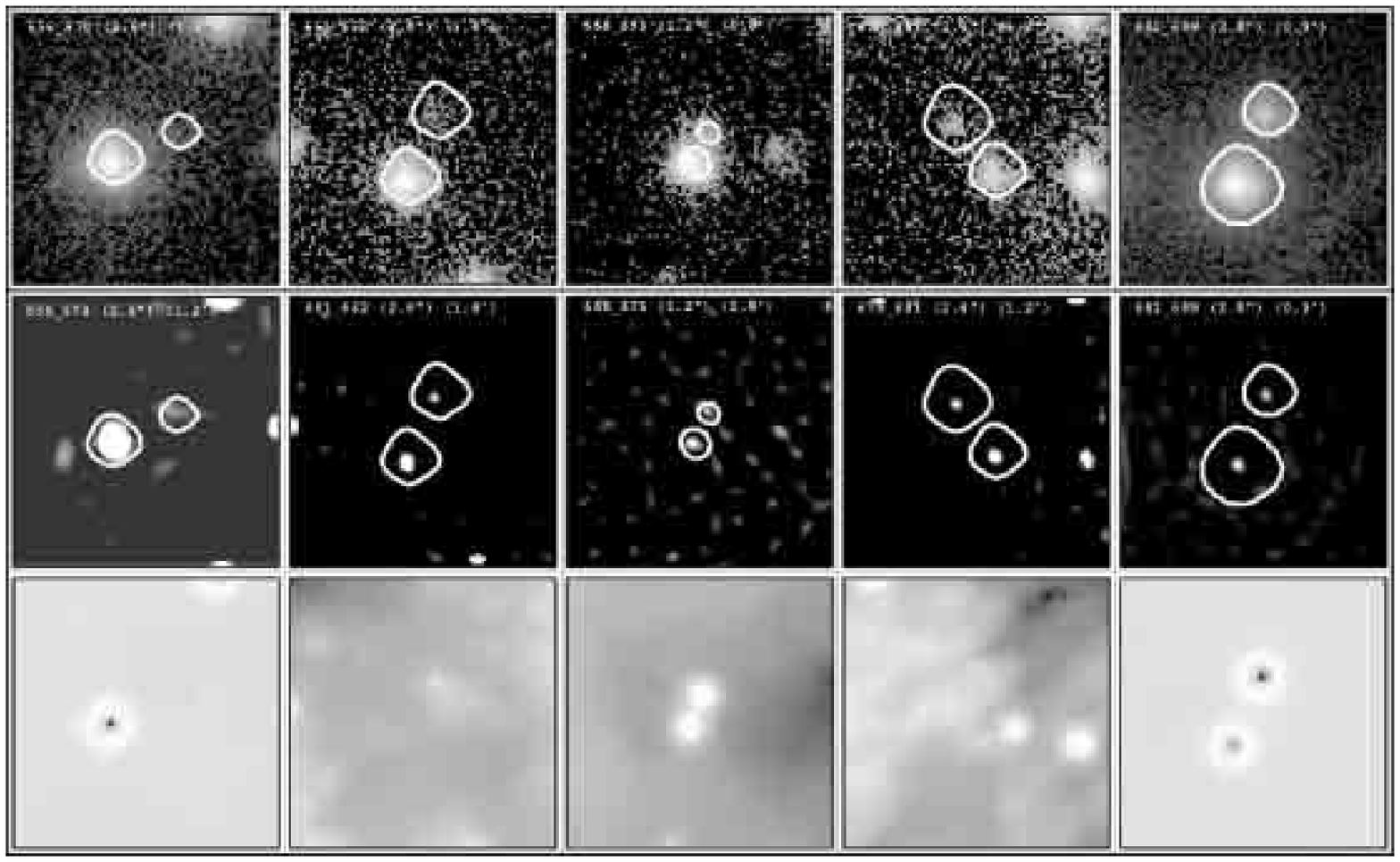} \hspace{0.00in}
\end{minipage}
\end{figure}

\clearpage
\newpage

\begin{figure}
\centering
\begin{minipage}[t]{1.0\textwidth}
  \centering
  \includegraphics[width=0.95\textwidth]{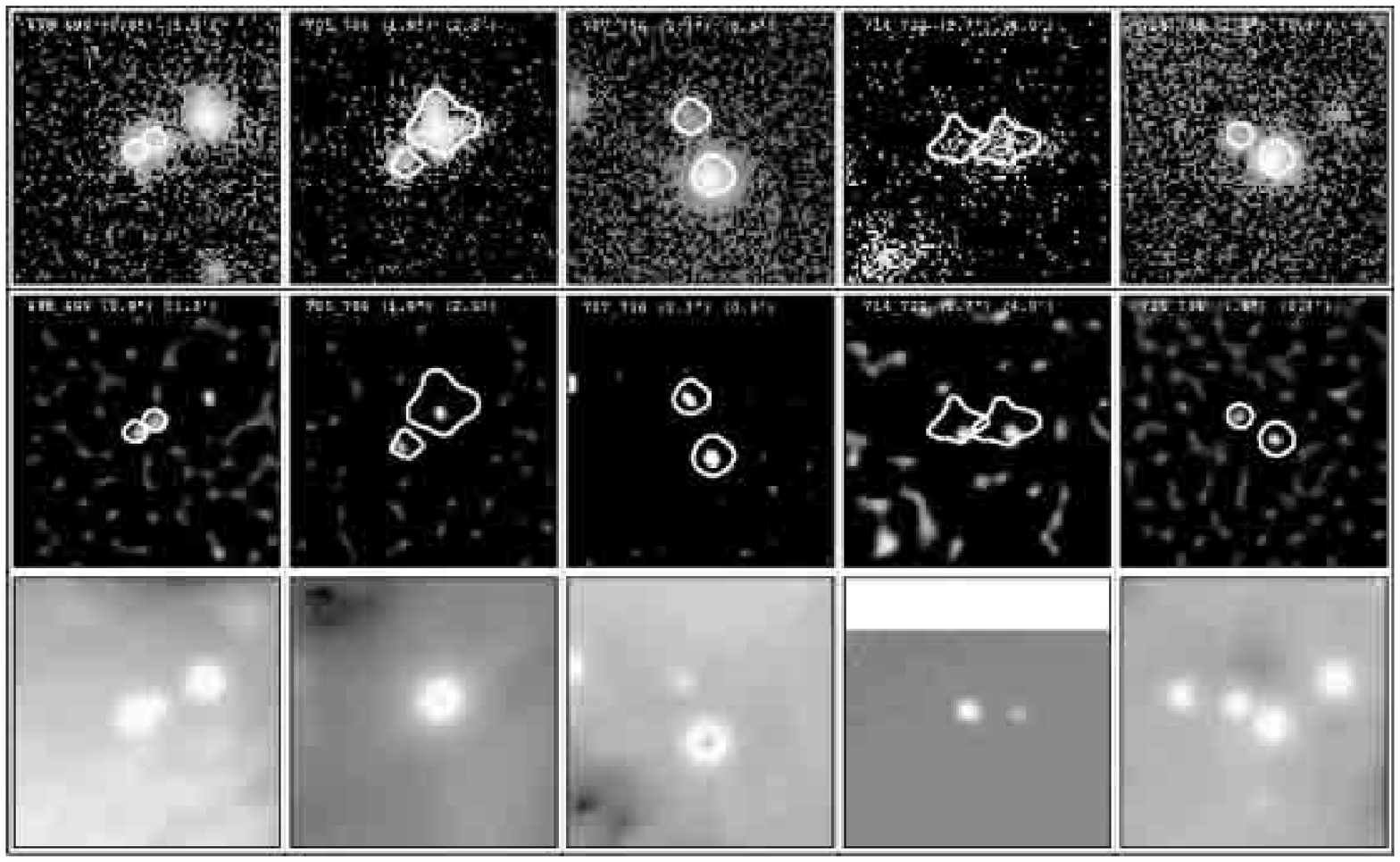} \hspace{0.00in}
\end{minipage}
\begin{minipage}[t]{1.0\textwidth}
  \centering
  \includegraphics[angle=0.,width=0.95\textwidth]{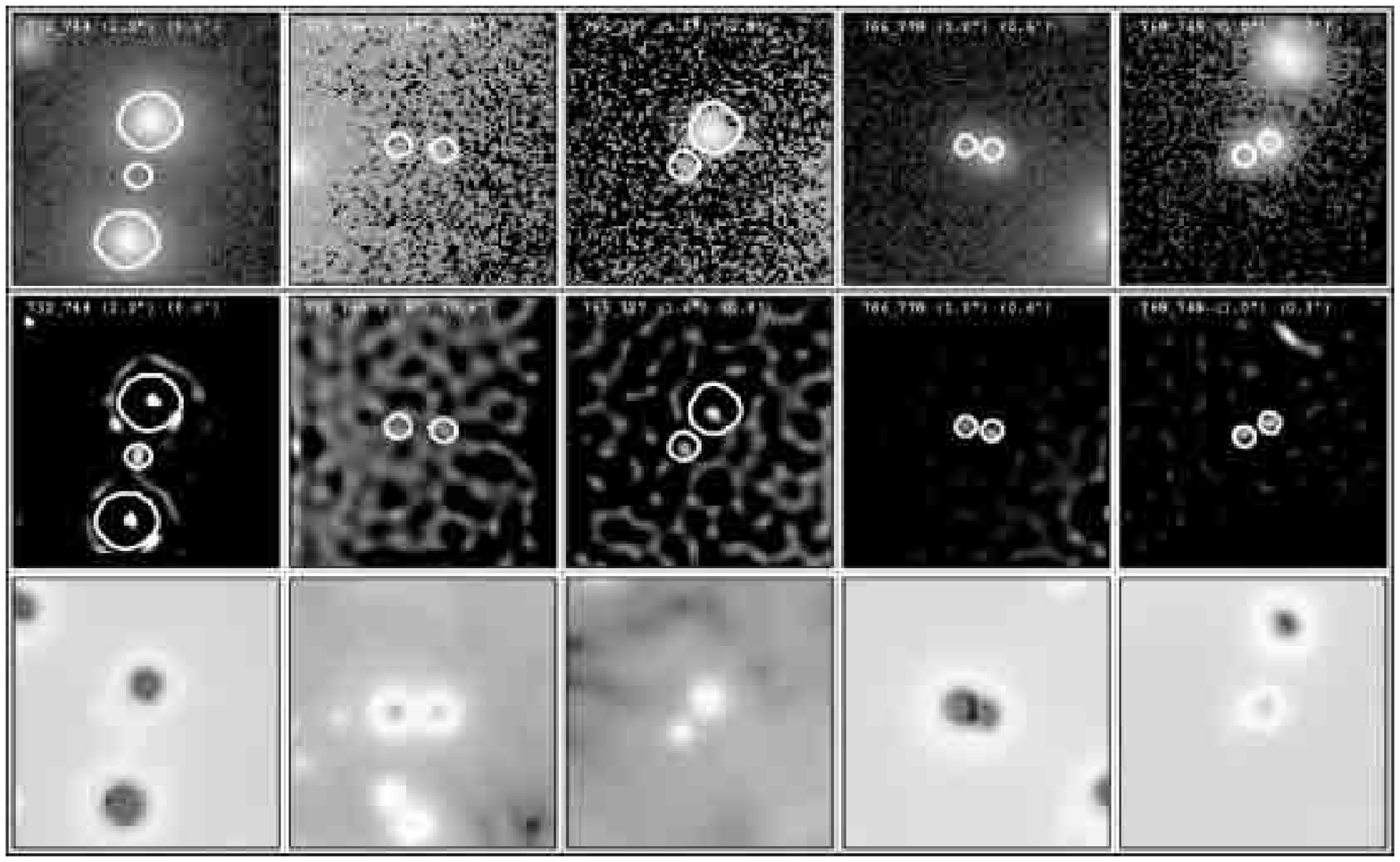} \hspace{0.00in}
\end{minipage}
\end{figure}

\clearpage
\newpage

\begin{figure}
\centering
\begin{minipage}[t]{1.0\textwidth}
  \centering
  \includegraphics[width=0.95\textwidth]{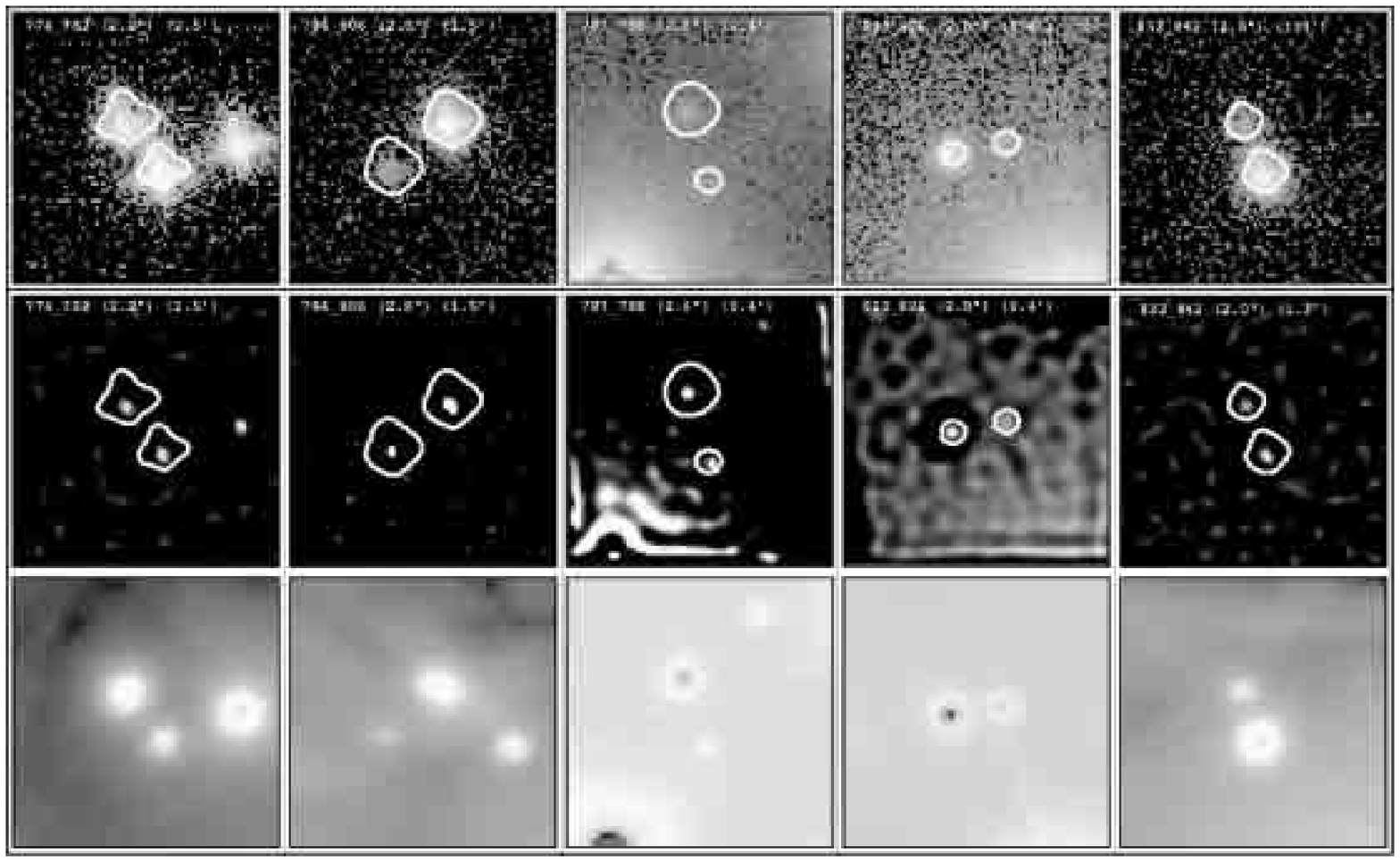} \hspace{0.00in}
\end{minipage}
\begin{minipage}[t]{1.0\textwidth}
  \centering
  \includegraphics[angle=0.,width=0.95\textwidth]{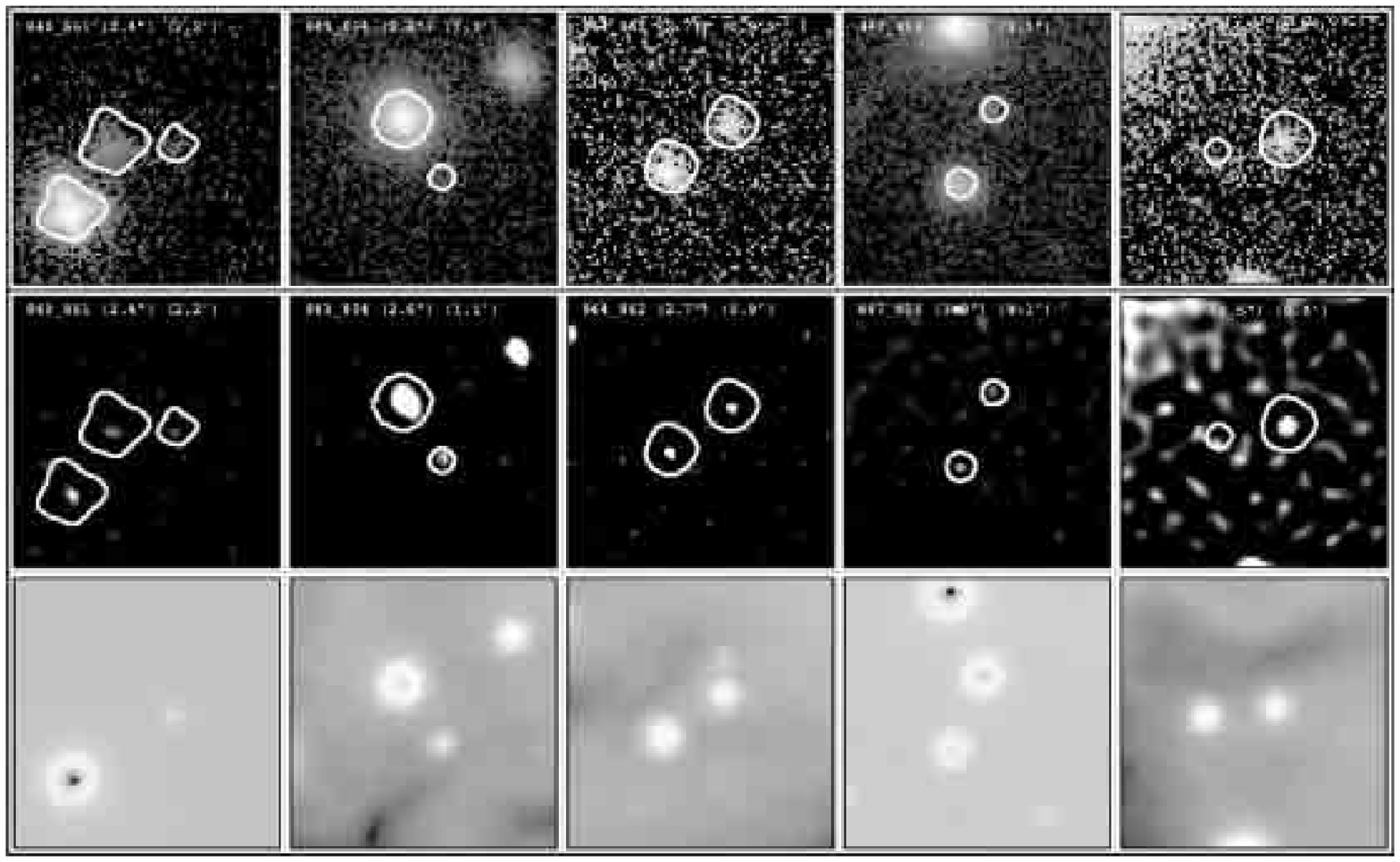} \hspace{0.00in}
\end{minipage}
\end{figure}

\clearpage
\newpage

\begin{figure}
\centering
\begin{minipage}[t]{1.0\textwidth}
  \centering
  \includegraphics[width=0.95\textwidth]{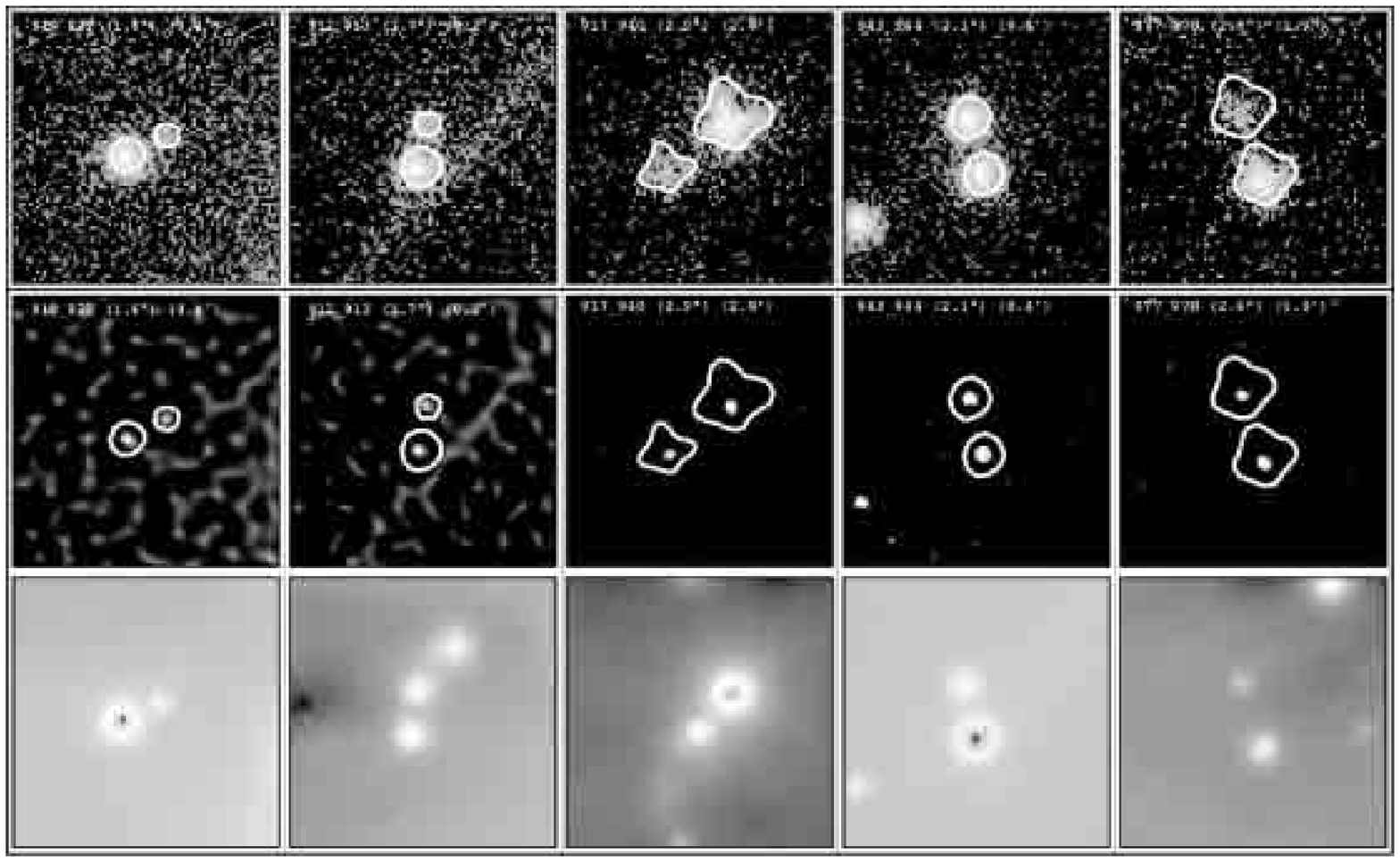} \hspace{0.00in}
\end{minipage}
\begin{minipage}[t]{1.0\textwidth}
  \centering
  \includegraphics[angle=0.,width=0.95\textwidth]{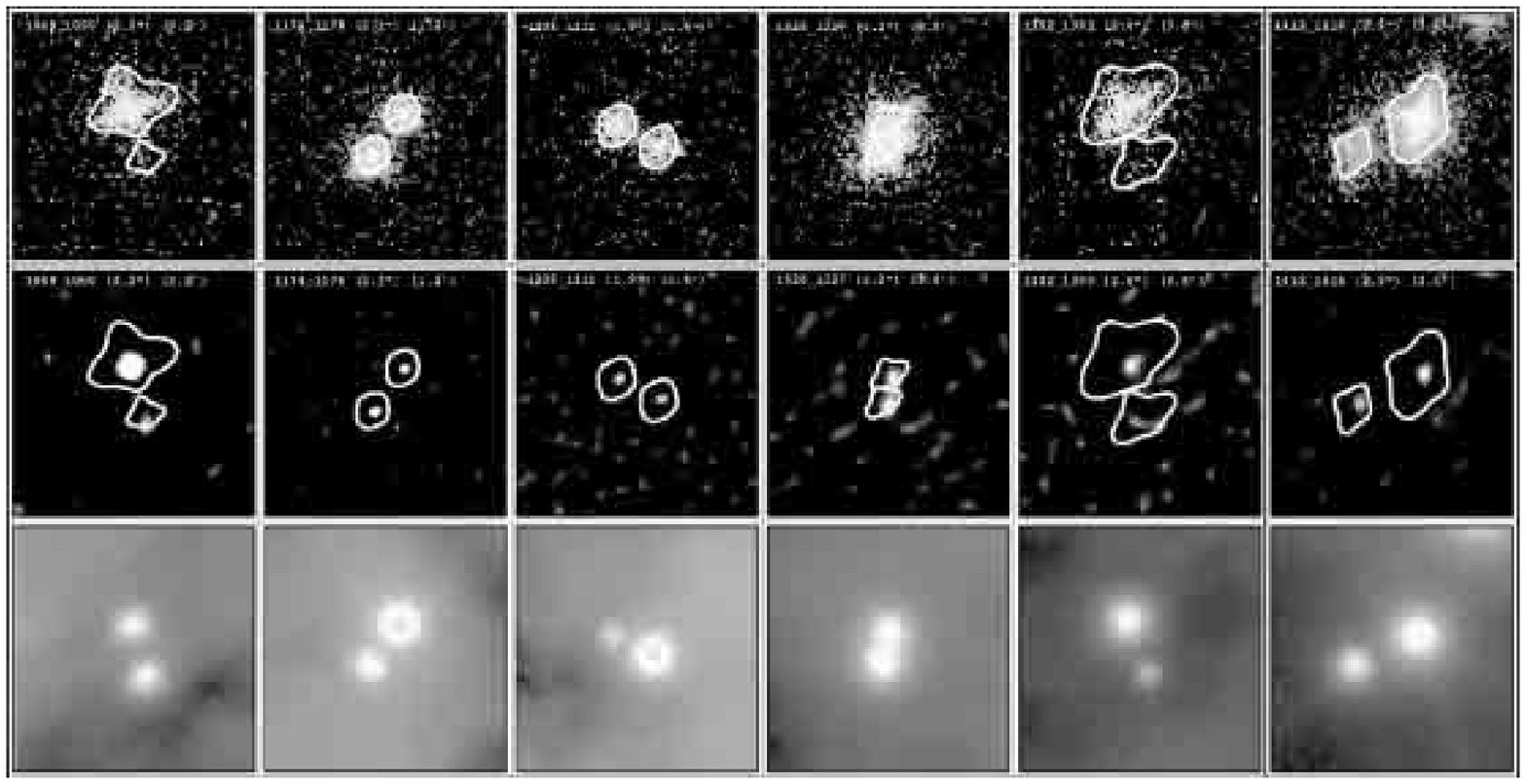} \hspace{0.00in}
\end{minipage}
\end{figure}

\clearpage
\newpage

\clearpage
\newpage

\clearpage
\newpage

\begin{deluxetable}{rrrrcccrrcc}
\centering
\tabletypesize{\scriptsize} \tablewidth{0pt}
\tablecolumns{11}

\tablecaption{Flaring COUP X-ray sources without optical or
near-infrared counterparts \label{flaring_table}}

\tablehead{

\colhead{Seq} & \colhead{COUP J} & \colhead{NC$_t$} &
\colhead{NC$_h$} & \colhead{PSF} & \colhead{MedE} &
\colhead{$N_H$} & \colhead{kT}
& \colhead{$F_h$} & \colhead{$N_H$ OMC} & \colhead{Mem} \\

\colhead{\#} && \colhead{cts} & \colhead{cts} &&
\colhead{keV} & \colhead{cm$^{-2}$}& \colhead{keV} &
\colhead{\tablenotemark{a}} & \colhead{cm$^{-2}$} & \colhead{Class}\\

\colhead{(1)} & \colhead{(2)} & \colhead{(3)} & \colhead{(4)} &
\colhead{(5)} & \colhead{(6)} & \colhead{(7)} & \colhead{(8)} &
\colhead{(9)} & \colhead{(10)} & \colhead{(11)}}

\startdata
\object[COUP 0261]{261} & 053506.1-052306 & 53.1 & 51.5 & 0.87 & 3.5 & 22.9 & 1.9 & 1.7 & 22.7 & OMC \\
\object[COUP 0288]{288} & 053507.4-052301 & 47.9 & 38.5 & 0.87 & 3.1 & 22.7 & 2.3 & 1.3 & 22.8 & OMC \\
\object[COUP 0377]{377} & 053510.4-052223 & 114.0 & 114.5 & 0.87 & 4.5 & 23.3 & 2.8 & 7.8 & 22.9 & OMC \\
\object[COUP 0425]{425} & 053511.5-052340 & 62.1 & 62.7 & 0.86 & 4.6 & 23.1 & 6.4 & 2.7 & 22.9 & OMC \\
\object[COUP 0436]{436} & 053511.6-052729 & 57.4 & 38.6 & 0.86 & 2.5 & 22.0 & 7.6 & 1.3 & 22.9 & OMC \\
\object[COUP 0471]{471} & 053512.2-052424 & 513.7 & 511.7 & 0.86 & 5.1 & 23.5 & 6.2 & 25.8 & 22.8 & OMC \\
\object[COUP 0487]{487} & 053512.6-052204 & 142.3 & 138.5 & 0.86 & 3.5 & 22.8 & 3.3 & 4.7 & 22.9 & OMC \\
\object[COUP 0510]{510} & 053512.9-052354 & 404.0 & 404.3 & 0.86 & 4.8 & 23.5 & 1.6 & 18.1 & 22.9 & OMC \\
\object[COUP 0511]{511} & 053512.9-052431 & 21.3 & 24.4 & 0.86 & 4.8 & 23.6 & 0.9 & 1.1 & 22.8 & OMC \\
\object[COUP 0532]{532} & 053513.2-052330 & 75.6 & 73.4 & 0.86 & 5.7 & 23.8 & 2.7 & 9.6 & 22.9 & OMC \\
\object[COUP 0574]{574} & 053513.7-052230 & 57.6 & 52.4 & 0.86 & 4.8 & 23.4 & 15.0 & 2.3 & 22.9 & OMC \\
\object[COUP 0577]{577} & 053513.8-052150 & 44.1 & 45.2 & 0.87 & 4.5 & 23.5 & 1.4 & 1.9 & 22.8 & OMC \\
\object[COUP 0582]{582} & 053513.8-052407 & 18.9 & 20.4 & 0.88 & 5.1 & 23.3 & 15.0 & 0.9 & 22.8 & OMC \\
\object[COUP 0591]{591} & 053513.9-052235 & 22.3 & 21.3 & 0.69 & 4.7 & 23.5 & 2.0 & 1.2 & 22.9 & OMC \\
\object[COUP 0594]{594} & 053513.9-052409 & 125.1 & 126.3 & 0.88 & 5.9 & 23.9 & 11.1 & 6.7 & 22.8 & OMC \\
\object[COUP 0598]{598} & 053514.0-052012 & 468.5 & 443.3 & 0.87 & 3.9 & 22.8 & 6.4 & 17.2 & 22.8 & OMC \\
\object[COUP 0599]{599} & 053514.0-052222 & 846.5 & 827.7 & 0.86 & 3.8 & 23.0 & 2.6 & 29.8 & 22.9 & OMC \\
\object[COUP 0615]{615} & 053514.2-052408 & 37.1 & 31.8 & 0.88 & 4.7 & 23.3 & 5.1 & 1.4 & 22.8 & OMC \\
\object[COUP 0617]{617} & 053514.2-052612 & 111.2 & 112.5 & 0.86 & 4.1 & 23.0 & 3.4 & 4.3 & 22.9 & OMC \\
\object[COUP 0625]{625} & 053514.3-052317 & 658.7 & 650.9 & 0.86 & 4.8 & 23.4 & 2.7 & 29.2 & 22.9 & OMC \\
\object[COUP 0641]{641} & 053514.5-052407 & 206.7 & 205.2 & 0.87 & 4.6 & 23.3 & 2.8 & 10.0 & 22.8 & OMC \\
\object[COUP 0647]{647} & 053514.6-052211 & 808.9 & 809.9 & 0.86 & 5.2 & 23.5 & 5.5 & 38.9 & 22.8 & OMC \\
\object[COUP 0656]{656} & 053514.7-052238 & 84.3 & 67.8 & 0.67 & 3.9 & 23.3 & 1.5 & 3.6 & 22.9 & OMC \\
\object[COUP 0662]{662} & 053514.8-052225 & 3270.8 & 3246.0 & 0.86 & 4.5 & 23.2 & 4.0 & 135.9 & 22.8 & OMC \\
\object[COUP 0667]{667} & 053514.8-052406 & 376.3 & 366.1 & 0.88 & 4.3 & 23.1 & 4.4 & 21.5 & 22.8 & OMC \\
\object[COUP 0678]{678} & 053515.0-052231 & 614.8 & 594.3 & 0.86 & 4.1 & 22.9 & 5.2 & 22.6 & 22.8 & OMC \\
\object[COUP 0680]{680} & 053515.1-052217 & 1384.3 & 1377.0 & 0.86 & 4.7 & 23.2 & 7.2 & 59.4 & 22.8 & OMC \\
\object[COUP 0681]{681} & 053515.1-052229 & 112.7 & 114.9 & 0.91 & 4.5 & 23.2 & 4.3 & 4.5 & 22.8 & OMC \\
\object[COUP 0683]{683} & 053515.1-052304 & 41.4 & 43.4 & 0.86 & 4.7 & 23.0 & 15.0 & 1.8 & 22.8 & OMC \\
\object[COUP 0706]{706} & 053515.4-052115 & 96.1 & 93.7 & 0.40 & 4.1 & 22.9 & 6.4 & 8.1 & 22.8 & OMC \\
\object[COUP 0721]{721} & 053515.6-051809 & 132.0 & 111.6 & 0.86 & 3.2 & 22.7 & 2.5 & 3.9 & 22.7 & OMC \\
\object[COUP 0730]{730} & 053515.7-051947 & 208.4 & 211.0 & 0.87 & 4.5 & 23.2 & 4.5 & 9.1 & 22.8 & OMC \\
\object[COUP 0760]{760} & 053516.0-051944 & 500.1 & 496.6 & 0.87 & 4.7 & 23.3 & 5.8 & 22.4 & 22.8 & OMC \\
\object[COUP 0797]{797} & 053516.3-052044 & 1716.0 & 1709.0 & 0.85 & 5.2 & 23.5 & 8.1 & 89.4 & 22.8 & OMC \\
\object[COUP 0811]{811} & 053516.5-052054 & 75.4 & 76.9 & 0.86 & 5.3 & 23.6 & 2.9 & 4.1 & 22.8 & OMC \\
\object[COUP 0861]{861} & 053517.1-052129 & 402.8 & 332.1 & 0.87 & 3.7 & 22.8 & 10.3 & 12.7 & 22.7 & OMC \\
\object[COUP 0872]{872} & 053517.3-052051 & 48.7 & 43.8 & 0.86 & 3.2 & 22.7 & 2.6 & 1.4 & 22.7 & OMC \\
\object[COUP 0940]{940} & 053518.0-052055 & 273.1 & 267.3 & 0.69 & 4.8 & 23.3 & 7.3 & 15.3 & 22.7 & OMC \\
\object[COUP 0984]{984} & 053518.6-051905 & 85.1 & 83.8 & 0.87 & 4.7 & 23.0 & 15.0 & 3.8 & 22.7 & OMC \\
\object[COUP 1051]{1051} & 053519.7-052110 & 170.9 & 166.3 & 0.86 & 4.2 & 22.8 & 15.0 & 6.7 & 22.6 & OMC \\
\object[COUP 1401]{1401} & 053527.9-051859 & 1565.5 & 1109.0 & 0.72 & 2.8 & 22.4 & 3.2 & 44.2 & 22.1 & OMC \\
\object[COUP 1465]{1465} & 053531.2-052215 & 56.3 & 54.1 & 0.87 & 3.8 & 23.0 & 2.4 & 2.1 & 22.4 & OMC \\
\enddata

\tablecomments{Column 1: COUP source number. Column 2: IAU
designation. Column 3: Total band ($0.5-8$ keV) net counts in
extracted area. Column 4: Hard band ($2-8$ keV) net counts in
extracted area. Column 5: Fraction of the point spread function
fraction in the extracted area. Column 6: Median energy of
extracted photons in the full band (corrected for background).
Column 7: Source hydrogen column density inferred from { it XSPEC}
spectral fit. Column 8: Source plasma energy from the spectral
fit. Column 9: Observed hard band flux in units of $10^{-15}$ erg
cm$^{-2}$ s$^{-1}$. Column 10: Average hydrogen column density in
OMC derived from the single-dish $^{13}$CO map by \citet{Bally87}.
Column 11: Membership class derived here (\S
\ref{candidates_section}). }

\tablenotetext{a}{in units of $10^{-15}$ erg s$^{-1}$ cm$^{-2}$ in
the $2-8$ keV band}

\end{deluxetable}

\clearpage
\newpage

\begin{deluxetable}{rrrrcccrrccc}
\centering
\tabletypesize{\scriptsize} \tablewidth{0pt} \tablecolumns{12}
\tablecaption{Non-flaring COUP X-ray sources without optical or
near-infrared counterparts \label{non_flaring_table}}

\tablehead{

\colhead{Seq} & \colhead{COUP J} & \colhead{NC$_t$} &
\colhead{NC$_h$} & \colhead{PSF} & \colhead{MedE} &
\colhead{$N_H$} & \colhead{kT} & \colhead{$F_h$} &
\colhead{$N_H$ OMC} & \colhead{Class} & \colhead{$P$}\\

\colhead{\#} && \colhead{cts} & \colhead{cts} &&
\colhead{keV} & \colhead{cm$^{-2}$}& \colhead{keV} &
\colhead{\tablenotemark{a}} & \colhead{cm$^{-2}$} & &\\

\colhead{(1)} & \colhead{(2)} & \colhead{(3)} &
\colhead{(4)} & \colhead{(5)} & \colhead{(6)} &
\colhead{(7)} & \colhead{(8)} & \colhead{(9)} &
\colhead{(10)} & \colhead{(11)} & \colhead{(12)}}

\startdata
\object[COUP 0004]{4} & 053434.9-052507 & 16.5 & 57.8 & 0.89 & 5.0 & 23.4 & 1.4 & 2.9 & 22.3 & EG & 0.11\\
\object[COUP 0005]{5} & 053438.2-052338 & 68.6 & 60.0 & 0.69 & 4.7 & 23.5 & 9.5 & 3.9 & 22.1 & EG & 0.21\\
\object[COUP 0008]{8} & 053439.8-052456 & 1129.7 & 706.9 & 0.90 & 2.6 & 22.0 & 15.0 & 25.4 & 22.1 & EG & 0.06\\
\object[COUP 0018]{18} & 053443.4-052059 & 49.8 & 41.3 & 0.88 & 4.2 & 22.6 & 15.0 & 2.0 & 22.3 & EG & 0.56\\
\object[COUP 0019]{19} & 053443.9-052208 & 39.0 & 25.1 & 0.89 & 2.6 & 22.6 & 2.3 & 0.8 & 22.1 & EG & 0.51\\
\object[COUP 0022]{22} & 053444.5-052548 & 96.8 & 69.5 & 0.89 & 2.9 & 22.4 & 15.0 & 3.1 & 22.1 & EG & 0.31\\
\object[COUP 0024]{24} & 053445.3-052239 & 53.6 & 45.3 & 0.88 & 3.3 & 22.6 & 2.8 & 2.0 & 22.3 & EG & 0.50\\
\object[COUP 0025]{25} & 053445.4-052212 & 172.5 & 113.2 & 0.89 & 2.7 & 22.0 & 15.0 & 4.0 & 22.3 & EG & 0.59\\
\object[COUP 0031]{31} & 053446.7-052632 & 14.5 & 10.7 & 0.88 & 4.2 & 22.2 & 6.8 & 0.4 & 22.1 & EG & 0.03\\
\object[COUP 0032]{32} & 053446.8-052345 & 134.7 & 136.6 & 0.88 & 4.2 & 23.1 & 5.0 & 5.6 & 22.3 & EG & 0.22\\
\object[COUP 0034]{34} & 053447.2-052206 & 35.1 & 19.5 & 0.88 & 2.3 & 21.8 & 15.0 & 0.8 & 22.3 & OMC or EG? & 0.70\\
\object[COUP 0035]{35} & 053447.2-052307 & 199.2 & 141.5 & 0.87 & 2.8 & 22.3 & 6.8 & 5.3 & 22.3 & EG & 0.46\\
\object[COUP 0036]{36} & 053447.5-052128 & 152.2 & 121.8 & 0.89 & 3.3 & 22.5 & 3.7 & 5.1 & 22.4 & OMC or EG? & 0.66\\
\object[COUP 0038]{38} & 053448.0-052054 & 133.7 & 125.7 & 0.88 & 3.6 & 22.7 & 5.0 & 4.7 & 22.4 & EG & 0.38\\
\object[COUP 0039]{39} & 053448.1-052151 & 19.2 & 13.2 & 0.68 & 2.7 & 22.7 & 0.6 & 0.6 & 22.4 & OMC or EG? & 0.67\\
\object[COUP 0042]{42} & 053448.8-052118 & 42.1 & 40.9 & 0.88 & 4.0 & 23.4 & 1.4 & 1.6 & 22.4 & EG & 0.53\\
\object[COUP 0048]{48} & 053449.4-052359 & 69.1 & 56.9 & 0.88 & 4.0 & 23.5 & 1.1 & 2.6 & 22.4 & EG & 0.17\\
\object[COUP 0051]{51} & 053450.2-051854 & 120.9 & 94.0 & 0.78 & 3.3 & 22.3 & 15.0 & 4.0 & 22.4 & EG & 0.09\\
\object[COUP 0052]{52} & 053450.3-052323 & 231.2 & 178.4 & 0.89 & 3.3 & 22.3 & 15.0 & 7.4 & 22.4 & EG & 0.46\\
\object[COUP 0053]{53} & 053450.3-052631 & 13.5 & 4.9 & 0.48 & 1.4 & 20.0 & 10.8 & 0.4 & 22.3 & ONC & \nodata\\
\object[COUP 0056]{56} & 053450.7-052112 & 64.4 & 51.9 & 0.88 & 3.4 & 22.1 & 15.0 & 1.9 & 22.6 & EG & 0.28\\
\object[COUP 0061]{61} & 053451.4-052658 & 12.1 & 9.7 & 0.50 & 5.2 & \nodata & \nodata & 1.0 & 22.3 & EG & 0.00\\
\object[COUP 0063]{63} & 053451.5-052618 & 111.6 & 95.4 & 0.89 & 3.3 & 22.5 & 8.3 & 3.9 & 22.4 & EG & 0.02\\
\object[COUP 0070]{70} & 053452.4-051835 & 18.9 & 17.7 & 0.88 & 2.6 & 22.5 & 15.0 & 0.7 & 22.4 & EG & 0.00\\
\object[COUP 0076]{76} & 053452.9-052616 & 58.7 & 39.0 & 0.88 & 3.0 & 22.1 & 10.9 & 1.4 & 22.4 & EG & 0.06\\
\object[COUP 0081]{81} & 053453.3-052628 & 138.8 & 88.2 & 0.68 & 2.8 & 22.3 & 4.1 & 4.5 & 22.4 & EG & 0.00\\
\object[COUP 0082]{82} & 053453.4-052219 & 32.6 & 34.0 & 0.88 & 3.4 & 23.0 & 1.0 & 1.1 & 22.6 & EG & 0.45\\
\object[COUP 0083]{83} & 053453.5-052651 & 224.0 & 161.5 & 0.89 & 3.2 & 22.1 & 15.0 & 6.6 & 22.4 & EG & 0.00\\
\object[COUP 0084]{84} & 053453.8-052636 & 41.9 & 16.0 & 0.89 & 1.8 & 20.0 & 15.0 & 0.5 & 22.6 & ONC & \nodata\\
\object[COUP 0087]{87} & 053454.0-052501 & 41.7 & 26.1 & 0.88 & 3.2 & 22.0 & 15.0 & 1.1 & 22.6 & EG & 0.26\\
\object[COUP 0091]{91} & 053454.3-052207 & 79.1 & 69.1 & 0.88 & 4.1 & 23.0 & 15.0 & 3.0 & 22.6 & EG & 0.29\\
\object[COUP 0104]{104} & 053455.8-052338 & 201.2 & 186.9 & 0.87 & 3.8 & 22.7 & 8.6 & 7.3 & 22.6 & EG & 0.34\\
\object[COUP 0111]{111} & 053456.2-052229 & 999.2 & 849.2 & 0.87 & 3.5 & 22.5 & 10.4 & 32.5 & 22.6 & EG & 0.09\\
\object[COUP 0120]{120} & 053457.5-052346 & 63.5 & 47.2 & 0.88 & 4.1 & 23.0 & 15.0 & 2.0 & 22.6 & EG & 0.00\\
\object[COUP 0121]{121} & 053457.7-052223 & 285.2 & 261.3 & 0.88 & 3.8 & 22.6 & 15.0 & 10.2 & 22.6 & EG & 0.07\\
\object[COUP 0135]{135} & 053459.4-052251 & 14.6 & 2.2 & 0.88 & 1.1 & 20.0 & 1.6 & 0.1 & 22.6 & ONC & \nodata\\
\object[COUP 0136]{136} & 053459.4-052615 & 287.2 & 219.7 & 0.88 & 3.1 & 22.3 & 15.0 & 8.4 & 22.6 & EG & 0.20\\
\object[COUP 0138]{138} & 053459.6-052032 & 35.7 & 35.2 & 0.90 & 3.7 & 22.9 & 1.9 & 1.3 & 22.4 & EG & 0.00\\
\object[COUP 0145]{145} & 053500.2-052549 & 116.9 & 89.7 & 0.88 & 3.2 & 22.2 & 15.0 & 3.1 & 22.6 & EG & 0.08\\
\object[COUP 0146]{146} & 053500.4-051754 & 79.1 & 66.4 & 0.90 & 3.5 & 22.5 & 15.0 & 2.5 & 22.4 & EG & 0.00\\
\object[COUP 0151]{151} & 053501.1-052241 & 20.6 & 4.1 & 0.88 & 1.4 & 22.0 & 0.6 & 0.1 & 22.6 & EG & 0.30\\
\object[COUP 0156]{156} & 053501.2-052524 & 16.7 & 1.0 & 0.86 & 1.0 & 21.6 & 0.5 & 0.1 & 22.6 & ONC & \nodata\\
\object[COUP 0163]{163} & 053501.5-052158 & 98.9 & 87.3 & 0.86 & 4.0 & 22.8 & 4.3 & 3.4 & 22.6 & EG & 0.15\\
\object[COUP 0178]{178} & 053502.5-052601 & 20.9 & 18.5 & 0.88 & 3.7 & 22.4 & 15.0 & 0.7 & 22.6 & EG & 0.00\\
\object[COUP 0184]{184} & 053502.8-052628 & 23.0 & 17.5 & 0.90 & 3.5 & 21.9 & 15.0 & 0.6 & 22.6 & Unc & \nodata\\
\object[COUP 0185]{185} & 053502.9-051551 & 339.8 & 118.4 & 0.87 & 1.6 & 21.8 & 1.3 & 8.8 & 22.6 & ONC & \nodata\\
\object[COUP 0186]{186} & 053502.9-053244 & 45.8 & 34.0 & 0.88 & 2.7 & 23.0 & 3.3 & 1.8 & 22.6 & EG & 0.00\\
\object[COUP 0191]{191} & 053503.2-053019 & 56.2 & 63.5 & 0.89 & 4.2 & 23.1 & 7.8 & 2.7 & 22.6 & EG & 0.00\\
\object[COUP 0196]{196} & 053503.5-052624 & 9.3 & 4.7 & 0.49 & 2.2 & \nodata & \nodata & 0.4 & 22.7 & Unc & \nodata\\
\object[COUP 0198]{198} & 053503.6-052705 & 13.0 & 13.8 & 0.88 & 3.5 & 22.7 & 4.3 & 0.9 & 22.7 & EG & 0.00\\
\object[COUP 0203]{203} & 053503.9-052940 & 834.2 & 624.9 & 0.88 & 3.1 & 22.3 & 14.2 & 22.5 & 22.6 & EG & 0.00\\
\object[COUP 0204]{204} & 053504.0-052056 & 9.6 & 6.1 & 0.87 & 2.4 & \nodata & \nodata & 0.4 & 22.6 & EG & 0.00\\
\object[COUP 0225]{225} & 053504.7-052551 & 6.6 & 3.5 & 0.49 & 1.7 & \nodata & \nodata & 0.3 & 22.7 & Unc & \nodata\\
\object[COUP 0229]{229} & 053504.8-052302 & 68.9 & 68.1 & 0.87 & 4.1 & 23.0 & 4.6 & 2.6 & 22.7 & EG & 0.02\\
\object[COUP 0257]{257} & 053505.8-053418 & 80.8 & 41.2 & 0.90 & 2.2 & 21.3 & 15.0 & 4.9 & 22.8 & ONC & \nodata\\
\object[COUP 0258]{258} & 053505.9-052550 & 16.0 & 12.1 & 0.90 & 3.3 & 22.5 & 2.6 & 0.4 & 22.7 & EG & 0.00\\
\object[COUP 0263]{263} & 053506.3-052336 & 197.7 & 184.9 & 0.87 & 4.0 & 22.7 & 15.0 & 7.1 & 22.7 & EG & 0.22\\
\object[COUP 0268]{268} & 053506.5-051624 & 70.0 & 55.4 & 0.88 & 2.9 & 22.1 & 15.0 & 2.2 & 22.6 & EG & 0.00\\
\object[COUP 0277]{277} & 053506.9-052244 & 15.6 & 16.3 & 0.87 & 3.5 & 22.8 & 1.4 & 0.6 & 22.7 & EG & 0.39\\
\object[COUP 0278]{278} & 053506.9-052421 & 7.4 & 2.1 & 0.50 & 1.9 & \nodata & \nodata & 0.1 & 22.8 & Unc & \nodata\\
\object[COUP 0282]{282} & 053507.1-052650 & 28.7 & 25.8 & 0.87 & 3.2 & 22.3 & 15.0 & 0.9 & 22.8 & EG & 0.00\\
\object[COUP 0284]{284} & 053507.2-052253 & 111.0 & 114.4 & 0.86 & 4.9 & 23.4 & 9.6 & 5.6 & 22.7 & EG & 0.43\\
\object[COUP 0295]{295} & 053507.7-052626 & 13.0 & 8.9 & 0.86 & 2.3 & 20.0 & 15.0 & 0.2 & 22.8 & Unc & \nodata\\
\object[COUP 0297]{297} & 053507.8-052029 & 86.2 & 82.6 & 0.87 & 4.1 & 22.8 & 15.0 & 3.3 & 22.6 & EG & 0.02\\
\object[COUP 0306]{306} & 053508.3-051502 & 13.8 & 9.1 & 0.88 & 3.5 & 21.8 & 15.0 & 0.8 & 22.6 & Unc & \nodata\\
\object[COUP 0317]{317} & 053508.5-052501 & 18.4 & 18.9 & 0.86 & 5.0 & 23.5 & 10.5 & 0.8 & 22.8 & EG & 0.30\\
\object[COUP 0330]{330} & 053509.2-051648 & 139.6 & 130.1 & 0.78 & 3.8 & 23.0 & 2.4 & 5.9 & 22.6 & EG & 0.00\\
\object[COUP 0354]{354} & 053510.0-052004 & 95.0 & 91.0 & 0.87 & 3.8 & 22.8 & 4.3 & 3.3 & 22.6 & EG & 0.21\\
\object[COUP 0356]{356} & 053510.0-052618 & 9.5 & 1.7 & 0.49 & 1.0 & \nodata & \nodata & 0.120 & 22.8 & Unc & \nodata\\
\object[COUP 0360]{360} & 053510.2-051719 & 53.3 & 49.1 & 0.88 & 5.1 & 23.6 & 3.6 & 2.5 & 22.6 & EG & 0.00\\
\object[COUP 0366]{366} & 053510.2-052623 & 6.4 & 0.1 & 0.49 & 0.9 & \nodata & \nodata & 0.0 & 22.9 & Unc & \nodata\\
\object[COUP 0381]{381} & 053510.5-052003 & 68.9 & 63.5 & 0.87 & 3.7 & 22.6 & 15.0 & 2.5 & 22.6 & EG & 0.16\\
\object[COUP 0388]{388} & 053510.5-053143 & 12.5 & 9.0 & 0.50 & 3.1 & 22.2 & 4.9 & 0.6 & 22.6 & EG & 0.00\\
\object[COUP 0400]{400} & 053510.8-052804 & 39.9 & 6.5 & 0.38 & 1.2 & 21.5 & 1.1 & 0.5 & 22.8 & ONC & \nodata\\
\object[COUP 0402]{402} & 053510.9-052227 & 80.6 & 74.4 & 0.87 & 4.1 & 22.8 & 8.9 & 4.0 & 22.9 & OMC or EG? & 0.81\\
\object[COUP 0406]{406} & 053510.9-053006 & 26.3 & 24.9 & 0.68 & 3.7 & 22.8 & 2.8 & 1.2 & 22.7 & EG & 0.00\\
\object[COUP 0412]{412} & 053511.1-052606 & 6.4 & 0.4 & 0.49 & 1.2 & \nodata & \nodata & 0.0 & 22.9 & Unc & \nodata\\
\object[COUP 0451]{451} & 053511.8-052403 & 4.5 & 2.5 & 0.50 & 2.2 & \nodata & \nodata & 0.2 & 22.9 & Unc & \nodata\\
\object[COUP 0456]{456} & 053511.9-052303 & 3.6 & -0.8 & 0.86 & 1.0 & \nodata & \nodata & \nodata & 22.9 & Unc & \nodata\\
\object[COUP 0463]{463} & 053512.1-052008 & 20.3 & 17.8 & 0.86 & 4.4 & 22.7 & 15.0 & 0.8 & 22.7 & EG & 0.32\\
\object[COUP 0473]{473} & 053512.2-052600 & 6.9 & 2.8 & 0.60 & 1.7 & \nodata & \nodata & 0.1 & 22.9 & Unc & \nodata\\
\object[COUP 0479]{479} & 053512.4-052031 & 32.9 & 28.0 & 0.87 & 4.4 & 20.0 & 15.0 & 1.2 & 22.8 & EG & 0.54\\
\object[COUP 0495]{495} & 053512.7-052254 & 16.0 & 4.3 & 0.86 & 1.4 & 20.0 & 15.0 & 0.1 & 22.9 & ONC & \nodata\\
\object[COUP 0496]{496} & 053512.7-052410 & 11.5 & 13.0 & 0.86 & 5.2 & 23.3 & 15.0 & 0.7 & 22.9 & OMC or EG? & 0.71\\
\object[COUP 0505]{505} & 053512.8-052247 & 12.7 & 7.8 & 0.86 & 3.8 & 20.0 & 15.0 & 0.4 & 22.9 & OMC or EG? & 0.84\\
\object[COUP 0506]{506} & 053512.8-052342 & 8.8 & 7.0 & 0.60 & 4.8 & \nodata & \nodata & 0.6 & 22.9 & OMC or EG? & 0.83\\
\object[COUP 0508]{508} & 053512.9-052254 & 5.8 & 6.0 & 0.50 & 4.3 & \nodata & \nodata & 0.4 & 22.9 & OMC or EG? & 0.89\\
\object[COUP 0509]{509} & 053512.9-052330 & 40.7 & 39.3 & 0.86 & 5.1 & 23.4 & 15.0 & 3.6 & 22.9 & OMC or EG? & 0.89\\
\object[COUP 0521]{521} & 053513.1-051809 & 16.3 & 1.2 & 0.87 & 1.2 & 21.6 & 0.8 & 0.1 & 22.7 & Unc & \nodata\\
\object[COUP 0525]{525} & 053513.1-052250 & 63.3 & 58.7 & 0.86 & 5.2 & 23.5 & 15.0 & 2.9 & 22.9 & OMC or EG? & 0.87\\
\object[COUP 0530]{530} & 053513.2-052254 & 539.4 & 534.2 & 0.86 & 5.2 & 23.5 & 5.5 & 25.9 & 22.9 & OMC or EG? & 0.89\\
\object[COUP 0568]{568} & 053513.6-053212 & 109.6 & 109.9 & 0.88 & 3.8 & 23.0 & 1.5 & 4.3 & 22.6 & EG & 0.00\\
\object[COUP 0571]{571} & 053513.7-052147 & 37.2 & 39.0 & 0.87 & 4.6 & 23.2 & 15.0 & 1.6 & 22.8 & OMC or EG? & 0.73\\
\object[COUP 0575]{575} & 053513.7-052438 & 12.4 & 11.2 & 0.86 & 4.1 & \nodata & \nodata & 0.9 & 22.8 & OMC or EG? & 0.73\\
\object[COUP 0581]{581} & 053513.8-052246 & 14.1 & 8.9 & 0.68 & 2.5 & 22.5 & 1.1 & 0.3 & 22.9 & OMC or EG? & 0.88\\
\object[COUP 0589]{589} & 053513.9-052229 & 60.0 & 57.6 & 0.68 & 3.7 & 23.3 & 1.3 & 2.6 & 22.9 & OMC or EG? & 0.91\\
\object[COUP 0601]{601} & 053514.0-052313 & 9.0 & 4.4 & 0.47 & 2.3 & \nodata & \nodata & 0.3 & 22.9 & Unc & \nodata\\
\object[COUP 0607]{607} & 053514.1-052357 & 10.8 & 6.6 & 0.49 & 4.4 & \nodata & \nodata & 1.0 & 22.8 & OMC or EG? & 0.80\\
\object[COUP 0608]{608} & 053514.1-052620 & 57.7 & 59.5 & 0.86 & 4.4 & 23.4 & 1.8 & 2.4 & 22.9 & EG & 0.00\\
\object[COUP 0611]{611} & 053514.2-051804 & 64.0 & 64.9 & 0.87 & 3.9 & 23.0 & 1.4 & 2.5 & 22.7 & EG & 0.22\\
\object[COUP 0613]{613} & 053514.2-052154 & 9.1 & 12.5 & 0.85 & 5.0 & 23.6 & 1.4 & 0.5 & 22.8 & OMC or EG? & 0.79\\
\object[COUP 0628]{628} & 053514.4-052230 & 93.2 & 88.4 & 0.67 & 5.0 & 23.4 & 6.0 & 5.5 & 22.9 & OMC or EG? & 0.90\\
\object[COUP 0633]{633} & 053514.4-052410 & 20.5 & 22.9 & 0.86 & 4.5 & 23.7 & 0.6 & 1.0 & 22.8 & OMC or EG? & 0.83\\
\object[COUP 0635]{635} & 053514.4-052541 & 3.4 & 3.1 & 0.50 & 2.4 & \nodata & \nodata & 0.119 & 22.9 & Unc & \nodata\\
\object[COUP 0637]{637} & 053514.5-051956 & 20.1 & 6.5 & 0.89 & 1.7 & 21.5 & 5.6 & 0.2 & 22.8 & Unc & \nodata\\
\object[COUP 0659]{659} & 053514.7-052412 & 175.8 & 162.5 & 0.88 & 5.0 & 23.6 & 2.2 & 7.6 & 22.8 & OMC or EG? & 0.86\\
\object[COUP 0675]{675} & 053514.9-052449 & 6.2 & 4.9 & 0.49 & 5.3 & \nodata & \nodata & 0.5 & 22.8 & OMC or EG? & 0.72\\
\object[COUP 0676]{676} & 053514.9-052734 & 12.7 & 0.6 & 0.87 & 1.3 & 20.0 & 2.9 & 0.0 & 22.9 & Unc & \nodata\\
\object[COUP 0692]{692} & 053515.2-052509 & 6.9 & 4.1 & 0.49 & 2.1 & \nodata & \nodata & 0.3 & 22.8 & Unc & \nodata\\
\object[COUP 0702]{702} & 053515.4-051934 & 79.6 & 52.6 & 0.86 & 2.7 & 22.3 & 3.2 & 1.6 & 22.8 & EG & 0.54\\
\object[COUP 0703]{703} & 053515.4-052040 & 32.0 & 4.6 & 0.87 & 0.9 & 21.9 & 0.1 & 0.101 & 22.8 & HH & \nodata\\
\object[COUP 0704]{704} & 053515.4-052045 & 24.3 & 2.6 & 0.85 & 0.9 & 20.0 & 0.3 & 0.091 & 22.8 & HH & \nodata\\
\object[COUP 0709]{709} & 053515.4-052507 & 69.1 & 65.7 & 0.86 & 4.8 & 23.2 & 15.0 & 3.0 & 22.8 & OMC or EG? & 0.70\\
\object[COUP 0723]{723} & 053515.6-052126 & 585.1 & 580.0 & 0.87 & 4.8 & 23.2 & 15.0 & 25.7 & 22.8 & OMC or EG? & 0.64\\
\object[COUP 0748]{748} & 053515.8-052417 & 8.7 & 4.3 & 0.87 & 2.3 & \nodata & \nodata & 1.690 & 22.8 & Unc & \nodata\\
\object[COUP 0749]{749} & 053515.8-052457 & 70.5 & 71.7 & 0.87 & 4.9 & 23.1 & 15.0 & 3.2 & 22.8 & OMC or EG? & 0.67\\
\object[COUP 0751]{751} & 053515.8-053005 & 337.6 & 204.3 & 0.87 & 2.3 & 22.4 & 0.9 & 5.9 & 22.6 & EG & 0.00\\
\object[COUP 0755]{755} & 053515.9-052200 & 319.2 & 322.4 & 0.87 & 4.7 & 23.3 & 3.3 & 14.0 & 22.8 & OMC or EG? & 0.71\\
\object[COUP 0767]{767} & 053516.0-052334 & 35.2 & 7.9 & 0.49 & 1.2 & 20.0 & 2.0 & 0.4 & 22.8 & Unc & \nodata\\
\object[COUP 0772]{772} & 053516.0-052720 & 7.8 & -1.2 & 0.86 & 1.2 & \nodata & \nodata & \nodata & 22.8 & Unc & \nodata\\
\object[COUP 0786]{786} & 053516.2-052301 & 15.6 & 5.4 & 0.36 & 1.6 & 20.0 & 15.0 & 0.3 & 22.8 & Unc & \nodata\\
\object[COUP 0793]{793} & 053516.2-053332 & 107.3 & 98.5 & 0.74 & 3.7 & 22.8 & 4.7 & 4.9 & 22.6 & EG & 0.00\\
\object[COUP 0805]{805} & 053516.4-051952 & 28.5 & 24.4 & 0.87 & 4.7 & 23.0 & 15.0 & 1.1 & 22.8 & OMC or EG? & 0.60\\
\object[COUP 0819]{819} & 053516.6-052241 & 42.6 & 37.6 & 0.86 & 5.0 & 23.5 & 15.0 & 1.8 & 22.7 & OMC or EG? & 0.82\\
\object[COUP 0824]{824} & 053516.7-052158 & 121.0 & 119.9 & 0.86 & 4.2 & 22.9 & 9.0 & 4.7 & 22.7 & OMC or EG? & 0.77\\
\object[COUP 0838]{838} & 053516.9-051946 & 8.6 & 9.2 & 0.87 & 4.5 & 22.9 & 15.0 & 0.4 & 22.8 & EG & 0.58\\
\object[COUP 0846]{846} & 053516.9-052302 & 14.2 & 1.4 & 0.36 & 1.2 & 21.1 & 1.2 & 0.2 & 22.7 & Unc & \nodata\\
\object[COUP 0868]{868} & 053517.2-052135 & 57.4 & 11.2 & 0.86 & 1.1 & 20.0 & 1.1 & 0.3 & 22.7 & Unc & \nodata\\
\object[COUP 0877]{877} & 053517.3-052240 & 17.2 & 12.3 & 0.90 & 2.1 & \nodata & \nodata & 0.2 & 22.7 & Unc & \nodata\\
\object[COUP 0895]{895} & 053517.5-052037 & 119.2 & 107.2 & 0.87 & 3.2 & 22.8 & 1.8 & 3.3 & 22.7 & EG & 0.53\\
\object[COUP 0905]{905} & 053517.6-052233 & 8.7 & 0.1 & 0.41 & 1.3 & \nodata & \nodata & 0.0 & 22.7 & Unc & \nodata\\
\object[COUP 0908]{908} & 053517.7-051928 & 45.8 & 9.7 & 0.48 & 1.5 & 21.4 & 2.2 & 0.7 & 22.8 & ONC & \nodata\\
\object[COUP 0911]{911} & 053517.7-052320 & 32.9 & 8.1 & 0.69 & 1.3 & 20.0 & 1.6 & 0.5 & 22.7 & ONC & \nodata\\
\object[COUP 0916]{916} & 053517.8-051843 & 22.0 & 13.4 & 0.87 & 2.4 & 22.5 & 1.2 & 0.5 & 22.8 & OMC or EG? & 0.62\\
\object[COUP 0923]{923} & 053517.8-052321 & 22.7 & 10.1 & 0.68 & 1.6 & 20.0 & 15.0 & 0.4 & 22.7 & Unc & \nodata\\
\object[COUP 0933]{933} & 053517.9-052326 & 18.1 & 4.2 & 0.68 & 1.2 & 21.7 & 0.6 & 0.2 & 22.7 & Unc & \nodata\\
\object[COUP 0959]{959} & 053518.2-052926 & 8.7 & 2.0 & 0.33 & 1.2 & \nodata & \nodata & 0.2 & 22.6 & Unc & \nodata\\
\object[COUP 0961]{961} & 053518.3-051501 & 119.8 & 77.4 & 0.87 & 2.6 & 22.1 & 15.0 & 3.4 & 22.6 & EG & 0.00\\
\object[COUP 0973]{973} & 053518.5-051531 & 140.8 & 106.1 & 0.48 & 3.3 & 22.4 & 13.6 & 7.6 & 22.6 & EG & 0.00\\
\object[COUP 0975]{975} & 053518.5-051929 & 14.8 & 11.4 & 0.87 & 4.2 & 22.1 & 15.0 & 0.5 & 22.7 & OMC or EG? & 0.75\\
\object[COUP 1015]{1015} & 053519.1-052112 & 143.1 & 131.3 & 0.87 & 3.6 & 22.8 & 2.3 & 4.5 & 22.6 & EG & 0.42\\
\object[COUP 1016]{1016} & 053519.1-052118 & 283.9 & 281.4 & 0.85 & 4.5 & 23.0 & 15.0 & 12.0 & 22.6 & EG & 0.45\\
\object[COUP 1020]{1020} & 053519.1-052528 & 3.1 & 1.9 & 0.47 & 2.3 & \nodata & \nodata & 0.1 & 22.8 & Unc & \nodata\\
\object[COUP 1031]{1031} & 053519.3-053419 & 38.4 & 31.5 & 0.49 & 3.8 & 22.2 & 15.0 & 2.7 & 22.4 & EG & 0.00\\
\object[COUP 1033]{1033} & 053519.4-052716 & 14.2 & 8.2 & 0.87 & 2.6 & 21.3 & 15.0 & 0.3 & 22.7 & ONC & \nodata\\
\object[COUP 1042]{1042} & 053519.6-052057 & 45.3 & 31.7 & 0.87 & 2.6 & 21.9 & 15.0 & 0.9 & 22.6 & ONC & \nodata\\
\object[COUP 1055]{1055} & 053519.8-051841 & 27.3 & 20.2 & 0.88 & 3.1 & \nodata & \nodata & 0.7 & 22.7 & EG & 0.57\\
\object[COUP 1057]{1057} & 053519.8-052440 & 14.8 & 13.7 & 0.87 & 3.7 & 22.5 & 15.0 & 0.5 & 22.6 & EG & 0.52\\
\object[COUP 1072]{1072} & 053520.0-052223 & 8.0 & 5.6 & 0.49 & 2.1 & \nodata & \nodata & 0.3 & 22.6 & EG & 0.45\\
\object[COUP 1078]{1078} & 053520.0-052916 & 10.7 & 4.5 & 0.33 & 1.9 & \nodata & \nodata & 0.4 & 22.4 & EG & 0.00\\
\object[COUP 1092]{1092} & 053520.2-052706 & 13.8 & 4.2 & 0.87 & 1.3 & 20.0 & 1.9 & 0.1 & 22.7 & ONC & \nodata\\
\object[COUP 1098]{1098} & 053520.4-051932 & 195.8 & 186.4 & 0.86 & 3.7 & 22.8 & 5.3 & 7.0 & 22.6 & OMC or EG? & 0.72\\
\object[COUP 1099]{1099} & 053520.4-052019 & 14.3 & 10.3 & 0.88 & 3.5 & 23.4 & 0.6 & 0.4 & 22.6 & OMC or EG? & 0.65\\
\object[COUP 1108]{1108} & 053520.5-052425 & 26.2 & 21.8 & 0.87 & 4.1 & 22.5 & 15.0 & 0.9 & 22.6 & EG & 0.48\\
\object[COUP 1109]{1109} & 053520.5-052632 & 38.5 & 32.7 & 0.87 & 3.7 & 23.0 & 3.1 & 1.2 & 22.7 & EG & 0.22\\
\object[COUP 1123]{1123} & 053520.9-052234 & 312.4 & 291.0 & 0.86 & 4.1 & 22.7 & 15.0 & 11.5 & 22.4 & EG & 0.33\\
\object[COUP 1133]{1133} & 053521.0-052905 & 24.0 & 16.6 & 0.87 & 4.9 & 23.3 & 15.0 & 1.1 & 22.4 & EG & 0.00\\
\object[COUP 1146]{1146} & 053521.3-051902 & 39.5 & 36.3 & 0.87 & 3.5 & 22.6 & 3.8 & 1.5 & 22.6 & OMC or EG? & 0.60\\
\object[COUP 1157]{1157} & 053521.6-051754 & 33.4 & 23.3 & 0.88 & 3.2 & 20.0 & 15.0 & 0.8 & 22.6 & EG & 0.46\\
\object[COUP 1176]{1176} & 053521.8-052441 & 18.8 & 15.9 & 0.85 & 4.0 & 22.6 & 15.0 & 0.6 & 22.6 & EG & 0.30\\
\object[COUP 1183]{1183} & 053522.0-051933 & 33.5 & 37.8 & 0.86 & 4.2 & 23.0 & 4.7 & 1.6 & 22.6 & EG & 0.54\\
\object[COUP 1189]{1189} & 053522.1-052129 & 4.7 & -2.4 & 0.87 & 1.1 & \nodata & \nodata & \nodata & 22.4 & Unc & \nodata\\
\object[COUP 1192]{1192} & 053522.1-052331 & 13.6 & -0.9 & 0.87 & 0.9 & 21.1 & 0.3 & \nodata & 22.4 & ONC & \nodata\\
\object[COUP 1221]{1221} & 053522.6-053338 & 76.7 & 75.3 & 0.59 & 3.7 & 22.7 & 4.5 & 4.6 & 22.1 & EG & 0.00\\
\object[COUP 1228]{1228} & 053522.7-052514 & 14.6 & 13.6 & 0.86 & 3.5 & 23.3 & 0.7 & 0.5 & 22.6 & EG & 0.23\\
\object[COUP 1237]{1237} & 053523.0-052459 & 799.9 & 56.0 & 0.86 & 1.0 & 20.0 & 0.6 & 1.7 & 22.6 & ONC & \nodata\\
\object[COUP 1238]{1238} & 053523.0-052836 & 26.7 & -4.0 & 0.68 & 0.8 & 20.8 & 0.3 & \nodata & 22.3 & ONC & \nodata\\
\object[COUP 1254]{1254} & 053523.5-051930 & 80.8 & 75.5 & 0.87 & 3.5 & 22.7 & 2.6 & 3.1 & 22.4 & EG & 0.57\\
\object[COUP 1272]{1272} & 053523.9-051913 & 6.2 & 4.3 & 0.48 & 3.2 & \nodata & \nodata & 0.311 & 22.4 & EG & 0.53\\
\object[COUP 1283]{1283} & 053524.3-052206 & 122.6 & 119.9 & 0.87 & 4.4 & 23.0 & 15.0 & 5.1 & 22.3 & EG & 0.23\\
\object[COUP 1294]{1294} & 053524.5-052526 & 6.4 & 2.0 & 0.47 & 1.3 & \nodata & \nodata & 0.102 & 22.6 & Unc & \nodata\\
\object[COUP 1304]{1304} & 053524.7-052759 & 38400.3 & 27991.4 & 0.87 & 3.0 & 22.2 & 15.0 & 979.6 & 22.3 & EG & 0.22\\
\object[COUP 1310]{1310} & 053525.0-052326 & 262.5 & 214.9 & 0.87 & 3.4 & 22.4 & 15.0 & 8.2 & 22.4 & EG & 0.03\\
\object[COUP 1315]{1315} & 053525.1-052524 & 15.8 & 15.0 & 0.86 & 4.5 & 23.4 & 1.6 & 0.6 & 22.6 & EG & 0.29\\
\object[COUP 1318]{1318} & 053525.2-052823 & 8.4 & 6.3 & 0.47 & 2.3 & \nodata & \nodata & 0.3 & 22.3 & EG & 0.00\\
\object[COUP 1325]{1325} & 053525.4-052012 & 280.6 & 279.2 & 0.87 & 4.3 & 23.1 & 5.0 & 13.2 & 22.3 & EG & 0.33\\
\object[COUP 1337]{1337} & 053525.8-051809 & 7.3 & 3.6 & 0.33 & 2.3 & \nodata & \nodata & 0.4 & 22.3 & EG & 0.24\\
\object[COUP 1339]{1339} & 053525.9-053049 & 80.7 & 74.3 & 0.89 & 4.1 & 22.7 & 15.0 & 3.2 & 22.3 & EG & 0.21\\
\object[COUP 1368]{1368} & 053526.7-052013 & 84.6 & 80.3 & 0.87 & 3.8 & 23.0 & 2.8 & 3.2 & 22.3 & EG & 0.17\\
\object[COUP 1375]{1375} & 053527.0-052532 & 52.3 & 51.9 & 0.87 & 4.6 & 23.1 & 15.0 & 2.6 & 22.6 & EG & 0.01\\
\object[COUP 1376]{1376} & 053527.1-053002 & 55.2 & 43.7 & 0.88 & 3.3 & 22.4 & 15.0 & 1.7 & 22.3 & EG & 0.34\\
\object[COUP 1377]{1377} & 053527.2-052615 & 100.0 & 79.2 & 0.86 & 3.2 & 22.3 & 15.0 & 3.1 & 22.4 & EG & 0.12\\
\object[COUP 1381]{1381} & 053527.3-052928 & 16.9 & 6.0 & 0.77 & 1.7 & 22.5 & 0.5 & 0.2 & 22.3 & EG & 0.43\\
\object[COUP 1383]{1383} & 053527.4-051744 & 43.4 & 26.3 & 0.88 & 2.8 & 21.7 & 15.0 & 1.0 & 22.1 & EG & 0.06\\
\object[COUP 1389]{1389} & 053527.6-051816 & 42.6 & 42.2 & 0.88 & 3.6 & 22.8 & 2.9 & 1.5 & 22.1 & EG & 0.31\\
\object[COUP 1395]{1395} & 053527.7-052617 & 16.9 & 6.2 & 0.89 & 1.9 & 22.1 & 2.2 & 0.2 & 22.4 & EG & 0.21\\
\object[COUP 1399]{1399} & 053527.8-053109 & 370.4 & 232.5 & 0.87 & 2.5 & 22.2 & 6.2 & 8.5 & 22.1 & EG & 0.00\\
\object[COUP 1420]{1420} & 053529.0-052116 & 9.0 & 11.3 & 0.87 & 5.1 & \nodata & \nodata & 0.8 & 22.4 & EG & 0.05\\
\object[COUP 1428]{1428} & 053529.5-052311 & 24.3 & 20.7 & 0.87 & 3.4 & 22.4 & 15.0 & 0.8 & 22.3 & EG & 0.00\\
\object[COUP 1437]{1437} & 053529.9-052858 & 289.9 & 217.9 & 0.88 & 3.4 & 22.3 & 15.0 & 8.2 & 22.1 & EG & 0.32\\
\object[COUP 1452]{1452} & 053530.7-052535 & 5.1 & 3.6 & 0.48 & 2.4 & \nodata & \nodata & 0.2 & 22.4 & EG & 0.11\\
\object[COUP 1453]{1453} & 053530.7-052541 & 4.4 & -0.4 & 0.50 & 1.4 & \nodata & \nodata & \nodata & 22.4 & Unc & \nodata\\
\object[COUP 1460]{1460} & 053531.0-052004 & 37.2 & 40.2 & 0.87 & 3.7 & 22.9 & 1.9 & 1.4 & 22.3 & EG & 0.12\\
\object[COUP 1461]{1461} & 053531.0-052428 & 8.6 & 4.8 & 0.50 & 2.8 & \nodata & \nodata & 0.4 & 22.4 & EG & 0.01\\
\object[COUP 1467]{1467} & 053531.2-052725 & 66.0 & 47.0 & 0.87 & 2.9 & 22.5 & 2.9 & 1.5 & 22.1 & EG & 0.27\\
\object[COUP 1471]{1471} & 053531.4-052136 & 375.2 & 222.1 & 0.87 & 2.3 & 22.2 & 2.6 & 8.2 & 22.4 & EG & 0.13\\
\object[COUP 1476]{1476} & 053531.5-052245 & 21.7 & 24.1 & 0.87 & 5.0 & 23.6 & 1.5 & 1.2 & 22.3 & EG & 0.16\\
\object[COUP 1482]{1482} & 053531.8-052230 & 27.8 & 22.7 & 0.87 & 3.0 & 22.9 & 1.2 & 0.7 & 22.3 & EG & 0.00\\
\object[COUP 1486]{1486} & 053532.0-052945 & 149.4 & 110.4 & 0.88 & 2.8 & 22.2 & 5.6 & 3.6 & 22.1 & EG & 0.47\\
\object[COUP 1491]{1491} & 053532.4-052822 & 223.1 & 150.4 & 0.88 & 2.8 & 22.0 & 15.0 & 6.1 & 22.1 & EG & 0.33\\
\object[COUP 1493]{1493} & 053532.6-053047 & 17.0 & 13.3 & 0.88 & 4.0 & \nodata & \nodata & 0.7 & 22.1 & EG & 0.51\\
\object[COUP 1494]{1494} & 053532.6-053125 & 19.8 & 14.9 & 0.20 & 3.4 & 22.3 & 15.0 & 2.7 & 21.6 & EG & 0.38\\
\object[COUP 1498]{1498} & 053532.8-052222 & 21.1 & 21.2 & 0.87 & 3.8 & 23.3 & 0.7 & 0.8 & 22.4 & EG & 0.00\\
\object[COUP 1502]{1502} & 053533.1-051846 & 37.9 & 17.9 & 0.87 & 2.0 & 21.9 & 1.5 & 0.9 & 22.1 & EG & 0.25\\
\object[COUP 1504]{1504} & 053533.3-051651 & 194.8 & 165.0 & 0.88 & 3.7 & 22.4 & 15.0 & 6.6 & 21.6 & EG & 0.00\\
\object[COUP 1505]{1505} & 053533.3-052720 & 54.3 & 26.7 & 0.88 & 2.0 & 21.4 & 15.0 & 1.2 & 22.1 & EG & 0.16\\
\object[COUP 1506]{1506} & 053533.4-052702 & 293.9 & 200.9 & 0.87 & 2.7 & 22.3 & 3.8 & 6.8 & 22.1 & EG & 0.27\\
\object[COUP 1509]{1509} & 053533.8-052244 & 7.4 & 4.7 & 0.68 & 2.5 & \nodata & \nodata & 0.2 & 22.4 & EG & 0.03\\
\object[COUP 1510]{1510} & 053534.1-052541 & 14.4 & 16.9 & 0.88 & 3.9 & 23.2 & 1.5 & 0.6 & 22.3 & EG & 0.15\\
\object[COUP 1515]{1515} & 053534.9-052836 & 62.5 & 42.6 & 0.88 & 2.8 & 21.9 & 15.0 & 1.5 & 22.1 & EG & 0.55\\
\object[COUP 1518]{1518} & 053535.0-053035 & 182.7 & 163.5 & 0.89 & 3.2 & 22.4 & 5.6 & 5.7 & 22.1 & EG & 0.52\\
\object[COUP 1527]{1527} & 053535.9-052531 & 42.1 & 35.8 & 0.87 & 3.5 & 22.5 & 15.0 & 1.4 & 22.3 & EG & 0.00\\
\object[COUP 1528]{1528} & 053536.1-051913 & 36.9 & 28.0 & 0.87 & 2.6 & 22.6 & 1.1 & 0.8 & 22.3 & EG & 0.36\\
\object[COUP 1530]{1530} & 053536.3-052427 & 7.6 & 6.2 & 0.49 & 4.9 & \nodata & \nodata & 0.6 & 22.3 & EG & 0.00\\
\object[COUP 1536]{1536} & 053537.2-052004 & 22.9 & 19.9 & 0.88 & 2.7 & 23.1 & 0.6 & 0.6 & 22.4 & EG & 0.41\\
\object[COUP 1538]{1538} & 053537.4-052324 & 20.8 & 15.6 & 0.89 & 2.9 & 22.2 & 5.1 & 0.7 & 22.4 & EG & 0.01\\
\object[COUP 1541]{1541} & 053537.8-051826 & 120.4 & 103.0 & 0.88 & 3.6 & 22.9 & 3.5 & 4.0 & 22.1 & EG & 0.43\\
\object[COUP 1545]{1545} & 053538.3-052540 & 21.2 & 18.9 & 0.88 & 3.4 & 22.6 & 2.7 & 0.7 & 22.3 & EG & 0.00\\
\object[COUP 1548]{1548} & 053538.7-051858 & 9.9 & 16.0 & 0.88 & 4.2 & \nodata & \nodata & 0.5 & 22.3 & EG & 0.30\\
\object[COUP 1551]{1551} & 053539.4-052854 & 154.3 & 97.6 & 0.88 & 2.6 & 22.2 & 5.8 & 4.5 & 22.1 & EG & 0.38\\
\object[COUP 1552]{1552} & 053539.5-052842 & 14.5 & 7.1 & 0.49 & 1.8 & 21.7 & 4.0 & 0.5 & 22.1 & EG & 0.24\\
\object[COUP 1554]{1554} & 053540.0-052016 & 221.7 & 225.5 & 0.87 & 4.2 & 22.9 & 15.0 & 9.7 & 22.4 & EG & 0.10\\
\object[COUP 1555]{1555} & 053540.3-052840 & 56.6 & 46.8 & 0.88 & 3.5 & 22.4 & 15.0 & 1.7 & 22.1 & EG & 0.23\\
\object[COUP 1556]{1556} & 053540.3-053013 & 120.8 & 88.2 & 0.87 & 2.9 & 22.5 & 2.8 & 6.3 & 22.1 & EG & 0.58\\
\object[COUP 1559]{1559} & 053540.6-052807 & 127.3 & 101.3 & 0.88 & 3.3 & 22.4 & 15.0 & 4.1 & 22.1 & EG & 0.25\\
\object[COUP 1565]{1565} & 053541.7-052015 & 385.3 & 300.5 & 0.88 & 2.9 & 22.5 & 2.5 & 10.0 & 22.4 & EG & 0.06\\
\object[COUP 1575]{1575} & 053542.8-051930 & 42.7 & 37.5 & 0.88 & 3.7 & 22.4 & 15.0 & 1.6 & 22.3 & EG & 0.26\\
\object[COUP 1578]{1578} & 053543.2-052759 & 154.2 & 104.4 & 0.87 & 2.6 & 22.2 & 4.3 & 3.6 & 22.1 & EG & 0.01\\
\object[COUP 1580]{1580} & 053543.5-052119 & 67.6 & 68.8 & 0.88 & 4.2 & 22.5 & 15.0 & 2.9 & 22.3 & EG & 0.26\\
\object[COUP 1581]{1581} & 053543.6-051928 & 63.3 & 49.2 & 0.89 & 3.8 & 22.6 & 15.0 & 2.6 & 22.3 & EG & 0.23\\
\object[COUP 1582]{1582} & 053543.6-052400 & 91.2 & 82.4 & 0.87 & 3.8 & 22.7 & 7.3 & 3.4 & 22.3 & EG & 0.00\\
\object[COUP 1583]{1583} & 053543.7-052637 & 12.5 & 10.2 & 0.49 & 4.0 & \nodata & \nodata & 1.0 & 22.3 & EG & 0.00\\
\object[COUP 1586]{1586} & 053545.0-052233 & 23.5 & 20.1 & 0.89 & 3.4 & 22.6 & 3.2 & 0.7 & 22.4 & EG & 0.14\\
\object[COUP 1589]{1589} & 053545.1-052255 & 31.3 & 34.0 & 0.89 & 4.5 & \nodata & \nodata & 1.5 & 22.4 & EG & 0.21\\
\object[COUP 1597]{1597} & 053548.2-052357 & 14.1 & 15.5 & 0.88 & 3.6 & 23.4 & 0.5 & 0.6 & 22.3 & EG & 0.00\\
\object[COUP 1598]{1598} & 053549.2-052417 & 115.3 & 114.6 & 0.88 & 4.0 & 23.0 & 2.4 & 4.8 & 22.3 & EG & 0.20\\
\object[COUP 1600]{1600} & 053550.2-052139 & 43.9 & 54.9 & 0.88 & 4.7 & 22.9 & 15.0 & 2.5 & 22.4 & EG & 0.18\\
\object[COUP 1601]{1601} & 053550.2-052211 & 19.3 & 13.7 & 0.68 & 2.9 & 22.0 & 15.0 & 0.8 & 22.4 & EG & 0.15\\
\object[COUP 1605]{1605} & 053552.8-052713 & 96.0 & 82.2 & 0.88 & 3.2 & 22.2 & 15.0 & 3.1 & 22.3 & EG & 0.00\\
\object[COUP 1606]{1606} & 053552.9-052305 & 50.4 & 38.4 & 0.89 & 3.7 & 22.8 & 4.6 & 1.8 & 22.3 & EG & 0.30\\
\object[COUP 1607]{1607} & 053553.2-052603 & 698.8 & 554.5 & 0.89 & 3.0 & 22.4 & 5.6 & 22.7 & 22.1 & EG & 0.00\\
\object[COUP 1609]{1609} & 053554.4-052436 & 346.0 & 258.1 & 0.89 & 3.1 & 22.2 & 15.0 & 10.3 & 22.1 & EG & 0.08\\
\object[COUP 1615]{1615} & 053557.4-052415 & 151.6 & 122.0 & 0.89 & 3.2 & 22.6 & 4.0 & 5.2 & 22.3 & EG & 0.15\\
\enddata
\tablecomments{Column 1: COUP source number. Column 2: IAU
designation. Column 3: Total band ($0.5-8$ keV) net counts in
extracted area. Column 4: Hard band ($2-8$ keV) net counts in
extracted area. Column 5: Fraction of the point spread function
extracted. Column 6: Median energy of photons in the total band
(corrected for background). Column 7: Source hydrogen column
density inferred from {\it XSPEC} spectral fit. Column 8: Source
plasma energy from spectral fit. Column 9: Observed hard band flux
in units of $10^{-15}$ erg cm$^{22}$ s$^{-1}$. Column 10: Average
hydrogen column density in OMC from the $^{13}$CO map by
\citet{Bally87}. Column 11: Membership class derived here(\S
\ref{candidates_section}). Column 12: Membership probability
derived here(\S \ref{EG_simul_section}).}

\tablecomments{Spectral results are unreliable for COUP
\object[COUP 84]{84}, \object[COUP 257]{257}, \object[COUP
495]{495}, and \object[COUP 1033]{1033} because the fitting
procedure failed due to substantial hard-band contamination.
Spectral results are unreliable for COUP \object[COUP 479]{479},
\object[COUP 505]{505}, and \object[COUP 1157]{1157} because the
fitting procedure failed due to substantial soft-band
contamination.}

\tablenotetext{a}{in units of $10^{-15}$ erg s$^{-1}$ cm$^{-2}$ in
the $2-8$ keV band}

\end{deluxetable}

\clearpage
\newpage

\newpage

\begin{deluxetable}{ccccccrrrrrccc}
\centering \rotate \tabletypesize{\scriptsize} \tablewidth{0pt}

\tablecolumns{14} \tablecaption{COUP double sources with
$<3\arcsec$ separations \label{binaries_table}}

\tablehead{

\multicolumn{6}{c}{X-ray sources} &&
\multicolumn{7}{c}{Optical/near-infrared sources} \\ \cline{1-6}
\cline{8-14}

\colhead{COUP(W)} & \colhead{COUP(E)} & \colhead{Sep} &
\colhead{Off-axis} & \colhead{MedE(W-E)} &  \colhead{$N_H$(W-E)}
&& \colhead{IR(W)} & \colhead{Opt(W)} & \colhead{IR(E)} &
\colhead{Opt(E)} & \colhead{Class(W)} & \colhead{Class(E)} &
Notes \\

\colhead{\#} & \colhead{\#} & \colhead{$\arcsec$}&
\colhead{$\arcmin$} & \colhead{keV} & \colhead{cm$^{-2}$}&&
& & & & & & \\

\colhead{(1)} & \colhead{(2)} & \colhead{(3)} & \colhead{(4)} &
\colhead{(5)} & \colhead{(6)} && \colhead{(7)} & \colhead{(8)} &
\colhead{(9)} & \colhead{(10)} & \colhead{(11)} & \colhead{(12)} &
\colhead{(13)}}

\startdata
\object[COUP 0057]{57} & \object[COUP 0059]{59} & 2.1 & 6.6 & 1.1-1.1 & 20.7-20.8 & & 2MASS & 116 & \nodata & 10281 & M & M & \tablenotemark{a} \\
\object[COUP 0074]{74} & \object[COUP 0075]{75} & 2.2 & 7.4 & 1.3-1.4 & 21.4-21.8 & & 2MASS & 137 & 2MASS & 136 & M & M & \tablenotemark{b} \\
\object[COUP 0123]{123} & \object[COUP 0124]{124} & 1.4 & 4.8 & 1.5-1.4 & 21.2-21.1 & & 2MASS & 176 & 2MASS & 176 & M & M & \tablenotemark{c} \\
\object[COUP 0213]{213} & \object[COUP 0214]{214} & 0.9 & 3.2 & 1.1-1.5 & 20.9-20.6 & & 51 & \nodata & 53 & 248 & M & M & \\
\object[COUP 0251]{251} & \object[COUP 0252]{252} & 2.3 & 3.1 & 1.0-1.3 & 20.0-20.9 & & 103 & 275b & 102 & 275a & M & M & \\
\object[COUP 0283]{283} & \object[COUP 0286]{286} & 3.0 & 2.7 & 1.3-1.0 & 21.1-20.0 & & 141 & 300 & 143 & 302 & M & M & \\
\object[COUP 0315]{315} & \object[COUP 0316]{316} & 1.7 & 2.3 & 2.7-1.1 & 22.5-21.3 & & 174 & \nodata & 172 & 322 & M & M & \tablenotemark{d} \\
\object[COUP 0374]{374} & \object[COUP 0375]{375} & 2.1 & 2.7 & 1.8-2.0 & 21.6-22.3 & & 235 & \nodata & 231 & \nodata & M & M & \\
\object[COUP 0378]{378} & \object[COUP 0384]{384} & 1.8 & 1.9 & 1.2-1.4 & 21.1-21.2 & & 238 & 345 & 244 & 350 & M & M & \\
\object[COUP 0403]{403} & \object[COUP 0408]{408} & 2.1 & 1.8 & 3.2-2.7 & 22.5-22.4 & & 261 & 355 & 266 & 10415 & M & M & \\
\object[COUP 0425]{425} & \object[COUP 0442]{442} & 2.8 & 1.4 & 4.6-1.1 & 23.1-20.0 & & \nodata & \nodata & 296 & 9008 & NM & M & \tablenotemark{e} \\
\object[COUP 0440]{440} & \object[COUP 0441]{441} & 3.0 & 1.3 & 4.8-3.9 & 23.2-23.0 & & 297 & \nodata & 299 & \nodata & M & M & \\
\object[COUP 0495]{495} & \object[COUP 0508]{508} & 2.7 & 1.3 & 1.4-4.3 & 20.0-\nodata & & \nodata & \nodata & \nodata & \nodata & NM? & NM? & \tablenotemark{f} \\
\object[COUP 0503]{503} & \object[COUP 0504]{504} & 0.7 & 2.3 & 3.4-3.2 & 22.7-22.6 & & 346 & \nodata & 344 & \nodata & M & M & \\
\object[COUP 0519]{519} & \object[COUP 0530]{530} & 2.4 & 1.3 & 1.0-5.2 & 20.0-23.5 & & 357 & 9030 & \nodata & \nodata & M & NM? & \tablenotemark{g} \\
\object[COUP 0526]{526} & \object[COUP 0534]{534} & 2.9 & 1.6 & 2.0-1.5 & 20.0-21.8 & & 360 & \nodata & 363 & 401 & M & M & \\
\object[COUP 0527]{527} & \object[COUP 0535]{535} & 1.6 & 4.3 & 1.4-3.0 & 21.2-22.5 & & 2MASS & 405 & 2MASS & 10438 & M & M & \tablenotemark{h} \\
\object[COUP 0528]{528} & \object[COUP 0537]{537} & 2.8 & 3.0 & 2.6-3.1 & 22.3-22.7 & & 364 & 400 & 373 & \nodata & M & M & \\
\object[COUP 0540]{540} & \object[COUP 0541]{541} & 1.0 & 0.9 & 1.2-2.0 & 21.1-21.5 & & 372 & 9037 & 375 & 9040 & M & M & \\
\object[COUP 0579]{579} & \object[COUP 0580]{580} & 2.2 & 1.8 & 1.7-2.4 & 21.7-22.2 & & 396 & 423 & 400 & \nodata & M & M & \\
\object[COUP 0583]{583} & \object[COUP 0603]{603} & 2.7 & 1.1 & 3.1-4.6 & 22.7-23.2 & & 403 & \nodata & \nodata & \nodata & M & M & \tablenotemark{i} \\
\object[COUP 0589]{589} & \object[COUP 0590]{590} & 2.1 & 1.4 & 3.7-3.1 & 23.3-22.6 & & \nodata & \nodata & 408 & \nodata & NM? & M & \tablenotemark{j} \\
\object[COUP 0621]{621} & \object[COUP 0628]{628} & 2.5 & 1.3 & 3.6-5.0 & 22.7-23.4 & & 428 & \nodata & \nodata & \nodata & M & NM? & \tablenotemark{k} \\
\object[COUP 0623]{623} & \object[COUP 0629]{629} & 1.7 & 1.0 & 2.8-2.3 & 22.5-22.5 & & 429 & 9062 & 431 & 9064 & M & M & \\
\object[COUP 0655]{655} & \object[COUP 0663]{663} & 2.8 & 1.3 & 3.6-1.4 & 22.9-21.5 & & 454 & \nodata & 463 & 452 & M & M & \\
\object[COUP 0656]{656} & \object[COUP 0670]{670} & 2.6 & 1.2 & 3.9-1.5 & 23.3-21.6 & & \nodata & \nodata & 466 & 454 & NM & M & \tablenotemark{l} \\
\object[COUP 0661]{661} & \object[COUP 0662]{662} & 2.6 & 1.4 & 1.5-4.5 & 20.0-23.2 & & 457 & 9074 & \nodata & \nodata & M & NM & \tablenotemark{m} \\
\object[COUP 0668]{668} & \object[COUP 0673]{673} & 1.2 & 0.8 & 1.2-1.4 & 21.6-21.2 & & 465 & 9077 & 467 & 9079 & M & M & \\
\object[COUP 0678]{678} & \object[COUP 0681]{681} & 2.4 & 1.2 & 4.1-4.5 & 22.9-23.2 & & \nodata & \nodata & \nodata & \nodata & NM & NM & \tablenotemark{n} \\
\object[COUP 0682]{682} & \object[COUP 0689]{689} & 2.8 & 0.9 & 2.1-1.3 & 21.9-20.9 & & 487 & 463 & 493 & 468 & M & M & \\
\object[COUP 0698]{698} & \object[COUP 0699]{699} & 0.8 & 1.3 & 1.4-1.5 & 21.1-21.6 & & 499 & 9096 & 505 & 9096 & M & M & \\
\object[COUP 0705]{705} & \object[COUP 0706]{706} & 1.9 & 2.5 & 2.1-4.1 & 22.1-22.9 & & 506 & 473 & \nodata & \nodata & M & NM & \tablenotemark{o} \\
\object[COUP 0707]{707} & \object[COUP 0716]{716} & 2.3 & 0.9 & 1.5-3.4 & 21.7-22.7 & & 511 & 476 & 516 & \nodata & M & M & \\
\object[COUP 0714]{714} & \object[COUP 0722]{722} & 1.7 & 4.0 & 1.2-1.3 & 21.5-21.4 & & 510 & \nodata & 520 & \nodata & M & M & \\
\object[COUP 0725]{725} & \object[COUP 0734]{734} & 1.6 & 0.3 & 2.9-3.2 & 22.5-22.6 & & 526 & \nodata & 533 & \nodata & M & M & \\
\object[COUP 0732]{732} & \object[COUP 0744]{744} & 2.2 & 0.6 & 1.3-1.3 & 21.0-21.3 & & 535 & 1864 & 538 & \nodata & M & M & \tablenotemark{p} \\
\object[COUP 0733]{733} & \object[COUP 0746]{746} & 1.6 & 0.4 & 2.5-2.2 & 22.5-22.1 & & 531 & \nodata & 542 & 488a & M & M & \\
\object[COUP 0765]{765} & \object[COUP 0777]{777} & 1.6 & 0.8 & 3.2-2.4 & 22.6-22.9 & & 567 & \nodata & 574 & 9128 & M & M & \\
\object[COUP 0766]{766} & \object[COUP 0778]{778} & 1.0 & 0.6 & 1.6-1.1 & 21.5-21.4 & & 565 & 1863b & 573 & 1863a & M & M & \tablenotemark{q} \\
\object[COUP 0768]{768} & \object[COUP 0769]{769} & 1.0 & 0.3 & 1.6-3.1 & 21.5-22.7 & & 559 & 503 & 562 & \nodata & M & M & \tablenotemark{r} \\
\object[COUP 0774]{774} & \object[COUP 0782]{782} & 2.2 & 2.5 & 1.5-1.6 & 21.6-22.0 & & 576 & 506 & 584 & 9135 & M & M & \\
\object[COUP 0784]{784} & \object[COUP 0806]{806} & 2.8 & 1.5 & 3.5-4.9 & 22.7-23.3 & & 586 & 511b & 599 & \nodata & M & M & \\
\object[COUP 0787]{787} & \object[COUP 0788]{788} & 2.6 & 0.4 & 2.4-1.5 & 22.4-22.1 & & 585 & 512 & 582 & \nodata & M & M & \\
\object[COUP 0820]{820} & \object[COUP 0826]{826} & 2.0 & 0.4 & 3.6-2.3 & 22.9-22.1 & & 611 & \nodata & 619 & 524 & M & M & \\
\object[COUP 0832]{832} & \object[COUP 0842]{842} & 2.0 & 1.3 & 1.3-1.3 & 21.5-21.3 & & 628 & 529 & 634 & 9158 & M & M & \\
\object[COUP 0840]{840} & \object[COUP 0861]{861} & 2.4 & 2.2 & 1.5-3.7 & 21.2-22.8 & & 636 & \nodata & \nodata & \nodata & M & NM & \tablenotemark{s} \\
\object[COUP 0843]{843} & \object[COUP 0854]{854} & 2.6 & 1.1 & 2.7-1.4 & 22.7-21.2 & & 630 & \nodata & 642 & 536 & M & M & \\
\object[COUP 0844]{844} & \object[COUP 0862]{862} & 2.7 & 0.9 & 1.6-1.2 & 22.0-20.5 & & 638 & 532 & 649 & 539 & M & M & \\
\object[COUP 0847]{847} & \object[COUP 0856]{856} & 3.0 & 0.1 & 1.4-1.5 & 21.6-21.6 & & 640 & 534 & 644 & 537 & M & M & \\
\object[COUP 0884]{884} & \object[COUP 0906]{906} & 2.5 & 0.8 & 4.0-1.7 & 23.0-\nodata & & 674 & 9181 & 691 & 9188 & M & M & \\
\object[COUP 0910]{910} & \object[COUP 0922]{922} & 1.6 & 0.5 & 1.1-1.3 & 21.3-21.2 & & 698 & 9194 & 705 & 562 & M & M & \\
\object[COUP 0912]{912} & \object[COUP 0913]{913} & 1.7 & 0.2 & 1.9-1.5 & 22.1-21.8 & & 700 & 9195 & 701 & 9196 & M & M & \\
\object[COUP 0917]{917} & \object[COUP 0940]{940} & 2.9 & 2.8 & 1.6-4.8 & 21.6-23.3 & & 706 & 561 & \nodata & \nodata & M & NM & \tablenotemark{t} \\
\object[COUP 0943]{943} & \object[COUP 0944]{944} & 2.1 & 0.4 & 1.4-2.3 & 21.4-22.4 & & 727 & 575 & 725 & 9210 & M & M & \\
\object[COUP 0977]{977} & \object[COUP 0978]{978} & 2.6 & 1.9 & 3.0-3.3 & 22.8-22.6 & & 762 & \nodata & 758 & \nodata & M & M & \\
\object[COUP 1049]{1049} & \object[COUP 1050]{1050} & 2.2 & 3.2 & 1.7-1.4 & 21.9-\nodata & & 817 & 10561 & 816 & 10560 & M & M & \\
\object[COUP 1174]{1174} & \object[COUP 1175]{1175} & 2.1 & 1.2 & 1.5-1.5 & 21.7-21.5 & & 922 & 698 & 928 & 9292 & M & M & \\
\object[COUP 1205]{1205} & \object[COUP 1211]{1211} & 1.9 & 1.4 & 2.4-1.1 & 22.3-20.0 & & 945 & 710 & 948 & 710h & M & M & \\
\object[COUP 1326]{1326} & \object[COUP 1327]{1327} & 1.2 & 3.0 & 1.5-1.6 & 21.7-21.8 & & 1068 & 777 & 1069 & 777 & M & M & \\
\object[COUP 1392]{1392} & \object[COUP 1393]{1393} & 2.4 & 3.6 & 3.0-2.5 & 22.6-22.2 & & 1127 & \nodata & 1125 & \nodata & M & M & \\
\object[COUP 1412]{1412} & \object[COUP 1416]{1416} & 2.9 & 3.1 & 1.2-1.3 & 21.2-21.4 & & 1146 & 830 & 1150 & 10667 & M & M & \\
\enddata

\tablecomments{Columns 1-2:  Source numbers of COUP sources within
3\arcsec\/ of each other.  Column 3:  Component separation in
arcsec. Column 4:  Off-axis angle in arcmin. Column 5:  Median
energy of binary components in keV corrected for background.
Column 6: Logarithm of column density of binary components in
$cm^{-2}$. Column 7: Near-infrared counterpart in the VLT catalog
of the first COUP component (McCaughrean et al.\ 2005). Column 8:
Optical counterpart of the first COUP component
\citep{Hillenbrand97, Herbst02}.  Column 9: Near-infrared
counterpart of the second COUP component.  Column 10: Optical
counterpart of the second COUP component.  Column 11: Proposed
class of the first COUP component (M = established Member, NM =
New Member based on variability (\S
\ref{flaring_members_section}), NM? = possible New Member based on
high source density (\S \ref{EG_simul_section}). Column 12:
Proposed class of the second COUP component.}

\end{deluxetable}

\clearpage
\newpage

{\bf Notes to Table \ref{binaries_table}}
\bigskip

$^a$ {COUP \object[COUP 0057]{57}-\object[COUP 0059]{59}. These
sources lie outside the VLT survey field of view. The
corresponding 2MASS source 05345071-0524014 is an unresolved blend
of these two stars.}

$^b$ {COUP \object[COUP 0074]{74}-\object[COUP 0075]{75}. These
sources lie outside the VLT survey field of view.  The
corresponding 2MASS source 05345275-0527545 is an unresolved blend
of these two stars.}

$^c$ {COUP \object[COUP 0123]{123}-\object[COUP 0124]{124}. These
sources lie outside the VLT survey field of view.  The
corresponding 2MASS source 05345766-0523522 is an unresolved blend
of these two stars.}

$^d$ {COUP \object[COUP 0315]{315}-\object[COUP 0316]{316}. These
sources are unlikely to be a physical binary due to the large
difference in X-ray absorption and optical brightness.}

$^e$ {COUP \object[COUP 0425]{425}-\object[COUP 0442]{442}. These
sources are unlikely to be a physical binary due to the large
difference in X-ray absorption and optical brightness. Neither
$K_s$-band nor optical band images detect the highly absorbed
component which is a flaring X-ray source (COUP \object[COUP
0425]{425}).}

$^f$ {COUP \object[COUP 0495]{495}-\object[COUP 0508]{508}. The
VLT survey did not detect either of these two faint COUP sources,
possibly due to strong nebula emission in the region.  The sources
are unlikely to be a physical binary due to the large difference
in X-ray absorption.}

$^g$ {COUP \object[COUP 0519]{519}-\object[COUP 0530]{530}. VLT
did not detect the X-ray brighter, but highly absorbed component
(COUP \object[COUP 0530]{530}). These sources are unlikely to be a
physical binary due to the large difference in X-ray absorption
and optical brightness.}

$^h$ {COUP \object[COUP 0527]{527}-\object[COUP 0535]{535}. These
sources lie outside the VLT survey field of view;  the
corresponding 2MASS source 05351324-0527541 is an unresolved blend
of these two stars. The COUP stars are detected and resolved in
optical catalogs.}

$^i$ {COUP \object[COUP 0583]{583}-\object[COUP 0603]{603}. The
highly absorbed component of this X-ray double (COUP \object[COUP
0603]{603}) was not detected in the VLT survey but is source 151
in the $L$-band survey of \citet{Lada04}. The VLT image has a
third source, VLT 403, located $\simeq 1$\arcsec\/ NE of COUP
\object[COUP 0583]{583}.}

$^j$ {COUP \object[COUP 0589]{589}-\object[COUP 0590]{590}. The
southern component COUP \object[COUP 0590]{590} lines in a bright
extended region in the VLT $K_s$-band image.}

$^k$ {COUP \object[COUP 0621]{621}-\object[COUP 0628]{628}. The
VLT did not detect the northern, highly absorbed component of this
visual binary (COUP \object[COUP 0628]{628}). Another $K_s$-band
object, possibly a nebular knot, located $\simeq 2$\arcsec\/ NW of
VLT 428 is not associated with any COUP source.}

$^l$ {COUP \object[COUP 0656]{656}-\object[COUP 0670]{670}. The
VLT did not detect the highly absorbed component of this visual
binary (COUP \object[COUP 0656]{656}). These sources are unlikely
to be a physical binary due to the large difference in X-ray
absorption and optical brightness.}

$^m$ {COUP \object[COUP 0661]{661}-\object[COUP 0662]{662}. The
VLT did not detect the highly absorbed, X-ray flaring component of
this visual binary (COUP \object[COUP 0662]{662}). These sources
are unlikely to be a physical binary due to the large difference
in X-ray absorption and optical brightness.}

$^n$ {COUP \object[COUP 0678]{678}-\object[COUP 0681]{681}. These
highly absorbed components lie in a region of bright $K_s$-band
nebulosity. \citet{Getman05} associate COUP \object[COUP
0678]{678} with the star VLT 476 with offset of $0.9\arcsec$ based
on their automated identification procedure. But, from examination
of the images here, we suggest that this offset is too large to be
instrumental and that the association is incorrect. We thus
include COUP \object[COUP 0678]{678}, which exhibits X-ray flares,
in Table \ref{flaring_table} as a new obscured member of the Orion
region.}

$^o$ {COUP \object[COUP 0705]{705}-\object[COUP 0706]{706}. VLT
did not detect the highly absorbed, X-ray flaring component of
this visual binary (COUP \object[COUP 0706]{706}). These sources
are unlikely to be a physical binary due to the large difference
in X-ray absorption and optical brightness.}

$^p$ {COUP \object[COUP 0732]{732}-\object[COUP 0744]{744}.
Together with COUP \object[COUP 0745]{745} $\simeq 3$\arcsec\/ SE
of \object[COUP 0744]{744}, these sources comprise an apparent
triple system with $K_s$-band counterparts VLT 535-538-540. VLT
resolved the NW component into another double, VLT 535(= COUP
\object[COUP 0732]{732}) and VLT 545 with separation $0.1$\arcsec,
too close to be resolved by $Chandra$.  These X-ray sources are
sufficiently bright that errors in the image reconstruction
technique are visible as rings in the middle panel of Figure
\ref{close_binaries_fig}.}

$^q$ {COUP \object[COUP 0766]{766}-\object[COUP 0778]{778}. They
correspond to $\theta^1$ Ori BE and $\theta^1$ Ori BW, and are
discussed in detail by \citet{Stelzer05}.}

$^r$ {COUP \object[COUP 0768]{768}-\object[COUP 0769]{769}. These
sources are unlikely to be a physical binary due to the large
difference in X-ray absorption and optical brightness.}

$^s$ {COUP \object[COUP 0840]{840}-\object[COUP 0861]{861}. The
VLT did not detect highly absorbed, X-ray flaring component of
this double (COUP \object[COUP 0861]{861}). These sources are
unlikely to be a physical binary due to the large difference in
X-ray absorption.}

$^t$ {COUP \object[COUP 0917]{917}-\object[COUP 0940]{940}.
Optical band and VLT $K_s$ band images did not detect the highly
absorbed, X-ray flaring component of this visual binary (COUP
\object[COUP 0940]{940}). However, the VLT resolves a third star
(VLT 712) between the two $Chandra$ sources. The automated
identification procedure of \citet{Getman05} associated COUP
\object[COUP 0940]{940} with $L$-band source FLWO 890
\citep{Muench02}. This is probably incorrect as FLWO 890 is more
likely associated with the $K_s$-band source VLT 712. Thus COUP
\object[COUP 0940]{940}, which exhibits X-ray flares, is
classified here as a new Orion cloud member without optical or IR
counterpart.}


\begin{deluxetable}{rrccrrrrrrrrcrrc}

\centering \rotate \tabletypesize{\scriptsize} \tablewidth{0pt}
\tablecolumns{16}

\tablecaption{COUP stars with discrepant proper motions
\label{33_possible_field_table}}

\tablehead{

\colhead{COUP} & \colhead{COUP J} & \colhead{$N_h$} &
\colhead{$L_t$\tablenotemark{a}} & \colhead{IR} & \colhead{$J$} &
\colhead{$H$} & \colhead{$K_s$} & \colhead{JW} & \colhead{$V$} &
\colhead{$I$} & \colhead{$A_v$} & \colhead{$\log L_t/L_{bol}$} &
\colhead{P} & \colhead{SpT} & \colhead{Field?}\\

\colhead{\#} & & \colhead{cm$^{-2}$} & \colhead{erg/s} & &
\multicolumn{3}{c}{mag} & & \multicolumn{3}{c}{mag} & & & &

}

\startdata
\object[COUP 0002]{2} & 053429.5-052354 & 21.2 & 30.34 & 05342924-0523567 & 10.19 & 9.78 & 9.61 & 5 & 12.0 & 11.0 & 0.3 & -3.89 & 1 & K1 & F\\
\object[COUP 0007]{7} & 053439.7-052425 & 20.9 & 31.01 & 05343976-0524254 & 8.85 & 8.10 & 7.95 & 45 & 11.4 & 9.9 & 0.8 & -3.77 & 0 & K1-K4 & F\\
\object[COUP 0127]{127} & 053458.0-052940 & 21.4 & 28.27 & 05345805-0529405 & 13.92 & 12.85 & 12.26 & 182 & \nodata & 16.6 & \nodata & \nodata & 48 & \nodata & \nodata\\
\object[COUP 0153]{153} & 053501.1-052955 & 21.2 & 27.82 & 05350116-0529551 & 12.87 & 12.24 & 12.03 & 209 & 16.8 & 14.2 & \nodata & \nodata & 0 & \nodata & F\\
\object[COUP 0188]{188} & 053503.0-053001 & 21.3 & 30.89 & 05350299-0530015 & 10.33 & 9.51 & 9.22 & 232 & 13.6 & 11.7 & 2.2 & -3.52 & 79 & K1-K2 & F\\
\object[COUP 0267]{267} & 053506.4-053335 & 20.0 & 30.50 & 05350644-0533351 & 11.01 & 9.68 & 8.72 & 295 & 15.7 & 13.9 & 0.0 & -2.53 & 0 & M0.5e-M2 & \nodata\\
\object[COUP 0378]{378} & 053510.4-052245 & 21.1 & 30.11 & 238 & 10.75 & 9.84 & 9.49 & 345 & 14.8 & 12.4 & 1.3 & -3.85 & 81 & M0-M2 & \nodata\\
\object[COUP 0492]{492} & 053512.7-052034 & 21.9 & 29.48 & 340 & 13.40 & 12.18 & 11.49 & 389 & 19.5 & 15.9 & 0.5 & -3.18 & 52 & M4.5-M6 & \nodata\\
\object[COUP 0561]{561} & 053513.6-051954 & 22.1 & 30.78 & 387 & 11.12 & 9.41 & 8.34 & 413 & \nodata & 14.6 & 0.0 & -1.94 & 0 & K5 & \nodata\\
\object[COUP 0645]{645} & 053514.6-052042 & 22.1 & 30.23 & 442 & 11.57 & 9.88 & 9.01 & 445a & 18.9 & 14.7 & \nodata & \nodata & 26 & cont & \nodata\\
\object[COUP 0669]{669} & 053514.9-052159 & 21.5 & 30.76 & 472 & 10.92 & 10.06 & 9.76 & 457 & 14.5 & 12.5 & 2.0 & -3.25 & 79 & K3-K4 & F\\
\object[COUP 0672]{672} & 053514.9-052339 & 21.4 & 30.46 & 470 & 10.65 & 9.80 & 9.43 & 456 & 14.6 & 12.2 & \nodata & \nodata & 4 & \nodata & F\\
\object[COUP 0743]{743} & 053515.8-052301 & 21.5 & 29.29 & 546 & 11.34 & 10.34 & 9.66 & 494 & 15.9 & 13.6 & 2.0 & -4.36 & 0 & K7 & \nodata\\
\object[COUP 0756]{756} & 053515.9-052221 & 21.6 & 29.33 & 552 & 12.28 & 10.92 & 9.95 & 496 & 17.0 & 14.2 & 3.0 & -4.30 & 0 & K7e & \nodata\\
\object[COUP 0813]{813} & 053516.5-052405 & 21.5 & 30.05 & 608 & 11.87 & 11.05 & 10.63 & 521 & 18.5 & 14.3 & 6.1 & -4.30 & 86 & M0 & F\\
\object[COUP 0922]{922} & 053517.8-052315 & 21.2 & 29.57 & 705 & 11.15 & 10.21 & 9.77 & 562 & 15.7 & 13.0 & 2.2 & -4.37 & 17 & M0 & \nodata\\
\object[COUP 0939]{939} & 053518.0-051613 & 22.1 & 30.58 & 05351799-0516136 & 11.49 & 9.97 & 8.86 & 566 & 18.7 & 14.9 & \nodata & \nodata & 0 & \nodata & \nodata\\
\object[COUP 0949]{949} & 053518.2-051306 & 22.1 & 30.11 & 05351822-0513068 & 11.48 & 10.30 & 9.64 & 572 & 17.2 & 13.9 & \nodata & \nodata & 81 & \nodata & \nodata\\
\object[COUP 0958]{958} & 053518.2-052535 & 20.0 & 27.57 & 736 & 13.23 & 12.65 & 12.39 & 583 & 16.2 & 14.5 & 0.0 & -5.25 & 0 & M1 & F\\
\object[COUP 0994]{994} & 053518.8-051445 & 21.5 & 30.15 & 05351883-0514455 & 12.21 & 11.31 & 10.89 & 597 & 16.8 & 14.0 & \nodata & \nodata & 0 & \nodata & F\\
\object[COUP 1111]{1111} & 053520.6-052353 & 21.4 & 30.37 & 869 & 11.84 & 11.07 & 10.66 & 659 & 18.2 & 14.0 & \nodata & \nodata & 84 & \nodata & F\\
\object[COUP 1122]{1122} & 053520.9-052150 & 21.5 & 30.05 & 882 & 11.82 & 10.66 & 9.89 & 665 & 15.5 & 13.7 & \nodata & \nodata & 9 & M:e & \nodata\\
\object[COUP 1158]{1158} & 053521.6-052147 & 21.9 & 30.49 & 914 & 11.90 & 10.54 & 9.70 & 687a & 17.7 & 14.7 & 2.7 & -2.90 & 54 & M1: & \nodata\\
\object[COUP 1326]{1326} & 053525.4-052134 & 21.7 & 29.60 & 1068 & 12.72 & 11.40 & 10.58 & 777 & 18.5 & 15.1 & 4.8 & -4.08 & 80 & K6 & \nodata\\
\object[COUP 1327]{1327} & 053525.4-052135 & 21.8 & 29.76 & 1069 & 12.04 & 10.96 & 10.60 & 777 & 18.5 & 15.1 & 4.8 & -3.92 & 80 & K6 & \nodata\\
\object[COUP 1336]{1336} & 053525.7-052641 & 21.8 & 29.86 & 1077 & 12.02 & 10.94 & 10.29 & 786 & 16.3 & 13.9 & 1.3 & -3.50 & 89 & M0-M1 & \nodata\\
\object[COUP 1373]{1373} & 053526.9-052448 & 21.1 & 28.75 & 1105 & 13.36 & 12.59 & 12.23 & 806 & \nodata & 15.2 & 0.0 & -4.28 & 0 & M6.5 & F\\
\object[COUP 1485]{1485} & 053532.0-051620 & 22.0 & 29.82 & 05353199-0516201 & 12.81 & 11.76 & 11.23 & 879 & 16.7 & 14.4 & \nodata & \nodata & 0 & $\leq$M1 & \nodata\\
\object[COUP 1512]{1512} & 053534.3-052659 & 20.6 & 29.29 & 05353437-0526596 & 13.03 & 12.33 & 11.96 & 902 & \nodata & 15.3 & \nodata & \nodata & 84 & \nodata & F\\
\object[COUP 1537]{1537} & 053537.3-052641 & 20.0 & 28.89 & 05353738-0526416 & 10.63 & 10.02 & 9.84 & 928 & 13.6 & 11.7 & \nodata & \nodata & 0 & \nodata & F\\
\object[COUP 1558]{1558} & 053540.5-052701 & 20.0 & 28.81 & 05354049-0527018 & 11.65 & 11.25 & 11.16 & 950 & 13.1 & 12.2 & \nodata & \nodata & 0 & \nodata & F\\
\object[COUP 1569]{1569} & 053542.0-052005 & 20.0 & 28.84 & 05354209-0520058 & 14.20 & 13.59 & 13.33 & 958 & 18.7 & 15.7 & 0.3 & -3.70 & 0 & M4 & F\\
\object[COUP 1613]{1613} & 053556.8-052527 & 21.8 & 28.14 & 05355672-0525263 & 13.15 & 12.53 & 12.48 & 1037 & 15.2 & 13.9 & 0.0 & -4.84 & 0 & K5 & F\\
\enddata

\tablecomments{Columns 1-2 give the COUP source number and COUP
CXO name.  Columns 3-4 give column density and observed source
luminosity (calculated assuming $d = 450$ pc which will be
incorrect for field stars) inferred from X-ray spectral fits.
Columns 5-8 present the NIR counterpart and its $JHK_s$
photometry. For the inner region, these are from the VLT merged
catalog of McCaughrean et al. (in preparation) while for the outer
region thay are from the 2MASS catalog.  Columns 9-12 give the
optical counterpart identifier from the \citet{Jones88} survey,
$VI$ photometry and visual extinction from \citet{Hillenbrand97}.
Column 13 gives the X-ray emissivity $\log L_t/L_{bol}$ in the
total $0.5-8$ keV band from \citet{Getman05}.  Column 14 lists the
proper motion membership probability from \citet{Jones88}. and
column 15 gives the spectral type from the optical spectroscopy of
\citet{Hillenbrand97} updated as described in \citet{Getman05}.
The final column gives the indicator of field star candidates
suggested by our analysis. $^a$ $L_t$ was calculated assuming
distance of 450 pc which will be incorrect for field stars.}

\end{deluxetable}


\begin{deluxetable}{rrrrrrrrrr} \centerline  \rotate
\tabletypesize{\scriptsize} \tablewidth{0pt}

\tablecolumns{10} \tablecaption{Suggested field stars undetected
in COUP\label{bright_K_undetected_tab}}

\tablehead{ \colhead{2MASS} & \colhead{$CR_u$} & \colhead{JW} &
\colhead{P} & \colhead{$V$} & \colhead{$I$} & \colhead{$J$} &
\colhead{$H$} & \colhead{$K_s$} & Remarks }

\startdata
 05343359-0523099 & 0.092 &  \nodata &\nodata&\nodata &\nodata & 11.28 & 10.78 & 10.72 & $0.7\arcsec$ from JW23 with P=0\\
 05350458-0528266 & 0.035 &      255 &   0   &  15.36 &  13.94 & 13.07 & 12.45 & 12.30 & Par 1738\\
 05352631-0515113 & 0.067 &      794 &   0   &  11.67 &  10.86 & 10.23 &  9.78 &  9.72 & Par 2021; G9IV-V\\
 05353053-0531559 & 0.076 &      858 &   0   &  18.03 &  15.60 & 13.82 & 13.07 & 12.72 & \nodata\\
 05353249-0529021 & 0.164 &      889 &   0   &  14.43 &  13.34 & 12.68 & 12.15 & 12.02 & Par 2098\\
 05354007-0527589 & 0.166 &      947 &  64   &  17.02 &  14.36 & 12.97 & 12.45 & 12.17 & M2:\\
\enddata

\tablecomments{Count rate upper limits $CR_u$ are in counts
ks$^{-1}$ from \citep{Getman05}. Membership probability are from
\citet{Jones88}.  $VI$ photometry are from \citet{Hillenbrand97},
$JHK_s$ photometry from 2MASS, and spectral types from
\citet{Hillenbrand97}.}

\end{deluxetable}

\end{document}